\documentclass{article}

\PassOptionsToPackage{numbers,sort&compress}{natbib}
\usepackage[preprint]{neurips_2026_vericode}

\usepackage[utf8]{inputenc}
\usepackage[T1]{fontenc}
\usepackage{tikz}
\usetikzlibrary{positioning,arrows.meta,calc}
\usepackage{pgfplots}
\pgfplotsset{compat=1.18}
\usepackage{subcaption}
\usepackage{hyperref}       % hyperlinks
\usepackage{url}            % simple URL typesetting
\usepackage{booktabs}       % professional-quality tables
\usepackage{amsmath}
\usepackage{amsfonts}       % blackboard math symbols
\usepackage{nicefrac}       % compact symbols for 1/2, etc.
\usepackage{microtype}      % microtypography
\usepackage{xcolor}         % colors
\usepackage{listings}      % for code listings
\usepackage{graphicx}
\usepackage{placeins}
\usepackage{float}

\newcommand{\code}{\lstinline}

\usepackage{titlesec}
\titlespacing*{\paragraph}{0pt}{1.5ex plus 0.5ex minus .2ex}{1em}

\usepackage{xcolor}
\definecolor{keywordcolor}{rgb}{0.7, 0.1, 0.1}   % red
\definecolor{tacticcolor}{rgb}{0.0, 0.1, 0.6}    % blue
\definecolor{commentcolor}{rgb}{0.4, 0.4, 0.4}   % grey
\definecolor{symbolcolor}{rgb}{0.0, 0.1, 0.6}    % blue
\definecolor{sortcolor}{rgb}{0.1, 0.5, 0.1}      % green
\definecolor{attributecolor}{rgb}{0.7, 0.1, 0.1} % red
\definecolor{varcolor}{rgb}{0.42, 0.20, 0.56}    % muted violet, pseudocode variables
\definecolor{fncolor}{rgb}{0.05, 0.35, 0.55}     % steel blue, pseudocode procedures
\definecolor{stringcolor}{rgb}{0.6, 0.35, 0.0}   % brown, string literals
\definecolor{numbercolor}{rgb}{0.0, 0.45, 0.45}  % teal, number literals
\definecolor{defcolor}{rgb}{0.55, 0.0, 0.45}     % magenta, top-level definition names
\definecolor{selfcolor}{rgb}{0.6, 0.5, 0.0}      % olive, Python's `self`

\makeatletter
\let\old@lstKV@SwitchCases\lstKV@SwitchCases
\def\lstKV@SwitchCases#1#2#3{}
\makeatother
\usepackage{lstlinebgrd}
\makeatletter
\let\lstKV@SwitchCases\old@lstKV@SwitchCases
\lst@Key{numbers}{none}{%
    \def\lst@PlaceNumber{\lst@linebgrd}%
    \lstKV@SwitchCases{#1}%
    {none:\\%
     left:\def\lst@PlaceNumber{\llap{\normalfont
                \lst@numberstyle{\thelstnumber}\kern\lst@numbersep}\lst@linebgrd}\\%
     right:\def\lst@PlaceNumber{\rlap{\normalfont
                \kern\linewidth \kern\lst@numbersep
                \lst@numberstyle{\thelstnumber}}\lst@linebgrd}%
    }{\PackageError{Listings}{Numbers #1 unknown}\@ehc}}
\makeatother

\lstdefinestyle{pseudo}{
  language={},
  basicstyle=\ttfamily\small,
  morekeywords={for,each,in,if,return,and,or,not,of,every,union},
  keywordstyle=\color{keywordcolor}\bfseries,
  emph={usedVars,boundVars,nonlocals,comprehensionAndLambdaTargets,params},emphstyle={\color{fncolor}\bfseries},
  emph=[2]{free,caps,mutated,ordered,threaded,inner,outer},
  emphstyle=[2]{\color{varcolor}},
  comment=[l]{//},commentstyle=\color{commentcolor}\itshape,
  mathescape=true,breaklines=true,columns=fullflexible,keepspaces=true,
  numbers=left,numberstyle=\tiny\color{gray},numbersep=8pt,xleftmargin=14pt
}
\lstdefinestyle{lean}{language=lean,basicstyle=\ttfamily\small,breaklines=true,columns=fullflexible,keepspaces=true,showstringspaces=false,extendedchars=true}
\lstdefinestyle{py}{
language=Python,
basicstyle=\ttfamily\small,
breaklines=true,
columns=fullflexible,
keepspaces=true,
showstringspaces=false,
sensitive=true,
keywordstyle=[1]{\ttfamily\color{keywordcolor}\bfseries},
keywordstyle=[2]{\ttfamily\color{sortcolor}},
morekeywords=[3]{self},
keywordstyle=[3]{\ttfamily\color{selfcolor}},
commentstyle={\ttfamily\itshape\footnotesize\color{commentcolor}},
stringstyle=\ttfamily\color{stringcolor},
moredelim=[is][\color{defcolor}]{@@}{@@},
literate=
{0}{{\color{numbercolor}0}}1
{1}{{\color{numbercolor}1}}1
{2}{{\color{numbercolor}2}}1
{3}{{\color{numbercolor}3}}1
{4}{{\color{numbercolor}4}}1
{5}{{\color{numbercolor}5}}1
{6}{{\color{numbercolor}6}}1
{7}{{\color{numbercolor}7}}1
{8}{{\color{numbercolor}8}}1
{9}{{\color{numbercolor}9}}1,
}

\newcommand{\leftsquigarrow}{\reflectbox{$\rightsquigarrow$}}

\title{PastaLean: Deterministic transpilation from Python to Lean for Verification}

\author{%
  Anirudh Gupta\\
  Indian Institute of Science\\
  Bengaluru \\
  \texttt{anirudhgupta@iisc.ac.in} \\
  \And
  Petros Markopoulos \\
  University of California San Diego \\
  La Jolla \\
  \texttt{pmarkopoulos@ucsd.edu} \\
  \AND
  Siddhartha Gadgil \\
  Indian Institute of Science \\
  Bengaluru \\
  \texttt{gadgil@iisc.ac.in} \\
}

\begin{document}

\maketitle

\begin{abstract}
 We present PastaLean, an extensible system for deterministic translation from Python to Lean via the Abstract Syntax Tree of Python. This allows for verification of Python code, especially auxiliary code used by AI. Every program is translated into a provable twin written in the form Lean's verification tooling can reason about, and a runnable twin that gives a fast and correct computation support. We also introduce our Type Inference engine TypeInfer which provides static typing support for untyped Python code. Our tool provides coverage of more than 92\% for LeetCode and HumanEval\textsuperscript{+} and more than 98\% for LiveCodeBench ground truth Python files. Further, at least 97\% of each dataset's (numerical) test cases pass, with 100\% on HumanEval\textsuperscript{+} and LiveCodeBench. The TypeInfer engine is the best-scoring tool on the TypeEvalPy static leaderboard, and on a repository-level TypyBench benchmark it scores above every competing repository-level inference tool.
\end{abstract}

\section{Introduction}

The rapid evolution of large language models (LLMs) has transformed them from systems primarily designed for natural-language understanding into increasingly capable programming agents that can generate, reason over, and manipulate executable code \cite{jiang2026codegeneration}. Agentic Artificial Intelligence (AI) systems make extensive use of Python for calculations, numerical experiments, data analysis, and other tasks. However, if the generated code is incorrect, the conclusions drawn from it will also likely be incorrect. Formal verification promises something stronger, a machine-checked guarantee that a program meets a specification on \emph{all} inputs. Yet the languages in which such guarantees are routinely established such as Lean~4~\cite{moura2021lean4}, Coq, Dafny~\cite{dafny}, F$^\star$~\cite{swamy2016fstar}, Verus~\cite{lattuada2023verus} are not the languages practitioners actually use. The code that needs checking is usually written in Python.

To address this issue, we introduce \emph{PastaLean}\footnote{The name ``PastaLean'' is derived from Py + AST + Lean.}, a system that provides \emph{deterministic} translation from Python to Lean. This enables verification of Python code.

PastaLean implements a recursive, extensible framework for transpiling Python code to Lean. Python code is first parsed into an Abstract Syntax Tree (AST) using Python's built-in \code{ast} module. The AST is then traversed recursively, and each node is translated into JSON using a set of translation functions with dynamic dispatch. The JSON is then converted into Lean code recursively using an extensible collection of translation functions based on Lean meta-programming. Each of these functions has output \emph{Typed Syntax} in Lean, so the compiler ensures that the output is syntactically correct. Further, a front end is used to verify that the output compiles. 

The specification is written as a contract in the Python source itself, by a developer or by a model, and the same deterministic translator turns it into a Lean theorem alongside the program. We write contracts in Python rather than directly in Lean because both authoring and auditing a contract is easier in Python than in Lean. Contracts are meant to be written by agentic loops as well as developers, and both are considerably more capable in Python than in Lean, especially smaller models; a contract authored directly in Lean is also harder to check against the intended property, since doing so means reading Lean rather than the language the developer already knows. Writing the contract in Python instead and passing it through the same fixed, independently verified translator used for the program body means that trusting the resulting Lean theorem reduces to trusting one translation step we have already vetted, rather than trusting freeform Lean text directly.

To ensure correctness of the Python code, it is crucial that the translation is deterministic -- an LLM may translate incorrect Python code to correct Lean code. The verification depends on the correctness of the translation code, but this is a fixed, open-source code base that can be verified independently. We also use differential testing to enhance trust in the translation. Further, an error in translation is highly unlikely to translate incorrect Python code to correct Lean code, and our main goal is to ensure that auxiliary code used by LLMs is correct.

Agentic AI systems typically generate and run a lot of exploratory code, only a small fraction of which is used in the final output. Only the code that is used in the final output needs to be verified. This means the verification adds only a small overhead to the overall computation. Furthermore, verification and exploration can be parallelized, so it does not slow down the overall computation.

Our use of Python, rather than directly generated Lean code, is motivated by its familiarity to both developers and LLMs. An alternative approach is to use a system like Dafny to generate code directly in a verified language. However, in that case if a specification is mistranslated, the generated code may be correct with respect to the specification, but incorrect with respect to the intended behavior.

\paragraph{Contributions.} These include: \textbf{A deterministic framework for translating Python into Lean}, emitting two definitions per function: a \emph{provable} twin in the form Lean's verification tooling reasons about, and a \emph{computable} twin that runs at speed and agrees with CPython; \textbf{Verification machinery for Python}, in which contracts written in the source become Hoare-triple~\cite{hoare1969axiomatic} theorems over the translated program; \textbf{TypeInfer}, a high performance type inference engine for untyped Python whose lattice and fixpoint are machine-checked in Lean; and \textbf{a type checker that shows its work}, emitting for every file a certificate of the evidence behind each verdict, re-checkable by a validator that runs none of the inference engine and backed by a Lean theorem saying what an accepted certificate means.

\paragraph{Limitations} We do not support all of Python during transpilation. Further, our proof automation is best-effort and may need manual interventions or may fail. Limitations are stated where they arise: Sections~\ref{sec:oop}, \ref{sec:proofautomation}; Appendices~\ref{app:twins}, \ref{app:contracts}, \ref{oop-appendix}, \ref{app:scope}, \ref{app:unsupported}, \ref{app:ti-proofs}, \ref{app:pyany}, \ref{app:ti-speed}, \ref{app:ti-limits}, \ref{app:cert-theorem}, \ref{app:cert-limits}, \ref{app:eval-cp}, \ref{app:conv-speed}, \ref{app:codegen-erros}, \ref{app:eval-type}.

% \begin{itemize}\itemsep2pt \parskip0pt
% \item \textbf{A deterministic framework for translating Python into Lean 4}, emitting two definitions per function: a \emph{provable} twin in the form Lean's verification tooling reasons about, and a \emph{computable} twin that runs at speed and agrees with CPython.
% \item \textbf{Verification machinery for ordinary Python}, in which contracts written in the source become Hoare-triple theorems over the translated program itself, discharged by loop invariants and a tactic portfolio rather than against a hand-written model.
% \item \textbf{TypeInfer}, a high performance type inference engine for untyped Python whose lattice and fixpoint are machine-checked in Lean, and which supplies the concrete types to Python codes, making the code more provable.
% \end{itemize}

\paragraph{Related Work.}

The work closest to PastaLean is Aeneas, which translates safe Rust, via the LLBC intermediate representation, into pure functional models for verification in Lean and backends~\cite{aeneas}. PastaLean follows the same broad strategy of verifying source programs through functional translation. However, Python lacks the static type and ownership information exploited by Aeneas and requires support for dynamic typing, mutation, exceptions, and its object model. The rocq-of-python project also translates Python into a proof assistant, targeting Rocq and using a dedicated representation of Python semantics~\cite{rocq-of-python}. PastaLean instead produces verification-friendly shallow embeddings in Lean, allowing generated definitions to use Lean and Mathlib~\cite{mathlib2020} libraries directly.

Other systems verify Python without translating it into an interactive theorem prover. Nagini translates statically typed Python into Viper and uses SMT-based automation for modular verification~\cite{nagini}, while CrossHair uses symbolic execution to search for counterexamples to Python contracts~\cite{crosshair}. Dafny takes the complementary approach of writing programs directly in a verification-aware language~\cite{dafny}. Strata, also built on Lean~4, instead compiles a Python front end through its intermediate Laurel language into verification conditions discharged by SMT solvers~\cite{strata}. PastaLean is distinguished by starting from ordinary Python and producing Lean code through a deterministic, extensible translator.

Inferring types for (untyped) Python is a classical problem: whole-program inference underpins the Python compilers Starkiller~\cite{salib2004starkiller} and Shed Skin~\cite{dufour2006shedskin} and PyPy's RPython annotator~\cite{rigo2006pypy}, all descendants of Agesen's Cartesian Product Algorithm~\cite{agesen1995cpa}, while recent work is based on learning rather than analytic~\cite{mir2022type4py} or call-graph driven~\cite{venkatesh2024headergen}. \code{TypeInfer} differs in being written and proved inside the prover it feeds. Recent LLM-based systems further motivate this design. PyVeritas uses an LLM to translate Python to C before applying bounded model checking~\cite{pyveritas}, while VeriBench evaluates agents that generate Lean implementations, specifications, and proofs from Python~\cite{veribench}. In PastaLean, translation of the executed program is deliberately kept outside the LLM: the specification and proof may come from a person or a model, drawing on the growing body of work on autoformalization and neural proof search~\cite{wu2022autoformalization,jiang2023draft,xin2024deepseekprover,ren2025deepseekproverv2}, but a fixed translator preserves the connection to the original Python code, and Lean checks the resulting proof.
\section{Transpilation and Verification}

Python's dynamic typing, mutable state, multiple returns, and object model make transpiling it to Lean nontrivial, and different translation choices affect how easy the result is to verify.

We outline our basic approach to transpilation, and then discuss how we address many of the issues.

\subsection{Recursive Transpilation Framework}\label{sec:framework}

Given Python code, we first parse it into an Abstract Syntax Tree (AST) using Python's built-in \code{ast} module. The AST is then traversed recursively, and each node is translated into JSON using a set of translation functions with dynamic dispatch.

The core of our approach is to recursively build Lean code using meta-programming in Lean~\cite{ullrich2019macros}, using functions registered in an \emph{environment extension}. Concretely, the Json corresponding to a Python expression, function or module is translated into Lean by calling the function \code{getCode (json: Json) (kind: SyntaxNodeKind) : PygenM <| TSyntax kind} which takes as input the Json and the \emph{kind} of syntax node to be generated, and returns in a monad the corresponding Lean code. 

The Monad \code{PygenM} is a state monad transformer over the \code{TermElabM} monad, which is the monad used for elaborating terms in Lean. The state contains information relevant to the generated Lean code. For instance, the first occurrence of a mutable variable \code{n} has to be translated (within a \code{do} block) into a \code{let mut n := ...} do element, while subsequent occurrences of the variable are translated into do elements of the form \code{n := ...}. The state keeps track of which variables have been declared as mutable, so that subsequent occurrences can be translated correctly.

The function \code{getCode} expects a string as the field \code{node_type} in the Json, which is used to look up the corresponding translation function in the environment extension \code{pygenExt}. A functions corresponding to the node type "type" is added to the environment extension by annotation with the attribute \code{[pygen type]}.
These functions can call \code{getCode} recursively to translate sub-expressions.

\paragraph{Transpilation targets.} \label{sec:transpilation-targets}

We map Python into a \emph{shallow embedding} in Lean, i.e., we use Lean's own types and functions to represent Python types and functions wherever possible. This allows the use of Lean's existing libraries for verification. Indeed, we use two shallow embeddings of Python in Lean, one for verifiability and one for computability, so in general we get two translations that differ slightly. One of these generates computable Lean code that can be executed, for instance using \code{Float} (which is not completely specified and does not have many theorems that are proved, or even true), while the other generates Lean code that is more amenable to verification, using \code{Real} in place of \code{Float}, preferring \code{List} over \code{Array} (Appendix~\ref{sec:array-vs-list}), and using standard Lean functions as far as possible. The former is used to verify the correctness of the latter, and to run the code for testing. Appendix~\ref{app:twins} lists the places they diverge, including the \code{List} versus \code{Array} choice of Appendix~\ref{sec:array-vs-list}, and Appendix~\ref{naming-convention} lists the naming conventions that keep the two apart.

Furthermore, we attempt to generate the most \emph{verification friendly} form of code possible. For instance, we prefer to use the functional style and recursion. Where we need to use a Monad only to support mutability, and \code{while} loops, we use the \code{Id} Monad. We use custom state and exception monads rather than \code{IO} where possible since \code{IO} supports too many side-effects.

\subsection{Supporting Python}

We sketch below some of the main issues in transpiling Python to Lean, and how we address them. Our goal is to support scripts such as those generated for auxiliary computations by LLMs, which are often small and self-contained, and to support verification of these scripts.

\paragraph{Arithmetic Operations.}\label{sec:arith}

Python's arithmetic operators are polymorphic, i.e., the same operator can work with different data types depending on the operands. For example, \code{+} performs numeric addition for integers and concatenation for strings, while Booleans behave like integers (True + True = 2, False + 3 = 3). 
Lean supports polymorphic operators through typeclasses. However, the semantics of Python's operators differ from Lean's. To ensure that semantics are preserved, we define a set of typeclasses that capture the behavior of Python's operators and corresponding syntax in Lean. For instance, we implement a \code{PyHAdd} typeclass for Python's addition operator, and syntax $+_p$ for addition with Python's semantics. A (default) instance delegates to Lean's native \code{HAdd} where applicable, allowing us to use theorems about addition in Lean.

Operations follow the numeric widening hierarchy (Section \ref{numeric-widening}), with the result taking the wider operand's type. We implement this with a single class \code{PyNumJoin} \ref{app:arith}, which records how any two numeric types convert into a common one.

\paragraph{Variables and Mutability.}
Variables in Python are mutable, meaning that their values can be changed after they are created. In Lean, variables are immutable by default, meaning that once a value is assigned to a variable, it cannot be changed. However, Lean does support mutable variables in monads~\cite{ulrich2022dounchained} through the use of the \code{let mut} construct, which allows us to create mutable variables.

In simple cases, mutability can be translated simply to reassignments in Lean giving functional, easy to verify, code. We generate such code whenever possible.
Otherwise, we use a monadic translation by generating a \code{do} block, i.e., a sequence of \emph{do elements}, if possible using just the \code{Id} monad.

Unlike Python, Lean has different syntax for first and subsequent assignments of mutable variables, which we handle as outlined in Section \ref{sec:framework}. In case of type mutation of a variable, we follow an SSA based approach of renaming the variables. See~\ref{sec:ssa} for more details.

\paragraph{Loops, State, Exceptions.}\label{sec:loops}

For Python code having \code{while} or \code{for} loops, state or exceptions, we use a monadic translation. We can build a monad supporting state and exceptions using the monad transformers \code{StateT} and \code{ExceptT}. We can then translate the Python code into a \code{do} block in that monad. Such a Monad is supported by Lean's \emph{mvcgen} and \emph{vcgen} tools~\cite{leanmvcgen2026}, which can be used to verify the code. We support Python's \code{try}, \code{except}, \code{finally} syntax for catching exceptions similarly.
We avoid using the \code{IO} monad, though it supports state and exceptions, as it makes verification hard.

\paragraph{Python's Standard functions and Libraries.}

Python's standard functions need to be translated to Lean code, which may not be part of Lean's standard libraries (including Mathlib). We implement the more commonly used functions in Lean, and map the Python function names to them. 
Our \code{PyAPI} library has over 200 Lean definitions covering the builtins (\code{len}, \code{sorted}, \code{abs}, \code{divmod}, \code{zip}, \code{sum}, \code{min}, \code{max}, \code{chr}, \code{ord} and the \code{int}, \code{str}, \code{float}, \code{list}, \code{set} and \code{dict} constructors, among others) together with over 60 string, list, dictionary and set methods such as \code{split}, \code{join}, \code{strip}, \code{append} and \code{pop}. These are idiomatic Lean implementations (and not foreign calls), so we can verify code that uses them.

Two lookup tables relate the two languages. One is a \code{match} on names, so \code{len(xs)} becomes \code{pyLen xs}, and the other does the same for method calls, so \code{s.split(",")} becomes \code{pyStringSplit s ","}. To add a new builtin, we add a Lean definition, a match arm in the name lookup, and tests using \code{#eval} that the Lean implementation coincides with the Python one (see Appendix~\ref{app:stdfuncs}).

Useful Python code needs many other libraries, like numpy, pandas etc. Just as with the standard library, we have implemented many functions in these in Lean and add them to the lookup tables (see Appendix~\ref{app:libraries}).

\paragraph{Polymorphic Functions.}\label{sec:poly}

There are two sources of Poylmorphism in Python: library functions and user-defined functions. Library functions such as \code{len}, \code{min} and \code{max} support a fixed, known set of types. These are implemented in Lean using type classes, with one instance per supporting type. These support proofs very well.
A user-written function has no fixed set of types for which to implement instances. We must instead use type inference, as we discuss in Section~\ref{sec:typeinfer}.

\paragraph{Classes and Python OOP.} \label{sec:oop}

While Lean does not have objects and methods, namespaces and the \emph{dot notation} can be used to emulate them.We translate a Python \code{Class} into a Lean \code{structure} paired with a \code{namespace} of functions. Class attributes map directly to structure fields, the constructor becomes \code{ClassName.new}, and methods translate to standalone functions that explicitly take \code{self} as an argument. To handle state changes in a purely functional way, we use \emph{value semantics}: modifying \code{self} simply returns an updated object, which is then rebound at the call site. Finally, Python dunder methods (e.g., \code{__add__}) are mapped to Lean's operator type classes, and single inheritance is achieved by embedding the parent structure. See Figure \ref{lst:oop-lean} as a concrete example of a converted code, and Appendix~\ref{oop-appendix} for the full translation rules. 

What we do not support is \emph{dynamic} dispatch: an overridden method is resolved statically, so a base-typed variable holding a subclass instance calls the base method. \code{super()} and metaclasses are also unsupported. Appendix~\ref{oop-appendix} lists what is translated and what is not. 

\paragraph{Nested Definitions and Closures.}

Unlike Python, Lean does not support nested functions. Hence, we take a nested \code{def} and lift it to a sibling \code{private partial def} with the closure's captured names as additional parameters. This is the standard \emph{lambda lifting} transformation~\cite{johnsson1985lambdalifting}. 

We use sibling definitions rather than \code{where} or \code{let rec}, since either would force the \emph{enclosing} \code{def} to be \code{partial}, which makes verification much harder (see Section \ref{app:lifting} for details).

The lifted function needs the correct set of parameters. We pass the helper exactly those variable names it reads that belong to the enclosing function and are not globals or builtins. We present details of the algorithms in Appendix \ref{app:lifting}, and the machine-checked scope-preservation result in Appendix~\ref{app:lifting-proof}.

\paragraph{Multiple Returns.}\label{sec:multiret}

A Python function may have multiple return statements, each with a potentially different type, and can also return nothing. We take the return type to be the join of the branch types over the lattice (Section \ref{sec:typeinfer}), so most disagreements resolve by numeric widening \ref{numeric-widening}. Branches mixing \code{int} and \code{float} widen to \code{float}. A branch returning nothing forces an \code{Option} type. Genuine lattice conflicts, such as \code{str} on one path and \code{int} on another, reach the top of the lattice and are boxed as \code{PyAny}, dispatched on its runtime tag (Section \ref{sec:pyany}). 

We will address other approaches: returning tagged unions or making the return type a function, in future work.

\paragraph{Reference Semantics and the Heap.}\label{sec:heap}
Python passes mutable values by reference. Class instances, lists, dictionaries and sets are handles to
a shared object, so \code{b = a} creates a second name for the same thing and a mutation through one is
visible through the other; only immutable values --- numbers, strings, tuples --- get copied. The translation of Section~\ref{sec:oop} takes the second reading for everything: an
object is a Lean structure, and a mutating method returns an updated copy that is rebound at the call
site. This is faithful whenever an object has a single handle, which is common in the code we
target, and it keeps the program in the pure fragment where Lean's equational reasoning applies
directly. It can be wrong if the program ever aliases a variable which holds a mutable object.

We therefore provide reference semantics as a second, opt-in translation tier (\code{--heap}), rather
than changing the default, so that a program pays for the heap only when it needs one. A heap is a
partial map from addresses to values, and a translated program that uses it runs in \code{HeapM}, a
monad carrying that heap alongside Python's exceptions. Every value Python passes by reference is
allocated in the heap and represented by a typed reference \code{Ref }$\alpha$; attribute access and attribute
assignment become reads and writes through that reference; and aliasing works out of the box, because
\code{b = a} now copies an address. A whole-program pass collects the types that can be stored into a
single value universe, and an interprocedural effect analysis decides which functions need the monad
at all, so pure helpers keep their pure translation. Pointer notation keeps the emitted code close to
its source. Appendix~\ref{app:heap} goes through an example in more detail.

\noindent
\begin{minipage}[t]{0.35\textwidth}
\begin{lstlisting}[style=py,caption={Python source.},label={lst:heap-py},breaklines=true]
def @@move@@(self, dx, dy):
    self.x = self.x + dx
    self.y = self.y + dy
\end{lstlisting}
\end{minipage}%
\hfill
\begin{minipage}[t]{0.65\textwidth}
\begin{lstlisting}[language=lean,caption={Generated \code{--heap} translation.},label={lst:heap-lean},breaklines=true,mathescape=true]
def @@Point.move@@ (self : Ref Point) (dx dy : Int) :=
  ((do self $\rightsquigarrow$ x $\leftsquigarrow$ ($\leftarrow$ (self $\rightsquigarrow$ x)) +$_p$ dx
       self $\rightsquigarrow$ y $\leftsquigarrow$ ($\leftarrow$ (self $\rightsquigarrow$ y)) +$_p$ dy)
   : HeapM Val Unit)
\end{lstlisting}
\end{minipage}

Specifications for heap programs are separation-logic Hoare triples over the points-to assertion
($\mapsto$) and the separating conjunction ($\ast$), and Lean's \code{vcgen} discharges them
with automatic framing, so an object the callee does not touch is carried across the call with no manual reasoning.

\paragraph{IO.}\label{sec:io}

Lean's \code{IO} is a grab-bag of functionalities. One monad covers files, processes, threads, mutable references,
randomness and foreign calls, all behind an opaque token standing for the real world. This generality helps ordinary programming but makes proving cumbersome. We implement a custom monad supporting only the IO operations we need, combining state with Python's exceptions. Appendices~\ref{app:twin-io} and~\ref{app:twin-exc} give the provable twin's treatment of both. 

\subsection{Examples of Transpilation}\label{sec:transpilation}

Appendix~\ref{app:full-example} follows one function through every stage of the pipeline, and Appendix~\ref{app:more-conversions} covers exceptions~(\ref{app:conv-exceptions}), IO~(\ref{app:conv-io}), decorators~(\ref{app:conv-decorators}), memoization~(\ref{app:conv-memo}) and generators~(\ref{app:conv-generators}). We give two examples showing different features of PastaLean. The first example is pure and recursive (Listing~\ref{lst:ex-quicksort}), with no mutations or side-effects. The corresponding program is pure, with comprehensions mapping to \code{List.filter} (see Section \ref{app:conv-comprehension}). \code{TypeInfer} soundly infers the types of variables and parameters from the program flow to annotate them correctly.

\noindent
\begin{minipage}[t]{0.40\textwidth}
\begin{lstlisting}[style=py,basicstyle=\ttfamily\scriptsize,caption={Gives a pure and recursive implementation.},label={lst:ex-quicksort}]
def @@quicksort@@(xs):
    if len(xs) <= 1:
        return xs
    pivot = xs[0]
    rest = xs[1:]
    lo = [x for x in rest if x < pivot]
    hi = [x for x in rest if x >= pivot]
    return quicksort(lo) + [pivot] \
         + quicksort(hi)

\end{lstlisting}
\end{minipage}\hfill
\begin{minipage}[t]{0.58\textwidth}
\begin{lstlisting}[style=lean,basicstyle=\ttfamily\scriptsize]
partial def @@quicksort@@ : List Int (*@\ensuremath{\rightarrow}@*) List Int :=
 fun (xs : List Int) =>
  if pyLen xs (*@\ensuremath{\le}@*) (1 : Int) then xs
  else
   let pivot := (xs(*@\ensuremath{[\![}@*)(0 : Int)(*@\ensuremath{]\!]}@*) : Int)
   let rest := (pySlice xs (some (1 : Int)) none none : List Int)
   let lo := (List.filter (fun (x : Int) => x < pivot)
                (pyIter rest) : List Int)
   let hi := (List.filter (fun (x : Int) => x (*@\ensuremath{\ge}@*) pivot)
                (pyIter rest) : List Int)
   quicksort lo +(*@\ensuremath{{}_{\mathrm p}}@*) [pivot] +(*@\ensuremath{{}_{\mathrm p}}@*) quicksort hi
\end{lstlisting}
\end{minipage}

% The second example is a Binary Search (Listing~\ref{lst:ex-bsearch}) which loops(\code{while}) until the element is found or returns \code{-1}.

% \noindent
% \begin{minipage}[t]{0.40\textwidth}
% \begin{lstlisting}[style=py,basicstyle=\ttfamily\scriptsize]
% def binary_search(xs, target):
%     lo = 0
%     hi = len(xs) - 1
%     while lo <= hi:
%         mid = (lo + hi) // 2
%         if xs[mid] == target:
%             return mid
%         if xs[mid] < target:
%             lo = mid + 1
%         else:
%             hi = mid - 1
%     return -1
% \end{lstlisting}
% \end{minipage}\hfill
% \begin{minipage}[t]{0.58\textwidth}
% \begin{lstlisting}[style=lean,basicstyle=\ttfamily\scriptsize,caption={A loop: mutation and early return inside a pure \code{Id} monad.},label={lst:ex-bsearch}]
% def binary_search := fun (xs : List Int) (*@\ensuremath{\mapsto}@*) fun (target : Int) (*@\ensuremath{\mapsto}@*)
%  Id.run
%   (do
%    let mut lo : Int := (0 : Int)
%    let mut hi : Int := pyLen xs -(*@\ensuremath{{}_{\mathrm p}}@*) (1 : Int)
%    while (lo (*@\ensuremath{\le}@*) hi) do
%      let mut mid : Int := pyFloorDiv (lo +(*@\ensuremath{{}_{\mathrm p}}@*) hi) (2 : Int)
%      if h_1 : xs(*@\ensuremath{[\![}@*)mid(*@\ensuremath{]\!]}@*) = target then
%        return mid
%      if h_2 : xs(*@\ensuremath{[\![}@*)mid(*@\ensuremath{]\!]}@*) < target then
%        lo := mid +(*@\ensuremath{{}_{\mathrm p}}@*) (1 : Int)
%      else
%        hi := mid -(*@\ensuremath{{}_{\mathrm p}}@*) (1 : Int)
%    let p'_ret_1 := -(1 : Int)
%    return p'_ret_1)
% \end{lstlisting}
% \end{minipage}

An example where exceptions are handled (Listing~\ref{lst:ex-exc}) is \code{linear_regression} from Python's \code{statistics} module. This guards its inputs with \code{raise} and turns a division by a constant column into a \code{ValueError}. A \code{raise} becomes a \code{throw} with message including the exception's class name, and \code{try}/\code{except} becomes a \code{try}/\code{catch} that tests that name and re-raises anything it does not handle. As raising is the only effect, so the function returns in \code{PyProofM}, an exception monad over pure state (not \code{IO}). We  have modified the standard library's code replacing \code{fsum} by \code{sum} and shortening messages.

\begin{figure}[t]
\noindent
\begin{minipage}[t]{0.46\textwidth}
\begin{lstlisting}[style=py,basicstyle=\ttfamily\fontsize{7.5}{9}\selectfont,breaklines=false]
def @@linear_regression@@(x, y):
    n = len(x)
    if len(y) != n:
        raise ValueError('lengths differ')
    if n < 2:
        raise ValueError('need two points')
    xbar = sum(x) / n
    ybar = sum(y) / n
    sxy = sum((xi - xbar) * (yi - ybar)
              for xi, yi in zip(x, y))
    sxx = sum((xi - xbar) * (xi - xbar)
              for xi in x)
    try:
        slope = sxy / sxx
    except ZeroDivisionError:
        raise ValueError('x is constant')
    intercept = ybar - slope * xbar
    return slope, intercept
\end{lstlisting}
\end{minipage}\hfill
\begin{minipage}[t]{0.52\textwidth}
\begin{lstlisting}[style=lean,basicstyle=\ttfamily\fontsize{7.5}{9}\selectfont,breaklines=false,caption={\code{statistics.linear\_regression}},label={lst:ex-exc}]
def @@linear_regression@@ :=
 fun (x : List Rat) (*@\ensuremath{\mapsto}@*) fun (y : List Rat) (*@\ensuremath{\mapsto}@*)
  ((do
    let mut n : Int := pyLen x
    if h_1 : pyLen y (*@\ensuremath{\ne}@*) n then
      throw (PyException.Raise "ValueError"
               "lengths differ")
    ...
    let mut slope := default
    try
      slope := sxy /(*@\ensuremath{{}_{\mathrm p}}@*) sxx
    catch caught =>
      if (caught).OfKind == "ZeroDivisionError"
      then
        throw (PyException.Raise "ValueError"
                 "x is constant")
      else
        throw caught
    ...
    return (slope, intercept)) :
   PyProofM _)
\end{lstlisting}
\end{minipage}
\end{figure}

\subsection{Dynamic Typing and Type Inference}
\label{sec:typeinfer}

Python's dynamic typing is easy to use for users, yet a nightmare for Software Verifiers. Python supports type mutations of a variable as well as multiple return types of a function. While types are dynamic, Python is strongly typed, i.e., types are strictly enforced at runtime.

We implement a Type inference engine (\code{TypeInfer}) which tries to resolve the Type from a given AST. The key idea is to assign to every Intermediate Representation (IR) node a \code{PyType} drawn from a lattice which resolves into a Python's numeric tower into Lean Types, or falls back to a dynamic type only when it must. Type inference is necessary for code generation since Lean needs a type for every binder. Falling back to dynamic types limits how much of the program can be proved: a concrete type such as \code{List Int} or $\mathbb{Q}$ has algebraic structures that allow tactics like \code{ring} and \code{omega} apply. A type cannot simply be read off a \code{def}, because Lean fixes a definition's parameter types before it looks at the body, so recovering them requires program analysis over the whole module. TypeInfer is a self-contained Lean library (see Appendix~\ref{app:typeinfer}) that takes the JSON IR of a module and returns the same IR with a type written onto every binder it could work out, recorded as an ordinary annotation; nothing else in the tree changes. Code generation then reads an inferred type in exactly the place it reads one the user wrote (see Appendix \ref{app:ti-io}).

\begin{lstlisting}[caption={The \texttt{PyType} lattice.},label={lst:pytype_lattice}, mathescape=true]
inductive @@PyType@@ where
  | unknown              -- $\bot$ : nothing known yet
  | any                  -- $\top$ : conflicting types, boxed as PyAny
  | int | bool | str | float | none
  | list  (elem : PyType) | set (elem : PyType)
  | tuple (elems : List PyType) | dict (key val : PyType)
  | opt   (inner : PyType)          -- Optional[T]
  | cls   (name : String)           -- a user class, e.g. TreeNode
  | fn    (args : List PyType) (ret : PyType)
\end{lstlisting}

We associate to each value a term in \code{PyType}. The term \code{unknown} is the bottom, representing no known constraint. The term \code{any} is the top, meaning the value could have any type. We define a \code{join} operation on \code{PyType}. To infer the type using this, we initially take the inferred type to be \code{unknown} and join in whatever can be inferred from the program, until nothing changes. The operation \code{join} is monotone, so the analysis is a monotone dataflow fixpoint~\cite{cousot1977abstract} and always terminates. Appendix \ref{app:ti-algo} gives details of the algorithm used for Type Inference.

To facilitate verification, it is important to avoid falling back to \code{PyType.any} by supporting the semantic of Python's types. Python promotes silently, so \code{True + 1} is \code{2} and a list holding both \code{5} and \code{float(`inf')} is a list of floats. To support this, we make \code{join} widen along Python's tower, $\mathtt{bool} < \mathtt{int} < \mathtt{float}$, so a mixed container keeps a concrete element type and the smaller operand is coerced up. A flat domain would send \code{join(int, float)} to the top and box the whole container (Appendix \ref{numeric-widening}). Types cross function boundaries both ways: return types flow out to the call sites, and the argument types at every call site are joined into the callee's parameters (Appendix \ref{app:ti-techn}).

In two cases a join is not sufficient. A rebinding can change a variable's type, for instance \code{x = "5"} followed by \code{x = int(x)}, and one Lean binder cannot hold both, so an SSA pass versions the variable at those points~\cite{cytron1991ssa} (Appendix \ref{sec:ssa}). The second case is when a value really has no single type. Here the join reaches the top and we box it as \code{PyAny}, the dynamic type of gradual typing~\cite{siek2006gradual}. This makes proving hard, so we only fall back to this finally (Appendix \ref{sec:pyany}); Appendix~\ref{app:pyany} covers how a boxed value computes~(\ref{app:pyany-runtime}) and what can still be proved about one~(\ref{app:pyany-cases}). Note that \code{TypeInfer} is not a heuristic that guesses the type -- it infers based on the code. As shown with HumanEval dataset examples, if the ambiguity between types to infer is real, then \code{TypeInfer} reports it honestly.

As our engine is written in Lean, we can prove that it has desired properties. We prove the important properties from the literature on lattice-based analyses: that the domain is a bounded join-semilattice and \code{join} is its least upper bound~\cite{davey2002lattices}; that \code{join} is monotone, so by Knaster-Tarski the least fixpoint exists and is unique, and the iteration reaches it in finitely many steps because a program mentions only finitely many types~\cite{tarski1955fixpoint}; that widening is sound, so a value of the inferred type still inhabits the widened one~\cite{milner1978polymorphism,wright1994soundness}; and that consistency for the dynamic type is reflexive, symmetric and \emph{not} transitive, which is what separates gradual typing from subtyping~\cite{siek2015refined}. Appendix \ref{app:ti-proofs} states each theorem and the paper that establishes it as the right requirement.

\code{TypeInfer} outperforms every other static tool on the TypeEvalPy micro-benchmark and leads the production type checkers on accuracy, speed and memory at the repository level, the more demanding of the two settings. (see Section \ref{results}). Appendix \ref{app:ti-speed} discusses reasons for the engine's speed.

\subsection{Type Checking with Certificates}\label{sec:typecheck}

Inference asks what type a slot has. Checking asks whether that slot's uses can be satisfied at all, and whether they agree with a type the user wrote. For a verification pipeline the second question carries as much weight as the first: a variable whose uses contradict each other has no Lean type to emit, and an annotation the body contradicts yields Lean that elaborates against a type the program never respects, so a proof over that definition would be a proof about the wrong program.

The judgment reads the same evidence as inference and combines it the other way round. Inference joins, because a value widens what a slot may hold; checking \emph{meets}, because a use narrows what a slot can be. We write the requirement domain as \code{Req = Option PyType}, with \code{none} for ``nothing required'' and \code{some unknown} for ``contradiction'', lift \code{meet} over that extra point, and prove the result a bounded meet-semilattice together with \code{hasType_commit}: committing the meet of a slot's requirements satisfies every one of its uses (Appendix~\ref{app:ti-proofs}). A verdict's independence from the order its uses were read is then a theorem rather than a property of the traversal. A \emph{slot} is a parameter, a return, a call argument, an annotated binder or a local variable, and each receives one of five verdicts under one of two judgments (Appendix~\ref{app:cert-shape}). Local variables are what let the checker look inside a function rather than only at its boundary, which is the case Listing~\ref{lst:cert-conflict} shows.

\begin{lstlisting}[style=pseudo,caption={A conflict the checker reports. Neither a signature nor an annotation mentions \texttt{buf}, so a boundary-only pass cannot see it.},label={lst:cert-conflict}]
# Python
def mangle(items):
    buf = items[0]
    k = buf.isalpha()   // buf must be a str
    m = buf << 1        // ... and an int. No type is both.

// the checker's output
real_conflict.py:2:5: error: unsatisfiable constraints
  mangle.buf: one use requires String (python: str), another requires
  Int (python: int); no type satisfies both
\end{lstlisting}

A verdict on its own is a claim the reader has to take on faith, so the checker writes a \emph{certificate}: a per-file record naming every slot it judged, the requirements it weighed at that slot, and the verdict it reached. The certificate is the witness and a Lean theorem is the guarantee, which is the verified-validator arrangement of translation validation and of proof-carrying code~\cite{necula1997pcc,pnueli1998translation,leroy2009compcert}. One theorem is proved once and every accepted certificate is an instance of it, so nothing is proved per file and the scheme still runs over a corpus of tens of thousands.

An asymmetry is what makes this cheap. \emph{Producing} a certificate needs the whole engine: the fixpoint, the interprocedural pass, the stamper. \emph{Checking} one needs none of them, because every judgment is a total function of a slot's recorded evidence, so replaying that function either reproduces the recorded verdict or refutes it. The validator is 280 lines and imports only the requirement algebra and the annotation reader, nothing from the engine's solver; keeping that import list short is the file's stated invariant. The trusted base for ``this module has no type error'' is therefore those 280 lines together with the requirement algebra they rest on, rather than the engine's 6{,}400.

Two theorems say what acceptance means, both \code{sorry}-free and resting on \code{propext}, \code{Classical.choice} and \code{Quot.sound} alone. \code{validate_sound}: wherever an accepted certificate says a slot's uses narrow to $t$, every runtime value of type $t$ satisfies every one of those uses. \code{validate_error_sound}: every reported error between two scalar types is a real error, in that no runtime value inhabits both. That second one is not ``the checker disagreed'' but a value-level impossibility, and it is exactly the claim a seeded-error corpus measures. Appendix~\ref{app:cert-theorem} states both and is explicit about their three limits --- the recorded evidence is taken as given rather than traced back to the expression it was read from, \code{HasType} is a model of Python's values and not Python itself, and no theorem connects the inference functions to that model. The certificate routes around the last gap rather than closing it.

\subsection{Verification}

As mentioned in Section~\ref{sec:transpilation-targets}, we generate the most verification-friendly form of code possible. We sketch the ingredients of the verification process here, and give more details in Appendix~\ref{app:contracts}.

\paragraph{Contracts and Monads.}

For monadic programs we use the \code{mvcgen} tactic \cite{lean-mvcgen} and its successor \code{vcgen}. Given a Hoare triple~\cite{hoare1969axiomatic} goal $\lbrace\!\lbrace P \rbrace\!\rbrace\ \mathit{prog}\ \lbrace\!\lbrace Q \rbrace\!\rbrace$, they reduce it to a set of \emph{verification conditions}. Users can register specifications so the tactics replace a call with its pre- and post-conditions instead of unfolding the definition. Control flow is handled through the tactics, leaving the user to provide loop invariants, termination measures, etc. wherever they are needed.

The tactics require the program's monad to carry a weakest-precondition interpretation, via a \code{WPMonad} instance. Lean supplies these for \code{Id}, \code{StateT}, \code{ReaderT}, \code{ExceptT} and \code{EStateM}, but users can provide custom instances for other monads. \code{vcgen} implements the frame rule, parameterised by a frame operator each monad registers along with a frame inference procedure. This makes separation logic usable for our heap-manipulating programs (Section~\ref{sec:heap}).

Appendix~\ref{app:contracts} gives the contract vocabulary in full. Loop \code{Invariant}s provided are parsed and become one \code{mvcgen} bullet over the loop state: this is essential for theorem proving. The approach applies broadly; we present five HumanEval problems proved end-to-end in Appendix~\ref{app:contracts}.

\begin{lstlisting}[language=lean,caption={A specification for \code{Point.move} of Listing~\ref{lst:heap-lean}, and a caller that frames.},label={lst:vcgen-spec},mathescape=true,breaklines=true]
@[spec]
theorem @@Point.move_spec@@ (self : Ref Point) (pt : Point) (dx dy : Int) :
    (*@\ensuremath{\{\!\{}@*) (self (*@\ensuremath{\mapsto}@*) pt : HProp Val) (*@\ensuremath{\}\!\}}@*) Point.move self dx dy
    (*@\ensuremath{\{\!\{}@*) fun _ => self (*@\ensuremath{\mapsto}@*) { pt with x := pt.x +(*@\ensuremath{{}_{\mathrm p}}@*) dx, y := pt.y +(*@\ensuremath{{}_{\mathrm p}}@*) dy } (*@\ensuremath{\}\!\}}@*) := by
  vcgen [Point.move, readRefM_spec, writeRefM_spec] simplifying_assumptions with finish

example (self other : Ref Point) (pt o : Point) (dx dy : Int) :
    (*@\ensuremath{\{\!\{}@*) (self (*@\ensuremath{\mapsto}@*) pt $\ast$ other (*@\ensuremath{\mapsto}@*) o : HProp Val) (*@\ensuremath{\}\!\}}@*) Point.move self dx dy
    (*@\ensuremath{\{\!\{}@*) fun _ => self (*@\ensuremath{\mapsto}@*) { pt with x := pt.x +(*@\ensuremath{{}_{\mathrm p}}@*) dx, y := pt.y +(*@\ensuremath{{}_{\mathrm p}}@*) dy } $\ast$ other (*@\ensuremath{\mapsto}@*) o (*@\ensuremath{\}\!\}}@*) := by
  vcgen [Point.move_spec] with finish
\end{lstlisting}

The precondition is the one cell the function touches, and the postcondition is that cell with both fields updated; the rest of the heap goes unmentioned, since a separation-logic specification describes a footprint, not a whole state. The proof is one invocation of \code{vcgen}, given the program, the specifications for reading and writing through a reference, and \code{finish} to close what remains.

The \code{@[spec]} attribute registers the result, so later calls to \code{move} can use it instead of getting unfolded. The \code{example} is such a caller, holding a second point \code{move} never touches; \code{vcgen} uses a frame rule to ignore it.

\paragraph{Proof Automation.}\label{sec:proofautomation}

PastaLean provides a best-effort tactic called \code{taste?} which uses the definition and important theorems related to the PyAPI of PastaLean. It uses Lean tactics like \code{grind}, \code{simp_all}, \code{nlinarith}, and \code{try?}, and is hence stronger than a simple \code{grind} tactic. We use this tactic as default search tactic. The \code{taste?} tactic is able to complete many simple non-monadic proofs.

% \subsection{Limitations}

% Naturally, not all of Python is supported. Programs that reshape themselves as they run --- \code{eval}, \code{exec}, monkey-patching, modifying a class after its definition, attribute names computed at runtime --- have no static meaning to translate and are not suited to our shallow embedding approach. \code{async}, \code{await} and threads were considered out of scope. We do not support multiple inheritance at all, and we are aware that even the semantics of single inheritance in Python and Lean are not the same. Regardless, as discussed in Section \ref{sec:oop} we do have a translation for single inheritance that works for simple cases. These limitations are not detrimental for the code we target as small scripts rarely use these features. \todo{This is definitely not comprehensive. We've sprinkled limitations throughout the paper, should this section restate previously mentioned stuff?}

\section{Results}\label{results}

We evaluate PastaLean on three axes: how much Python it translates into Lean that elaborates, whether the translated program computes the same answers as the original, and how well type inference recovers types. Appendix~\ref{app:methodology} gives the full methodology for code generation~(\ref{app:eval-cp}) and type inference~(\ref{app:eval-type}).

A program counts as \emph{elaborated} when its generated Lean compiles and type-checks, and a test as \emph{passed} when the computable twin returns CPython's answer on the dataset's own input. PastaLean elaborates 94.3\% of programs and passes 99.0\% of tests on average. Every elaborated HumanEval\textsuperscript{+} and LiveCodeBench program agrees with CPython on all of its test cases, and 97.0\% of LeetCode tests pass (Table~\ref{tab:codegen}). The full end-to-end conversion itself is fast: one warm backend converts about 100 Leetcode solution files in 1 second. Since the size of files may vary, we use another metric: the number of IR nodes generated per second. On average, the backend generates about 12K IR nodes, per second, with time linear in program size (Figure~\ref{fig:conv-scaling}). See Appendix \ref{app:conv-speed} for more details on the speed of PastaLean conversion. PastaLean also performs significantly better than LLM translations (Appendix~\ref{app:llm-baseline})

\begin{table}[t]
\centering
\begin{tabular}{@{}lrrrr@{}}
\toprule
Dataset & Problems & Elaborated (\%) & Tests & Passed (\%) \\
\midrule
HumanEval\textsuperscript{+} & 164   & 151\textsuperscript{$\dagger$} (92.1)   & 7{,}007   & 7{,}007 (100.0) \\
LeetCode                    & 2{,}589 & 2{,}391 (92.4) & 215{,}126 & 208{,}736 (97.0) \\
LiveCodeBench               & 476   & 469 (98.5)   & 10{,}213  & 10{,}213 (100.0) \\
\bottomrule
\end{tabular}
\caption{Code generation and execution. \emph{Elaborated} counts programs whose Lean type-checks; \emph{Tests} and \emph{Passed} are counted only over those. $\dagger$ the thirteen remaining HumanEval programs elaborate \textbf{164/164} with one type hint added (Appendix~\ref{app:codegen-erros}).}
\label{tab:codegen}
\end{table}

\begin{table}[t]
\centering
\setlength{\tabcolsep}{3.5pt}
\begin{subtable}[t]{0.505\textwidth}
\centering\footnotesize
\begin{tabular}{@{}lrrrrr@{}}
\toprule
Tool & Return & Param & Var & All & Match (\%) \\
\midrule
\textbf{TypeInfer} & \textbf{16{,}220} & \textbf{404} & \textbf{39{,}560} & \textbf{56{,}184} & \textbf{72.7} \\
HeaderGen          & 14{,}086 & 346 & 36{,}370 & 50{,}802 & 65.7 \\
Jedi               & 13{,}160 & 0   & 15{,}403 & 28{,}563 & 37.0 \\
Scalpel            & 15{,}383 & 171 & 18       & 15{,}572 & 20.2 \\
Type4Py            & 3{,}143  & 38  & 2{,}243  & 5{,}424  & 7.0 \\
\bottomrule
\end{tabular}
\subcaption{Per-file: static tools on the TypeEvalPy autogen set (77{,}268 facts).}
\label{tab:static-auto}
\end{subtable}
\hfill
\begin{subtable}[t]{0.485\textwidth}
\centering\footnotesize
\begin{tabular}{@{}lrrrrr@{}}
\toprule
Tool & TypeSim & Exact & Cover & Time & Mem \\
\midrule
\textbf{TypeInfer} & \textbf{0.284} & \textbf{0.249} & \textbf{0.380} & \textbf{135\,s} & 0.70 \\
pyrefly            & 0.243 & 0.236 & 0.281 & 228\,s & 3.09 \\
pyre               & 0.239 & 0.228 & 0.272 & 259\,s & 1.84 \\
pytype             & 0.158 & 0.137 & 0.208 & 2{,}015\,s & 1.27 \\
hityper            & 0.105 & 0.084 & 0.168 & 11{,}652\,s & \textbf{0.19} \\
\bottomrule
\end{tabular}
\subcaption{Repository-level: TypyBench (50 projects, 67{,}897 slots). Mem in GiB.}
\label{tab:typybench}
\end{subtable}
\caption{Type inference against other tools: exactly-matching type facts (\subref{tab:static-auto}) and TypyBench slot-weighted metrics (\subref{tab:typybench}). Bold marks the best per column.}
\label{tab:typeinfer-results}
\end{table}

For the \emph{type inference} task we check whether the inferred type equals the expected one exactly, comparing against the static tools on the TypeEvalPy leaderboard. Those numbers date from 2024, so we also ran the current production checkers (pyrefly, ty, zuban, mypy, pyright) on the same data. On the micro-benchmark \code{TypeInfer} leads every tool on every sub-task and on the total, by 3.2\% over HeaderGen, the closest (Tables~\ref{tab:static-micro} and~\ref{tab:checkers-micro}). On the autogen set it leads all static tools (Table~\ref{tab:static-auto}); against the production checkers it leads on returns and parameters, while pyrefly recovers more variable types and so takes the total by 6.1\% (Table~\ref{tab:checkers-auto}). \code{TypeInfer} is also the fastest tool on both sets.

Both TypeEvalPy sets are single-file, so they never exercise inference across imports. We therefore also run TypyBench~\cite{jiang2025typybench}: fifty real projects with their annotations stripped, where a type must flow across files to be recovered at all. Four other tools can be scored there, since they synthesise annotations into an untyped project in place: \code{pyrefly}, \code{pyre}, \code{pytype} and the research tool \code{hityper}; the remaining checkers verify existing annotations rather than synthesise them. \code{TypeInfer} leads on every accuracy metric and is also the fastest (Table~\ref{tab:typybench}), peaking at 0.70\,GiB on the largest project against 1.84\,GiB for \code{pyre}, 1.27\,GiB for \code{pytype} and 3.09\,GiB for \code{pyrefly}, because inference only needs a small read-only table of module signatures, which its worker processes share, whereas the checkers each load a full type-checking state per worker (Appendix~\ref{app:ti-repo}); only \code{hityper} is lighter still (0.19\,GiB), but it recovers the fewest types and slow. A lightweight transpiler-side recovery lifts every metric further (Appendix~\ref{app:ti-collab}).
% \clearpage

\begin{table}[t]
\centering
\begin{tabular}{@{}lrrr@{}}
\toprule
Tool & Caught (of 204) & Flagged clean (of 204) & Time \\
\midrule
\textbf{PastaLean} & \textbf{169 (82.8)} & \textbf{0 (0.0)} & 0.8\,s \\
ty                 & 149 (73.0) & 47 (23.0) & \textbf{0.1\,s} \\
pyright            & 143 (70.1) & 54 (26.5) & 5.5\,s \\
pyrefly            & 140 (68.6) & 56 (27.5) & 0.5\,s \\
zuban              & 132 (64.7) & 42 (20.6) & 1.6\,s \\
mypy               & 119 (58.3) & 80 (39.2) & 1.5\,s \\
\bottomrule
\end{tabular}
\caption{Differential type-checking recall over 204 seeded mutants of 77 files, all six tools on the same 16-core machine; percentages in parentheses. \emph{Caught} credits a tool only when it flags the mutant \emph{and} leaves the untouched original alone. \emph{Flagged clean} counts the originals it reports anyway, which are known-good by construction. Bold marks the best per column.}
\label{tab:typecheck}
\end{table}

For \emph{type checking} we measure recall differentially. A mutation script flips one annotation in known-good code to seed a single type error, and a tool is credited only when it flags the mutant \emph{and} leaves the untouched original alone, so nothing is scored for pre-existing noise. Over 204 mutants of 77 files PastaLean catches 169 (82.8\%) and flags none of the originals, ahead of ty (149), pyright (143), pyrefly (140), zuban (132) and mypy (119), each of which flags between a fifth and two fifths of the unmutated files as well (Table~\ref{tab:typecheck}). The second column has to be read with the first, and with one caveat: those originals are \code{example_scripts}, the corpus PastaLean's own filters were written against, so the zero there has home advantage. The differential rule protects the \emph{caught} column from that advantage; it does not protect the \emph{flagged clean} one. On a corpus of sixteen hand-seeded type errors the checker misses none, and every certificate it wrote over both corpora is accepted by the validator. Precision at scale we measure on 158 \code{mypy_primer} projects, every file of which already type-checks in the project's own CI, so each diagnostic is a candidate false positive: 294 diagnostics over 68{,}542 files read, 4.29 per thousand (Appendix~\ref{app:cert-eval}). Certifying costs nothing measurable, because producing the record is the same pass that reached the verdict --- within $\pm 3\%$ of checking without it at every job count.

\section{Conclusion}

PastaLean's deterministic translator carries over 92\% of LeetCode and HumanEval\textsuperscript{+} and over 98\% of LiveCodeBench into Lean with at least 97\% test-case agreement, and its \code{TypeInfer} engine leads both the TypeEvalPy and TypyBench benchmarks. Its checker leads the production type checkers on differential recall over seeded type errors, and reports each verdict with a certificate that a 280-line validator re-checks without running the engine, so a reader can audit the verdict rather than trust it. It does not yet handle all of Python, including for example dynamic dispatch, and multiple inheritance, and proof automation remains best-effort; we are actively working to close these gaps (Sections~\ref{sec:oop}, \ref{sec:pyany}). PastaLean is well suited to agentic verification: LLMs already write Python well, so translating from a language they succeed in is more promising than asking them to write Lean directly, and because the embedding is shallow, the resulting proofs draw on Lean's existing tooling and on Mathlib rather than a bespoke verification stack.   

\FloatBarrier

\small
\bibliographystyle{plain}
\bibliography{references}

\newpage

\appendix

\section{Appendices}

\subsection{Naming Conventions used in PastaLean}\label{naming-convention}

It is important to know few naming conventions used by PastaLean. To accommodate Python's naming convention and to avoid possible naming conflicts, we try use Python's own naming convention which is supported by Lean. However, any extra variables/functions introduced by PastaLean use \code{'} since its usage is supported by Lean. For example, local variables introduced by us begin/end with \code{p'} , for example \code{p'_ret_1}, \code{x'v1}, etc.

For a Python function \code{foo}, we name the verifiable function as \code{foo} and computable version as \code{foo'rn} suggesting it can be run. This avoids any of our naming schemes to clash with that of Python's.

While writing the theorem name's for proving a function, we use an auxiliary function proving the Hoare Triple verification as mentioned before. We name this function \code{foo_spec}. The final correctness theorem is simply \code{foo_correct}.

\subsection{Fully Converted Example}\label{app:full-example}

\begin{figure}[h]
\centering
\begin{tikzpicture}[
  font=\small,
  node distance=14mm,
  box/.style={draw, rounded corners=2pt, minimum height=8mm, align=center, inner xsep=5pt},
  ar/.style={-{Stealth[length=2mm]}, thick},
]
\node[box] (py)   {Python};
\node[box, right=of py]  (ast)  {AST};
\node[box, right=of ast] (ti)   {TypeInfer};
\node[box, right=of ti]  (lean) {Lean};
\node[box, right=of lean, draw, dashed] (prf) {Proof};

\draw[ar] (py)   -- node[above,font=\scriptsize] {parse} (ast);
\draw[ar] (ast)  -- node[above,font=\scriptsize] {annotate} (ti);
\draw[ar] (ti)   -- node[above,font=\scriptsize] {generate} (lean);
\draw[ar, dashed] (lean) -- node[above,font=\scriptsize] {if contracts} (prf);

\node[box, below=9mm of py, fill=black!4] (ctr) {Add contracts};
\draw[ar] (ctr) -- (py);
\end{tikzpicture}
\caption{The pipeline, left to right. Contracts are attached to the Python before it is parsed, and only a file that carries them reaches the proof stage.}
\label{fig:pipeline}
\end{figure}
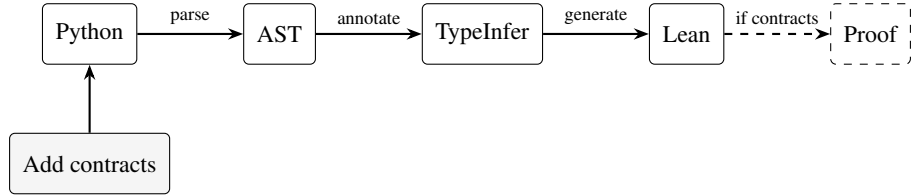

\begin{figure}[t]
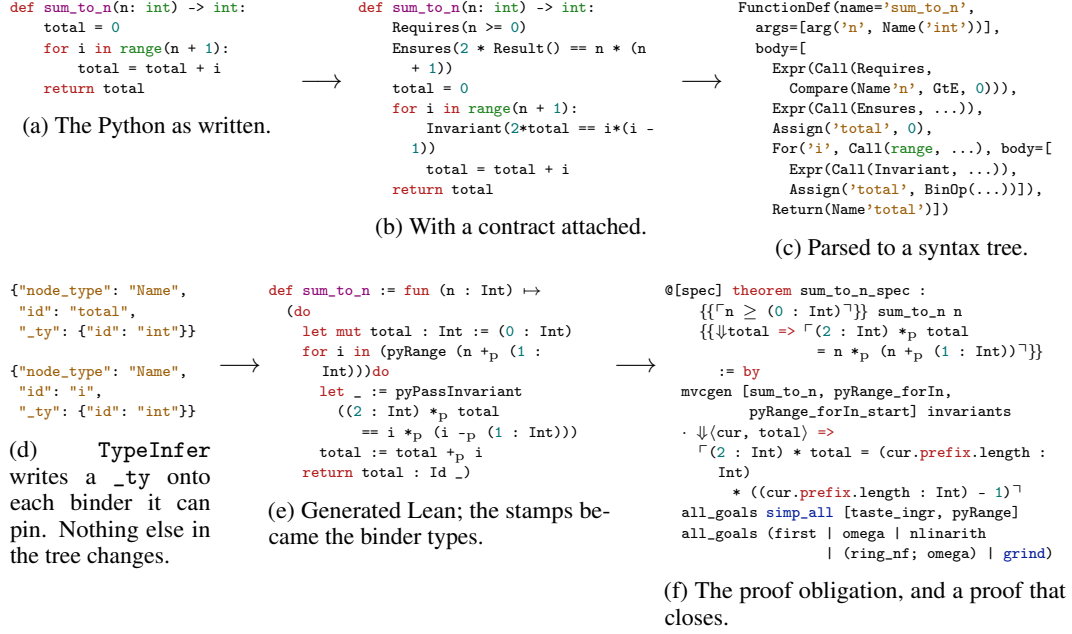

\centering
\begin{subfigure}[t]{0.26\textwidth}
\begin{lstlisting}[style=py,basicstyle=\ttfamily\fontsize{6.4}{7.6}\selectfont,aboveskip=0pt,belowskip=0pt,breaklines=true,breakautoindent=true]
def @@sum_to_n@@(n: int) -> int:
    total = 0
    for i in range(n + 1):
        total = total + i
    return total
\end{lstlisting}
\caption{The Python as written.}
\end{subfigure}%
\hfill\makebox[0.045\textwidth]{\raisebox{-3.4em}{$\longrightarrow$}}\hfill%
\begin{subfigure}[t]{0.29\textwidth}
\begin{lstlisting}[style=py,basicstyle=\ttfamily\fontsize{6.4}{7.6}\selectfont,aboveskip=0pt,belowskip=0pt,breaklines=true,breakautoindent=true]
def @@sum_to_n@@(n: int) -> int:
    Requires(n >= 0)
    Ensures(2 * Result() == n * (n + 1))
    total = 0
    for i in range(n + 1):
        Invariant(2*total == i*(i - 1))
        total = total + i
    return total
\end{lstlisting}
\caption{With a contract attached.}
\end{subfigure}%
\hfill\makebox[0.045\textwidth]{\raisebox{-3.4em}{$\longrightarrow$}}\hfill%
\begin{subfigure}[t]{0.31\textwidth}
\begin{lstlisting}[style=py,basicstyle=\ttfamily\fontsize{6.4}{7.6}\selectfont,aboveskip=0pt,belowskip=0pt,breaklines=true,breakautoindent=true]
FunctionDef(name='sum_to_n',
  args=[arg('n', Name('int'))],
  body=[
    Expr(Call(Requires,
      Compare(Name'n', GtE, 0))),
    Expr(Call(Ensures, ...)),
    Assign('total', 0),
    For('i', Call(range, ...), body=[
      Expr(Call(Invariant, ...)),
      Assign('total', BinOp(...))]),
    Return(Name'total')])
\end{lstlisting}
\caption{Parsed to a syntax tree.}
\end{subfigure}

\medskip

\begin{subfigure}[t]{0.19\textwidth}
\begin{lstlisting}[style=py,basicstyle=\ttfamily\fontsize{6.4}{7.6}\selectfont,aboveskip=0pt,belowskip=0pt,breaklines=true,breakautoindent=true]
{"node_type": "Name",
 "id": "total",
 "_ty": {"id": "int"}}

{"node_type": "Name",
 "id": "i",
 "_ty": {"id": "int"}}
\end{lstlisting}
\caption{\code{TypeInfer} writes a \code{\_ty} onto each binder it can pin. Nothing else in the tree changes.}
\end{subfigure}%
\hfill\makebox[0.04\textwidth]{\raisebox{-3.4em}{$\longrightarrow$}}\hfill%
\begin{subfigure}[t]{0.32\textwidth}
\begin{lstlisting}[style=lean,basicstyle=\ttfamily\fontsize{6.4}{7.6}\selectfont,aboveskip=0pt,belowskip=0pt,breaklines=true,breakautoindent=true]
def @@sum_to_n@@ := fun (n : Int) (*@\ensuremath{\mapsto}@*)
  (do
    let mut total : Int := (0 : Int)
    for i in (pyRange (n +(*@\ensuremath{{}_{\mathrm p}}@*) (1 : Int)))do
      let _ := pyPassInvariant
        ((2 : Int) *(*@\ensuremath{{}_{\mathrm p}}@*) total
           == i *(*@\ensuremath{{}_{\mathrm p}}@*) (i -(*@\ensuremath{{}_{\mathrm p}}@*) (1 : Int)))
      total := total +(*@\ensuremath{{}_{\mathrm p}}@*) i
    return total : Id _)
\end{lstlisting}
\caption{Generated Lean; the stamps became the binder types.}
\end{subfigure}%
\hfill\makebox[0.04\textwidth]{\raisebox{-3.4em}{$\longrightarrow$}}\hfill%
\begin{subfigure}[t]{0.38\textwidth}
\begin{lstlisting}[style=lean,basicstyle=\ttfamily\fontsize{6.4}{7.6}\selectfont,aboveskip=0pt,belowskip=0pt,breaklines=true,breakautoindent=true]
@[spec] theorem sum_to_n_spec :
    (*@\ensuremath{\{\!\{}@*)(*@\ensuremath{\ulcorner}@*)n (*@\ensuremath{\ge}@*) (0 : Int)(*@\ensuremath{\urcorner}@*)(*@\ensuremath{\}\!\}}@*) sum_to_n n
    (*@\ensuremath{\{\!\{}@*)(*@\ensuremath{\Downarrow}@*)total => (*@\ensuremath{\ulcorner}@*)(2 : Int) *(*@\ensuremath{{}_{\mathrm p}}@*) total
                  = n *(*@\ensuremath{{}_{\mathrm p}}@*) (n +(*@\ensuremath{{}_{\mathrm p}}@*) (1 : Int))(*@\ensuremath{\urcorner}@*)(*@\ensuremath{\}\!\}}@*) := by
  mvcgen [sum_to_n, pyRange_forIn,
          pyRange_forIn_start] invariants
  (*@\ensuremath{\cdot}@*) (*@\ensuremath{\Downarrow}@*)(*@\ensuremath{\langle}@*)cur, total(*@\ensuremath{\rangle}@*) =>
    (*@\ensuremath{\ulcorner}@*)(2 : Int) * total = (cur.prefix.length : Int)
        * ((cur.prefix.length : Int) - 1)(*@\ensuremath{\urcorner}@*)
  all_goals simp_all [taste_ingr, pyRange]
  all_goals (first | omega | nlinarith
                   | (ring_nf; omega) | grind)
\end{lstlisting}
\caption{The proof obligation, and a proof that closes.}
\end{subfigure}
\caption{One function through the pipeline of Figure~\ref{fig:pipeline}. \code{Requires} becomes the precondition, \code{Ensures} the postcondition, and \code{Invariant} the loop's invariant, restated over the loop cursor because the counter is not in scope at that point.}
\label{fig:stages}
\end{figure}

Figure~\ref{fig:pipeline} is the whole pipeline, and Figure~\ref{fig:stages} is one function carried through every stage of it. A Python file is parsed to an abstract syntax tree, the type inference engine annotates that tree, the code generator turns the annotated tree into Lean, and a file carrying contracts additionally yields a theorem and a proof. Contracts are the only optional input: without them the pipeline stops at Lean that compiles and runs, and with them it continues into a proof obligation. Every fragment shown is real output rather than a rendering of it.

\subsection{Provable and Computable Twin}\label{app:twins}

One Python function \code{foo} becomes two Lean definitions. \code{foo} is the provable twin and \code{foo'rn} is the runnable one. They are generated from the same IR by the same generators and differ only where being provable and being fast pull apart as previously described in section \ref{sec:transpilation-targets}. The provable twin may sometimes be computable in case the function is simple (because \code{Rat} is computable) and does not enough complex non-computable functions, etc. Though it may be slower than the runnable twin. This appendix lists the three places that happens.

They are two Lean definitions, so a theorem about \code{foo} is not literally a theorem about \code{foo'rn}: the proof is stated over $\mathbb{Q}$ or $\mathbb{R}$ and \code{List}, while the code that runs uses \code{Float} and \code{Array}. We do not prove that the runnable twin refines the provable one. The two are tied empirically instead, by running the runnable twin against CPython on the dataset's own tests (Appendix~\ref{app:eval-cp}), and making that tie a proof is future work.

\subsubsection{Different Programs Example for the Same Function}

Listing \ref{lst:twin-py} is ordinary Python; Listings \ref{lst:twin-prove} and \ref{lst:twin-run} are the two twins generated from it. The highlighted lines are the only places they differ: the sequence type and its operations, and how the print is interpreted.

\begin{lstlisting}[style=py,caption={The Python source.},label={lst:twin-py}]
def @@stack_sum@@(n: int) -> int:
    st = []
    for i in range(n):
        st.append(i)
    total = 0
    while st:
        total += st.pop()
    return total

def @@main@@():
    print(stack_sum(5))
\end{lstlisting}

\noindent
\begin{minipage}[t]{0.49\textwidth}
\begin{lstlisting}[style=lean,caption={Provable twin.},label={lst:twin-prove},basicstyle=\ttfamily\scriptsize,linebackgroundcolor={\ifnum\value{lstnumber}=4 \color{yellow!30}\fi\ifnum\value{lstnumber}=6 \color{yellow!30}\fi\ifnum\value{lstnumber}=9 \color{yellow!30}\fi\ifnum\value{lstnumber}=10 \color{yellow!30}\fi\ifnum\value{lstnumber}=15 \color{yellow!30}\fi}]
def @@stack_sum@@ := fun (n : Int) (*@\ensuremath{\mapsto}@*)
  Id.run
    (do
      let mut st : List Int := []
      for i in (pyRange n) do
        st := pyAppend st i
      let mut total : Int := 0
      while (pyTruthy st) do
        let v := pyPopValue st
        st := pyPopRest st
        total := total +(*@\ensuremath{{}_{\mathrm p}}@*) v
      return total)

def @@main'@@ :=
  pyPrintProof [pyPrintArg (stack_sum 5)]
\end{lstlisting}
\end{minipage}%
\hfill
\begin{minipage}[t]{0.49\textwidth}
\begin{lstlisting}[style=lean,caption={Runnable twin.},label={lst:twin-run},basicstyle=\ttfamily\scriptsize,linebackgroundcolor={\ifnum\value{lstnumber}=4 \color{yellow!30}\fi\ifnum\value{lstnumber}=6 \color{yellow!30}\fi\ifnum\value{lstnumber}=9 \color{yellow!30}\fi\ifnum\value{lstnumber}=10 \color{yellow!30}\fi\ifnum\value{lstnumber}=15 \color{yellow!30}\fi}]

def @@stack_sum'rn@@ := fun (n : Int) (*@\ensuremath{\mapsto}@*)
  Id.run
    (do
      let mut st : Array Int := #[]
      for i in (pyRange n) do
        st := pyArrayAppend st i
      let mut total : Int := 0
      while (pyTruthy st) do
        let v := pyArrayPopValue st
        st := pyArrayPopRest st
        total := total +(*@\ensuremath{{}_{\mathrm p}}@*) v
      return total)

def @@main''rn@@ :=
  pyPrintIO [pyPrintArg (stack_sum'rn 5)]
\end{lstlisting}
\end{minipage}

Everything outside the highlighted lines is character-for-character the same. The control flow, the arithmetic operators, the loop structure and the names are shared, which is what makes the two twins worth comparing: a reader can check that the runnable twin computes what the provable one is proved to compute by reading the difference, and the difference is four lines.

\subsubsection{Sequences: \code{List} for proving, \code{Array} for running} \label{sec:array-vs-list}

Python's \code{list} is a dynamic array: indexing and appending are both constant time. Lean's \code{List} is a linked list, so \code{xs ++ [v]} is $O(n)$ and \code{xs[i]} is $O(i)$. Competitive-programming and data-analysis code appends and indexes constantly, so a loop that is $O(n)$ in Python becomes $O(n^2)$ in Lean. For programs over large inputs this decides whether the translation is usable at all. Lean's \code{Array} gives us the interface Python's \code{list} has: $O(1)$ indexing, and an append that is amortised $O(1)$ because Lean updates the buffer destructively when it holds the only reference to it. Measured on an append-and-index loop at $n = 30{,}000$, the \code{List}-backed version runs about $134\times$ slower than the \code{Array}-backed one. That is the difference between passing a test suite and timing out on it. On the LeetCode set, array-backing recovered about eleven problems that had exhausted the time limit, several of them algorithmically fine to begin with: \code{count_primes}, an $O(n \log \log n)$ sieve, went from 1 of 32 test cases to 32 of 32, because every \code{sieve[j] = False} had been copying the whole list.

We cannot simply use \code{Array} everywhere, because the provable twin needs Mathlib's lemmas, and those are stated about \code{List}. So the provable twin keeps \code{List} and the runnable twin uses \code{Array}, whose \code{push} and \code{set!} are constant time \emph{when the array is uniquely owned}. Note that Lean's \code{List} and \code{Array} are equivalent theoretically however only low level implementation distinction holds. Uniqueness is what makes this sound rather than merely fast: code generation always threads a mutation as \code{xs := pyArrayAppend xs v}, reassigning the same binder, so the old reference dies at the assignment and Lean's reference counting mutates in place instead of copying~\cite{ullrich2019beans,reinking2021perceus}. This is also why the runnable twin is worth taking seriously as a program and not only as a witness: Lean compiled this way is competitive with systems languages, and a verified Lean implementation of DEFLATE has been reported to outrun a pure-Rust one on both speed and compression ratio~\cite{morrison2026leanrust}. Every array operation mirrors the semantics of its list counterpart, including negative indices and \code{IndexError} on an out-of-range access, so the two twins agree on results and only differ in cost. Some codes might still use List in computable twin and that is primarily because some other referenced variable or function may use List and does not provide an Array twin for the same. Hence, we may greedily interconvert from Arrays to Lists for that operation alone. Insertion, slicing and searching are inherently linear, so rather than reimplement them we convert: \code{pyArrayInsert} is \code{xs.toList}, then the list insert, then \code{.toArray}, and \code{pyCount} counts over \code{xs.toList}. Crucially this does not show up in the generated program. In Listing~\ref{lst:arr-roundtrip} the variable is an \code{Array} from its first line to its last, and \code{pyArrayInsert} has type \code{Array} $\to$ \code{Array}; the conversion is inside that operation's definition in the runtime. The program's types never mention \code{List}, so a linear operation costs a pass there and back inside one call and leaves everything around it constant-time.

\begin{lstlisting}[style=lean,caption={The generated code stays \code{Array} throughout; the conversion lives inside the runtime operation it calls.},label={lst:arr-roundtrip}]

# Python
def @@build@@(n: int) -> int:
    xs = [0] * n
    for i in range(n): xs[i] = i        # n writes
    xs.insert(0, -1)                    # one linear operation
    for i in range(n): xs[i] = xs[i]+1  # n more writes
    return xs[0] + len(xs)

-- generated runnable twin: xs is an Array from start to finish
def @@build'rn@@ := fun (n : Int) (*@\ensuremath{\mapsto}@*) Id.run (do
  let mut xs : Array Int := pyArrayRepeat #[0] n
  for i in (pyRange n) do
    xs := pySetItem xs i i
  xs := pyArrayInsert xs 0 (-1)
  for i in (pyRange n) do
    xs := pySetItem xs i (xs(*@\ensuremath{[\![}@*)i(*@\ensuremath{]\!]}@*) +(*@\ensuremath{{}_{\mathrm p}}@*) 1)
  return xs(*@\ensuremath{[\![}@*)0(*@\ensuremath{]\!]}@*) +(*@\ensuremath{{}_{\mathrm p}}@*) pyLen xs)

-- the conversion is here, in the runtime, not in the generated program:
def @@pyArrayInsert@@ (xs : Array (*@\ensuremath{\alpha}@*)) (idx : Int) (elem : (*@\ensuremath{\alpha}@*)) : Array (*@\ensuremath{\alpha}@*) :=
  (pyListInsert xs.toList idx elem).toArray

instance : PyCount (Array (*@\ensuremath{\alpha}@*)) (*@\ensuremath{\alpha}@*) where
  pyCount xs elem := pyListCount xs.toList elem
\end{lstlisting}

This conversion is done keeping in mind the time complexity of the algorithm. Each round trip is $O(n)$ and happens once per such call, whereas keeping the variable a \code{List} to avoid it would make every indexed write $O(n)$, so a loop of $n$ writes becomes $O(n^2)$. In Listing~\ref{lst:arr-roundtrip}, the runnable twin pays $2n$ constant-time writes plus one linear insert, while the provable twin pays $2n$ linear writes which is slower.

\subsubsection{Printing and input in the provable twin}\label{app:twin-io}

A program that prints cannot be run in the provable twin, because \code{IO} has no equational theory to reason against. The obvious move is to drop the output, and a \code{pyPrintNoop} exists for paths where nothing better is available. It is not what we do by default. The provable twin instead runs in
\code{PyProofM} $\alpha$ = \code{ExceptT PyException (StateM IOState)} $\alpha$,
where \code{IOState} carries the input as a stream of lines and the output as a list of lines. A \code{print} renders its arguments through exactly the same formatting code the runnable twin uses and appends the result to that list, and \code{input()} reads from the stream. Nothing is discarded, and the effects stay in a pure state monad, so what a program prints is a value a theorem can talk about.

\subsubsection{Exceptions in the provable twin}\label{app:twin-exc}

The two twins raise the same exceptions but carry them in different monads. Both use \code{ExceptT PyException}, and only the layer underneath changes: \code{IO} in the runnable twin, and \code{Id} or \code{StateM IOState} in the provable one. The consequence is that a function which raises but does no input or output stays pure in the provable twin, and so stays provable, which would not be true if raising forced it into \code{IO}. Being an \code{ExceptT} over a state monad rather than over \code{IO} is also what lets \code{mvcgen} generate verification conditions for it. Where the provable twin cannot thread its state monad at all, effects degrade to silent no-ops: \code{pyPrintNoop} accepts the same arguments as a real print and discards them, so the line still type-checks and the surrounding proof is unaffected. Section~\ref{app:more-conversions} describes the exception representation itself.

\subsection{Arithmetic: Correctness, Priorities, and Non-finite values}\label{app:arith}

Each Python operator is a single class with many instances, and Lean chooses among them from the types of the operands: $+_p$ on two strings finds the concatenation instance, on two lists the one that appends, and on an integer and a boolean the one that reads the boolean as $0$ or $1$. The result type is an \code{outParam}, so it is computed from the operands rather than supplied by the caller, and \code{1 + 2.5} yields a rational with no annotation anywhere in the generated code.

Some Python operators disagree with the instance Lean would otherwise pick, so the order in which instances are tried is itself part of the semantics. Python's \code{+} on lists concatenates, but Mathlib already defines \code{Add (List }$\alpha$\code{)} pointwise; if that instance is reached first, \code{[1,2] + [3,4]} elaborates to \code{[4,6]} \ref{lst:arith-priority}, a wrong answer that no type error catches. Instances are therefore placed in one of three bands according to \emph{when} they should be consulted. A \code{high} priority instance is tried before Lean's generic ones, which is how we override cases like list concatenation. A \code{default_instance} is consulted only at the end of elaboration, when an operand's type is still open and Lean has to commit to one; ranking matters there too, because both the rational and the integer instance apply and the rational one would otherwise win, pinning an integer comprehension to $\mathbb{Q}$ and dragging the element type of the list along with it. A \code{low} priority instance is tried last, which is where the derived instances of the next paragraph sit, so that they fire only for a mixed pair nothing else covers. Listing \ref{lst:arith-priority} shows both hazards.

\begin{lstlisting}[style=lean,caption={Two places where instance ordering decides the answer.},label={lst:arith-priority}]

-- Reached first, it silently turns [1,2] + [3,4] into [4,6].
instance (priority := high) {(*@\ensuremath{\alpha}@*)} : PyHAdd (List (*@\ensuremath{\alpha}@*)) (List (*@\ensuremath{\alpha}@*)) (List (*@\ensuremath{\alpha}@*)) where
  hAdd := ((*@\ensuremath{\cdot}@*) ++ (*@\ensuremath{\cdot}@*))

-- In `[x + y for x in xs for y in ys]` the operator elaborates before the
-- type of `x` is known, so a default instance decides. Both apply, and
-- without the higher rank the (*@\ensuremath{\mathbb{Q}}@*) one wins, forcing `xs : List (*@\ensuremath{\mathbb{Q}}@*)`.
@[default_instance]       instance : PyHAdd Rat Rat Rat where hAdd a b := a + b
@[default_instance 10001] instance : PyHAdd Int Int Int where hAdd a b := a + b
\end{lstlisting}

Python mixes numeric types freely, so every operator must accept every pair of them, and writing one instance per operator per pair does not scale. The conversions involved do not depend on the operator, so we declare them once: \texttt{PyNumJoin} $\alpha\;\beta\;\gamma$ records that $\alpha$ and $\beta$ both convert into a common $\gamma$, with one instance per pair in the tower $\mathbb{N} \subset \mathbb{Z} \subset \mathbb{Q}$ together with \texttt{Float}. A single generic instance per operator then reads those conversions off the class (Listing \ref{lst:pynumjoin}). Floor division and the bitwise operators sit outside this scheme, the first because Python's \texttt{//} fixes its own result type, the second because it must model unbounded integers.

\begin{lstlisting}[style=lean,caption={The numeric tower, declared once and lifted to every operator.},label={lst:pynumjoin}]

class PyNumJoin ((*@\ensuremath{\alpha}@*) (*@\ensuremath{\beta}@*) : Type) ((*@\ensuremath{\gamma}@*) : outParam Type) where
  coeL : (*@\ensuremath{\alpha}@*) (*@\ensuremath{\rightarrow}@*) (*@\ensuremath{\gamma}@*)
  coeR : (*@\ensuremath{\beta}@*) (*@\ensuremath{\rightarrow}@*) (*@\ensuremath{\gamma}@*)

instance : PyNumJoin Int Rat Rat := .mk2 (fun n => (n : Rat)) id

-- one generic instance per operator covers every pair in the tower
instance (priority := low) {(*@\ensuremath{\alpha}@*) (*@\ensuremath{\beta}@*) (*@\ensuremath{\gamma}@*)} [j : PyNumJoin (*@\ensuremath{\alpha}@*) (*@\ensuremath{\beta}@*) (*@\ensuremath{\gamma}@*)] [Add (*@\ensuremath{\gamma}@*)] : PyHAdd (*@\ensuremath{\alpha}@*) (*@\ensuremath{\beta}@*) (*@\ensuremath{\gamma}@*) where
  hAdd a b := j.coeL a + j.coeR b
\end{lstlisting}

\subsubsection{Implementing \texttt{float('inf')}} \label{floatinf}

Python has \code{inf} to represent infinity, i.e. no integer or float is bigger than it. Programmers often use \code{float('inf')} to initialise a running minimum (correspondingly \code{float('-inf')} for a running maximum), in the idiom \code{best = inf} followed by \code{best = min(best, ...)}. In the default exact mode a Python float is a $\mathbb{Q}$, and $\mathbb{Q}$ is a field with no infinities, so there is no value for the literal to denote. For plain execution we represent it as a large finite sentinel of magnitude $10^{30}$, letting the type of the slot it lands in select the representation through a class \code{PyNonFinite}, so that a \code{Float} slot receives the genuine IEEE infinity while a $\mathbb{Q}$ or $\mathbb{Z}$ slot receives the sentinel. That runs the idiom correctly, but a finite value can never satisfy $\forall x,\ x \le \mathit{inf}$, so it carries none of infinity's order structure and supports no proof.

For verification we therefore give \code{inf} a genuine representation. \code{XRat}, defined as \code{WithBot (WithTop }$\mathbb{Q}$\code{)} from \emph{Mathlib}, adjoins a greatest and a least element to the rationals. Because it is built from \code{Option} it stays computable, unlike \code{EReal}, which pulls in $\mathbb{R}$ and is noncomputable. \code{INF} and \code{NEGINF} are then $\top$ and $\bot$, finite values coerce in along $\uparrow$, and \code{min}, \code{max}, $\le$ and $+$ are inherited from Mathlib. The order laws the idiom depends on become theorems rather than assumptions (Figure \ref{lst:xrat}): \code{le_INF} says every value lies below \code{INF}, \code{min_INF} that \code{INF} is absorbed by \code{min}, and \code{minReduce_le} that the generated reduction is a genuine lower bound of the list it folds over. The last of these is exactly the postcondition the finite sentinel made unprovable.

\begin{lstlisting}[style=lean,caption={A provable \code{inf}: extended rationals and the laws a proof gets.},label={lst:xrat}]

abbrev @@XRat@@ := WithBot (WithTop (*@\ensuremath{\mathbb{Q}}@*))   -- Option-based, hence computable
def @@INF@@    : XRat := (*@\ensuremath{\top}@*)
def @@NEGINF@@ : XRat := (*@\ensuremath{\bot}@*)

@[simp] theorem le_INF  (x : XRat) : x (*@\ensuremath{\le}@*) INF := le_top
@[simp] theorem min_INF (x : XRat) : min x INF = x := by simp

-- the generated `best = inf; for x in xs: best = min(best, x)` reduction
theorem @@minReduce_le@@ (xs : List XRat) : (*@\ensuremath{\forall}@*) x (*@\ensuremath{\in}@*) xs, minReduce xs (*@\ensuremath{\le}@*) x
\end{lstlisting}

\subsubsection{Correctness of the operators} \label{arith-correct}

Each custom instance for Python is definitionally Lean's own arithmetic, so we ship a proof of that correspondence alongside it. Figure \ref{lst:arith-proof} shows few examples of such correctness theorems. The proofs are trivial, but they are very important for correctness and they are what lets a tactic rewrite a generated operator back into ordinary arithmetic and then finish the goal with \code{ring}.

\begin{lstlisting}[style=lean,caption={Arithmetic Operation Proofs},label={lst:arith-proof}]

@[simp] theorem pyAdd_int (a b : (*@\ensuremath{\mathbb{Z}}@*)) : a +(*@\ensuremath{{}_{\mathrm p}}@*) b = a + b := rfl
@[simp] theorem pySub_int (a b : (*@\ensuremath{\mathbb{Z}}@*)) : a -(*@\ensuremath{{}_{\mathrm p}}@*) b = a - b := rfl
@[simp] theorem pyMul_rat (a b : (*@\ensuremath{\mathbb{Q}}@*)) : a *(*@\ensuremath{{}_{\mathrm p}}@*) b = a * b := rfl
@[simp] theorem pyDiv_rat (a b : (*@\ensuremath{\mathbb{Q}}@*)) : a /(*@\ensuremath{{}_{\mathrm p}}@*) b = a / b := rfl

\end{lstlisting}

\subsection{Why nested definitions are lifted rather than kept nested}\label{app:lifting}

Consider a nested helper whose recursion is not structural, the shape any grid or graph search takes (Listing \ref{lst:lift-why}). Lean offers three ways to keep such a helper nested, and each one fails for a different reason.

\begin{itemize}
\item A plain \code{let dfs := fun i j }$\mapsto$\code{ ...} is not recursive, so the \code{dfs} appearing in its own body is an unknown identifier.
\item \code{let rec} is recursive and does see the enclosing locals, but it cannot be marked \code{partial}, and the recursion descends toward a bound rather than on a structural subterm, so termination checking rejects it.
\item A \code{where} clause may be \code{partial}, but it inherits that \code{partial} from the enclosing definition. The outer function itself then becomes \code{partial} and loses both its unfolding and its \code{[simp]} tag, so nothing about it can be proved by unfolding. A \code{where} clause also sees only the definition's binders, and since we emit functions in the form \code{def f := fun n }$\mapsto$\code{ ...}, the parameters are lambda binders it cannot see at all.
\end{itemize}

Lifting confines \code{partial} to the helper. The enclosing function stays an ordinary \code{def}, keeps its \code{[simp]} tag, and remains both evaluable and provable.

\begin{lstlisting}[style=lean,caption={A non-structural nested recursion, and the lifted form.},label={lst:lift-why}]

def @@count_paths@@(grid):
    def @@dfs@@(i, j):
        if i == len(grid) - 1: return 1
        return dfs(i + 1, j) + dfs(i, j + 1)   # not structural
    return dfs(0, 0)

-- lifted: `partial` is confined to the helper, `grid` is passed as a capture
private partial def _count_paths'dfs := fun (i j : Int) (grid : List (List Int)) (*@\ensuremath{\mapsto}@*) ...
def @@count_paths@@ := fun (grid : List (List Int)) (*@\ensuremath{\mapsto}@*) _count_paths'dfs 0 0 grid
\end{lstlisting}

\subsubsection{What Python and Lean change about lambda lifting}

Textbook lambda lifting~\cite{johnsson1985lambdalifting} is short: take the free variables of a nested function, make them extra parameters, move the function to the top level, and pass the captured values at every call. The transformation has itself been formalised in Lean 4 and proved correct~\cite{levy2026simplelambdalifting}; our concern is the converse, using it to \emph{produce} Lean whose enclosing definitions stay provable. That is written for a language with immutable bindings and nothing but lexical scope. Python breaks the first assumption, and its scope rules are subtle enough to have needed a tested formal semantics of their own~\cite{politz2013fullmonty}, while Lean adds requirements of its own, so we keep the shape of the transformation and change four things about how the free set is computed and two about what is emitted. Listing \ref{lst:lift-algo} shows the modified computation. 

We define two sets of variable names over a function \code{foo}. \code{usedVars(foo)} is every variable name that appears anywhere inside \code{foo}, whether it is read or written. \code{boundVars(foo)} is every variable name that \code{foo} introduces itself, that is, its parameters together with every variable it assigns to. A variable name that is used but not bound is one \code{foo} expects to get from somewhere outside itself, which is what makes it a candidate for capture. We write $A \setminus B$ for the names in $A$ that are not in $B$, $A \cap B$ for the names in both, and $A \cup B$ for their union. We write $A \mathbin{+\!\!+} B$ for $A$ followed by $B$ as a list rather than a set, which we need where the order matters: the extra parameters must be supplied at every call site in exactly the order they were added. Two smaller sets also appear in the listing: \code{comprehensionAndLambdaTargets(foo)} is the variable names that a comprehension or a lambda inside \code{foo} binds for itself, such as the \code{v} in \code{[v*2 for v in xs]}, and \code{nonlocals(foo)} is the names \code{foo} declares \code{nonlocal}.

\begin{lstlisting}[style=pseudo,caption={The free-variable computation, with the four Python-driven edits marked.},label={lst:lift-algo}]

free $\gets$ usedVars(inner) $\setminus$ boundVars(inner) // the textbook rule

// (1) a comprehension or lambda binds its own targets
free $\gets$ free $\setminus$ comprehensionAndLambdaTargets(inner)
// (2) a `nonlocal` name is assigned, so the rule above dropped it
free $\gets$ free $\cup$ nonlocals(inner)
// (3) keep only what the enclosing function itself binds
caps $\gets$ free $\cap$ boundVars(outer)
// (4) a capture the helper writes cannot travel as a parameter alone
mutated  $\gets$ every c in caps that inner rebinds or mutates in place
ordered  $\gets$ (caps $\setminus$ mutated) ++ mutated // passed in, in this order
threaded $\gets$ params(inner) mutated in place ++ mutated // handed back
\end{lstlisting}

Edit (3) is the one that keeps globals, builtins and imports out. In a purely lexical language every free name belongs to some enclosing scope, but a Python function also sees module globals and builtins, and those stay visible to a sibling top-level definition, so only a name the enclosing function itself binds would go out of scope and needs passing. Edits (2) and (4) are both consequences of Python allowing a nested function to write to an enclosing variable, which Johnsson's setting does not. A capture that is only read stays an ordinary parameter; one that is written is also handed back on the return so the caller can rebind it, which is how the write becomes visible outside while the helper stays a pure function of its inputs.

Lean forces two further changes. First, a non-structural recursion needs \code{partial}, and the nested forms spread it: a \code{where} clause inherits \code{partial} from the definition it sits in, so the \emph{enclosing} function would become \code{partial} and lose its provability. We therefore emit the helper as a sibling \code{private partial def}, which confines \code{partial} to it. Second, mutually recursive helpers have to be emitted inside one \code{mutual} block, so the nested definitions are grouped by which of them call each other and a group shares a single capture set. Lifting also runs outermost-first, since a helper nested several levels deep must capture the parameters its parent has just acquired, and each lifted parameter carries the type inferred for it in the enclosing scope, without which Lean's instance resolution has nothing to work from.

Shapes the transformation cannot express are refused rather than approximated: a helper that captures, or mutates a capture, and is then used as a value rather than called; a call inside a comprehension or lambda that would have to rebind threaded state; and a threading helper that can fall off the end without returning.

\subsubsection{Closures}

A closure is a nested function that keeps using variables from the function around it. If it only reads them we are already done, because those variables became parameters when we lifted it. Returning the function is then just calling the lifted version with the captured values filled in and the rest left open. A \code{make_adder} that closes over \code{n}, for instance, becomes the lifted helper applied to \code{n}: what comes back still takes \code{x}.

If the nested function \emph{changes} one of those variables, passing it in is not enough, because the outer function needs to see the new value. Python manages this because a closure there is a live object that keeps its variables alive, while Lean is functional and a function cannot hold state of its own, so the only way a change gets out is to return it. So the helper gives it back. It returns its own result together with the new value of the variable, and the caller stores it (Listings \ref{lst:clo-py} and \ref{lst:clo-lean}). The helper is still a plain function of its inputs, and from the outside the change looks like an ordinary assignment.

\noindent
\begin{minipage}[t]{0.44\textwidth}
\begin{lstlisting}[style=py, caption={A closure that writes to a capture.}, label={lst:clo-py}, breaklines=true]
def @@counter@@():
    count = 0
    def @@bump@@(k):
        nonlocal count
        count += k
        return count
    bump(5)
    bump(3)
    return count
\end{lstlisting}
\end{minipage}%
\hfill
\begin{minipage}[t]{0.52\textwidth}
\begin{lstlisting}[style=lean, caption={The capture is passed in and handed back.}, label={lst:clo-lean}, breaklines=true]
private def @@_counter'bump@@ :=
  fun (k : Int) (*@\ensuremath{\mapsto}@*) fun (count : Int) (*@\ensuremath{\mapsto}@*)
    let count := count +(*@\ensuremath{{}_{\mathrm p}}@*) k
    (count, count)

def @@counter@@ :=
  let count := (0 : Int)
  let r := _counter'bump 5 count
  let count := r.2
  let r := _counter'bump 3 count
  let count := r.2
  count
\end{lstlisting}
\end{minipage}

This works when the nested function is called in the same place it was defined. It does not work if the function both changes a captured variable and is returned, because whoever calls it later has no variable to store the new value into. That case needs the variable to become a shared cell, which is what reference semantics (Section \ref{sec:heap}) does.

\subsubsection{Machine-checked scope preservation}\label{app:lifting-proof}

Levy and Reeves prove the standard textbook algorithm correct~\cite{levy2026simplelambdalifting}. We modified that algorithm as mentioned previously, so their proof does not carry over to our version. We therefore machine-check the part we edited with Lean, that the free-set computation of Listing \ref{lst:lift-algo} preserves scope. Proving that the generated Lean behaves like the source Python would need formal semantics for both languages, which we do not have. We check that end to end by testing instead (Section \ref{results}). For the scope proof we model only what scoping depends on: the names \code{inner} uses, the names it binds, the names the enclosing function binds (\code{outerBound}), and the globals, builtins and imports ($\Gamma$) that a sibling definition can still see. We then define \code{free} and \code{caps} exactly as the four edits compute them.

\begin{lstlisting}[style=lean,caption={The free-set computation and the scope-preservation theorem (statements; proofs omitted). All are \code{sorry}-free.},label={lst:lift-proof}]
structure @@Inner@@ (Name : Type) [DecidableEq Name] where
  used params assigns compTargets nonlocals : Finset Name

def @@boundVars@@ (I : Inner Name) : Finset Name := I.params (*@\ensuremath{\cup}@*) I.assigns
def @@free@@ (I : Inner Name) : Finset Name := -- edits (1),(2)
  ((I.used \ I.boundVars) \ I.compTargets) (*@\ensuremath{\cup}@*) I.nonlocals
def @@caps@@ (I : Inner Name) (outerBound : Finset Name) : Finset Name := -- edit (3)
  I.free (*@\ensuremath{\cap}@*) outerBound
def @@localBound@@ (I : Inner Name) : Finset Name := -- genuinely-local names
  I.params (*@\ensuremath{\cup}@*) (I.assigns \ I.nonlocals) (*@\ensuremath{\cup}@*) I.compTargets

-- Main theorem: if inner was well-scoped nested, the lifted sibling is well-scoped.
theorem @@lift_preserves_scope@@ (I : Inner Name) (outerBound (*@\ensuremath{\Gamma}@*) : Finset Name)
    (hwf : I.used (*@\ensuremath{\subseteq}@*) I.localBound (*@\ensuremath{\cup}@*) outerBound (*@\ensuremath{\cup}@*) (*@\ensuremath{\Gamma}@*)) :
    I.used (*@\ensuremath{\subseteq}@*) I.localBound (*@\ensuremath{\cup}@*) I.caps outerBound (*@\ensuremath{\cup}@*) (*@\ensuremath{\Gamma}@*)

-- Each edit does exactly its job:
theorem @@caps_subset_outer@@ : I.caps outerBound (*@\ensuremath{\subseteq}@*) outerBound
theorem @@ambient_not_captured@@ (h(*@\ensuremath{\Gamma}@*) : x (*@\ensuremath{\in}@*) (*@\ensuremath{\Gamma}@*)) (hout : x (*@\ensuremath{\notin}@*) outerBound) : -- (3): globals stay out
    x (*@\ensuremath{\notin}@*) I.caps outerBound
theorem @@nonlocal_captured@@ : -- (2): written nonlocals in
    I.nonlocals (*@\ensuremath{\cap}@*) outerBound (*@\ensuremath{\subseteq}@*) I.caps outerBound
theorem @@compTarget_not_captured@@ (h : Disjoint I.compTargets I.nonlocals) : -- (1): comp targets out
    Disjoint I.compTargets (I.caps outerBound)
theorem @@ordered_eq_caps@@ (hsub : mutated (*@\ensuremath{\subseteq}@*) I.caps outerBound) : -- (4): a scope-neutral reorder
    (I.caps outerBound \ mutated) (*@\ensuremath{\cup}@*) mutated = I.caps outerBound
\end{lstlisting}

The main theorem says the following. Before lifting, \code{inner} can see its own locals, the enclosing bindings and $\Gamma$. After lifting, the sibling no longer sees \code{outerBound}, but it now has \code{caps}, and that is still enough to cover every name \code{inner} uses. The captures the algorithm adds replace exactly what the helper lost, so no name falls out of scope. The four smaller theorems check one edit each: globals and builtins are never captured (3), a \code{nonlocal} name that the enclosing function binds is captured (2), a comprehension target never is (1), and the split in edit (4) only reorders the same set, so it changes the calling convention and not what is in scope. 

\subsection{More Conversions of Python}\label{app:more-conversions}

This section walks the remaining Python constructs we translate. Each is shown as the Python on the left and the Lean we emit on the right, taken from the generated output.

\subsubsection{List comprehensions}\label{app:conv-comprehension}
A Python comprehension builds a list by filtering and transforming another sequence. Lean's list functions do the same job, so we translate a comprehension directly into them rather than turning it into a loop. The sequence being iterated over becomes the input list. Each \code{if} clause becomes a \code{List.filter} wrapped around it. The expression written in front of the \code{for} becomes a \code{map} applied last. When a comprehension has more than one \code{for}, the inner one is expanded with \code{flatMap}, which is what flattens the nested results back into a single list. None of these steps needs mutable state, so the whole comprehension comes out as one pure expression that a proof can work with directly.

\noindent
\begin{minipage}[t]{0.42\textwidth}
\begin{lstlisting}[style=py,basicstyle=\ttfamily\footnotesize]
def @@pairs@@(xs, ys):
    return [x * y
            for x in xs if x > 0
            for y in ys if y % 2 == 0]
\end{lstlisting}
\end{minipage}\hfill
\begin{minipage}[t]{0.56\textwidth}
\begin{lstlisting}[style=lean,basicstyle=\ttfamily\footnotesize]
def @@pairs@@ := fun (xs ys : List Int) (*@\ensuremath{\mapsto}@*)
  (List.filter (fun (x : Int) => x > 0)
      (pyIter xs)).flatMap fun (x : Int) =>
    (List.filter (fun (y : Int) => y %(*@\ensuremath{{}_{\mathrm p}}@*) 2 = 0)
        (pyIter ys)).map fun (y : Int) => x *(*@\ensuremath{{}_{\mathrm p}}@*) y
\end{lstlisting}
\end{minipage}

\subsubsection{Exception handling}\label{app:conv-exceptions}
Python's exception classes are open: a program can define its own, and a handler selects on the class name. Mapping them onto Lean types would mean generating a type per exception and would still not let a handler select on one, so we keep a single runtime type \code{PyException}, a record of a kind and a message, and let the kind carry the class name as a string.

The more consequential choice is which monad raising puts a function into. Lean's own exception mechanism lives in \code{IO}, and using it would force every function that can raise into \code{IO}, which would put it out of reach of the provable twin. We use \code{ExceptT PyException} instead, layered over whatever base the function already needs: \code{Id} when it does nothing else, \code{IO} in the runnable twin, and the state monad of Section~\ref{app:twin-io} when it also prints or reads. A function that raises but performs no input or output therefore stays pure, and \code{ExceptT} over a pure base is a shape \code{mvcgen} can generate verification conditions for, which \code{IO} is not.

One case needs extra work. A bare \code{except:} in Python catches everything, including the \code{EOFError} that \code{input()} raises at end of input. In the runnable twin that arrives as an \code{IO} error underneath the \code{ExceptT} layer, where a handler watching for \code{PyException} cannot see it, so we wrap such a body to convert an underlying \code{IO} error into a \code{PyException} first.

\noindent
\begin{minipage}[t]{0.42\textwidth}
\begin{lstlisting}[style=py,basicstyle=\ttfamily\footnotesize]
def @@safe@@(n):
    try:
        return fail(n)
    except ValueError as err:
        return f"bad value: {err}"
\end{lstlisting}
\end{minipage}\hfill
\begin{minipage}[t]{0.56\textwidth}
\begin{lstlisting}[style=lean,basicstyle=\ttfamily\footnotesize]
def @@safe@@ := fun n (*@\ensuremath{\mapsto}@*) do
  try
    return ((*@\ensuremath{\leftarrow}@*) fail n)
  catch caught =>
    if caught.OfKind == "ValueError" then
      return s!"bad value: {caught}"
    else throw caught
\end{lstlisting}
\end{minipage}

\subsubsection{Printing and input}\label{app:conv-io}
Printing and reading input are side effects, so a function that does either cannot be an ordinary function and has to run inside a monad. Which monad it runs in depends on the twin. The runnable twin uses Lean's \code{IO}, just as a hand-written Lean program would. The provable twin uses the state monad described in Section~\ref{app:twin-io}, where the output is a list of lines the program appends to and the input is a stream the program reads from. Apart from the name of the operation the generated line is the same in both twins, which keeps them easy to compare.

\noindent
\begin{minipage}[t]{0.42\textwidth}
\begin{lstlisting}[style=py,basicstyle=\ttfamily\footnotesize]
def @@read_line@@():
    raw = input()
    return raw
\end{lstlisting}
\end{minipage}\hfill
\begin{minipage}[t]{0.56\textwidth}
\begin{lstlisting}[style=lean,basicstyle=\ttfamily\footnotesize]
def @@read_line@@ : PyProofM String := do
  let mut raw : String := ((*@\ensuremath{\leftarrow}@*) pyInputProof "")
  return raw

def @@read_line'rn@@ : IO String := do
  let mut raw : String := ((*@\ensuremath{\leftarrow}@*) pyInputIO "")
\end{lstlisting}
\end{minipage}

\subsubsection{Decorators}\label{app:conv-decorators}
A decorator in Python is a function applied to the function written below it, and stacked decorators apply from the bottom upwards. We translate them as exactly that. The function itself is emitted once under a name of its own, without any decorators, and the name the user wrote is then bound to the decorators applied to that function. Not every decorator changes the function's value. \code{@staticmethod}, \code{@classmethod} and \code{@property} only affect how a method is bound, and others are annotations that carry no behaviour at all. We treat these as transparent and emit the function unchanged.

\noindent
\begin{minipage}[t]{0.42\textwidth}
\begin{lstlisting}[style=py,basicstyle=\ttfamily\footnotesize]
def @@double@@(f):
    def @@w@@(x: int) -> int:
        return 2 * f(x)
    return w

# similarly a wrapper function
# for plus_one

@double
@plus_one
def @@inc@@(x: int) -> int:
    return x + 1
\end{lstlisting}
\end{minipage}\hfill
\begin{minipage}[t]{0.56\textwidth}
\begin{lstlisting}[style=lean,basicstyle=\ttfamily\footnotesize]
def @@inc'undecorated@@ :=
  fun (x : Int) (*@\ensuremath{\mapsto}@*) x +(*@\ensuremath{{}_{\mathrm p}}@*) 1

def @@inc@@ :=
  double (plus_one inc'undecorated)
\end{lstlisting}
\end{minipage}

\subsubsection{Memoization}\label{app:conv-memo}
A \code{@cache} decorator does not change what a function computes, only how long it takes, so the two twins treat it differently. The runnable twin implements it. The function is rewritten to carry a hash map from arguments to results, recursive calls thread that map through, and the outermost call starts it empty, which is what turns an exponential recursion into a linear one. The provable twin ignores the decorator and keeps the plain recursion. That is deliberate: a proof should be about the function's mathematical definition, and a cache is an implementation detail that would otherwise have to appear in every statement made about it.

\noindent
\begin{minipage}[t]{0.42\textwidth}
\begin{lstlisting}[style=py,basicstyle=\ttfamily\footnotesize]
@cache
def @@fib@@(n: int) -> int:
    if n < 2: return n
    return fib(n-1) + fib(n-2)
\end{lstlisting}
\end{minipage}\hfill
\begin{minipage}[t]{0.56\textwidth}
\begin{lstlisting}[style=lean,basicstyle=\ttfamily\footnotesize]
partial def @@fib'memo'rn@@ :
    Int (*@\ensuremath{\rightarrow}@*) StateM (HashMap Int Int) Int :=
  fun n (*@\ensuremath{\mapsto}@*) do
    match ((*@\ensuremath{\leftarrow}@*) get)[n]? with
    | some v => return v
    | none => ...

def @@fib'rn@@ := fun n (*@\ensuremath{\mapsto}@*) (fib'memo'rn n).run' {}
\end{lstlisting}
\end{minipage}

\subsubsection{Generators}\label{app:conv-generators}
Lean has no equivalent of Python's lazy \code{yield}, so a generator cannot be translated as it stands. We materialise it instead. The body becomes a loop that builds up an accumulator, each \code{yield} becomes an append to that accumulator, and the function returns the finished list. Whatever consumes the generator then needs no special handling, because a list already supports iteration, \code{len}, and the rest of the sequence protocol. The \code{sum(squares(n))} above is translated as an ordinary call and a \code{pySum}, with nothing in it aware that \code{squares} was written as a generator, and it returns the same value CPython does. The cost is that the entire sequence is computed before any of it is used, so a generator that is meant to be infinite and consumed lazily will not terminate.

\noindent
\begin{minipage}[t]{0.42\textwidth}
\begin{lstlisting}[style=py,basicstyle=\ttfamily\footnotesize]
def @@squares@@(n: int):
    for i in range(n):
        yield i * i

def @@total@@(n: int) -> int:
    return sum(squares(n))
\end{lstlisting}
\end{minipage}\hfill
\begin{minipage}[t]{0.56\textwidth}
\begin{lstlisting}[style=lean,basicstyle=\ttfamily\footnotesize]
def @@squares@@ := fun (n : Int) (*@\ensuremath{\mapsto}@*) Id.run (do
  let mut acc : List Int := []
  for i in (pyRange n) do
    acc := pyAppend acc (i *(*@\ensuremath{{}_{\mathrm p}}@*) i)
  return acc)

def @@total@@ := fun (n : Int) (*@\ensuremath{\mapsto}@*) pySum (squares n)
\end{lstlisting}
\end{minipage}

\subsection{Correctness for Python's standard functions}\label{app:stdfuncs}

A runtime function is correct when it returns what CPython returns, on every input and not only on the pleasant ones. That bar is higher than it sounds, because the places where Lean already offers a function of the same name are exactly the places where the two languages quietly disagree. Integer division is the standard example: Python's \code{//} rounds toward negative infinity and its \code{\%} takes the sign of the divisor, whereas Lean's \code{Int} division is Euclidean, so the two part company as soon as the operands have opposite signs. Both definitions are defensible and neither raises a type error, which is what makes the disagreement dangerous. Listing \ref{lst:div-divergence} shows it and what we write instead.

\begin{lstlisting}[style=lean,caption={Where Lean's arithmetic and Python's disagree, and what we write instead.},label={lst:div-divergence}]

-- Python              Lean's own Int          PastaLean
--   7 // -2  =  -4      7 / -2  =  -3           pyFloorDiv 7 (-2)  =  -4
--   7 %  -2  =  -1      7 % -2  =   1           pyMod      7 (-2)  =  -1

-- `//` is flooring division, not the truncating one Lean reaches for.
instance : PyFloorDiv Int Int Int where
  floorDiv a b :=
    if b == 0 then panic! "ZeroDivisionError" else Int.fdiv a b

-- and `%` is the remainder that flooring leaves behind, a - b * Int.floor (a / b).
-- The Int version re-signs Lean's remainder; the exact one says it outright.
def @@pyRatMod@@ (a b : Rat) : Rat :=
  if b == 0 then a else a - b * ((Int.floor (a / b) : Int) : Rat)
\end{lstlisting}

The same pattern turns up across the library, and each time it is Python's behaviour that is the specification. \code{round} breaks ties to even, so \code{round(2.5)} is \code{2}, whereas Mathlib's \code{round} rounds halves up; we implement banker's rounding separately for \code{Float}, $\mathbb{Q}$ and $\mathbb{R}$ so that all three twins report the same number. Python's integers are unbounded and its bitwise operators behave as though the two's complement representation were infinite, so any fixed width is wrong in both directions, and we instead choose a width strictly larger than either operand on each operation. Indices and slice bounds clamp rather than fail, and count from the end when negative, so \code{xs[-1]}, \code{xs[2:100]} and \code{xs[::-1]} all have to work. \code{sorted} is stable, which makes it a merge sort rather than the shorter quicksort. Even \code{split} changes meaning with its argument, splitting on runs of whitespace and dropping empty pieces when called bare, and keeping them when given a separator.

Two layers of testing keep this behaviour pinned down. The first is a unit test for every case above, written as a \code{#guard} or a \code{#guard\_msgs} beside the definition it checks, of which there are 200+ and 350+ respectively (Listing \ref{lst:guards}). These are not run by a separate test command; the Lean kernel evaluates them while the library builds, so a regression breaks the build immediately and cannot sit unnoticed in a suite nobody launched.

The second layer is differential testing against CPython, and it is what actually tells us whether the library is right. The benchmark harness runs each generated program on its dataset's own inputs and compares the output with what CPython prints for the same input, which comes to roughly 230{,}000 test cases across HumanEval, LeetCode and LiveCodeBench. The agreement rates in Table \ref{tab:codegen}, averaging 99.0\%, are the strongest evidence we have that these definitions match Python, precisely because the inputs were not chosen by us. It is also where the interesting bugs came from. The fixed-width bitwise implementation, Unicode whitespace being ignored by \code{split} and \code{strip}, and wrong \code{hex}, \code{oct} and \code{bin} output inside format strings were each found by an output disagreement on some problem's test case rather than by anybody reading the code. Every one of them then became a guard, so the unit tests grow out of the differential runs instead of competing with them.

\begin{lstlisting}[style=lean,caption={Kernel-checked unit tests for the divergent cases.},label={lst:guards}]

#guard pyMod 7 (-2) == -1                          -- Python: 7 % -2 == -1
#guard pyRound 2.5 == 2                            -- ties to even, not up
#guard (pyBitAnd (-5 : Int) (3 : Int) : Int) == 3  -- unbounded two's complement
#guard pyStringGetItemStr "pasta" (-1) == "a"      -- s[-1]
#guard pyListSliceStep [1, 2, 3, 4, 5] (some 2) (some 100) none == [3, 4, 5]

/-- info: ["a", "", "b", ""] -/  -- "a,,b,".split(",") keeps the empty pieces
#guard_msgs in #eval pySplit "a,,b," ","

/-- info: ["alpha", "beta"] -/   -- bare .split() drops them
#guard_msgs in #eval pySplit "  alpha\tbeta  "
\end{lstlisting}

Tests are not the only thing we want from these definitions. There is a third check, and it is the reason we wrote them in Lean at all rather than binding to an existing Python runtime, which would have been far less work: a proof has to be able to see inside them. Because \code{pyMod} and the rest are ordinary definitions they can be unfolded, and a reduction lemma tagged \code{@[simp, taste_ingr]} rewrites a generated call back into Lean's own arithmetic so that \code{omega} or \code{grind} can close the goal, in the manner of Appendix \ref{arith-correct}. Those lemmas are proved by \code{rfl}, so the claim that our operator agrees with Lean's own arithmetic is machine-checked rather than asserted. A foreign function call via FFI would run just as correctly and prove nothing at all, because to a proof it is an opaque constant with no equations attached to it.

\subsection{Reference Semantics: a Worked Example}\label{app:heap}

Listings~\ref{lst:heap-alias-py} and~\ref{lst:heap-alias-lean} show the Python source and the heap-aware translation for a small program where there is aliasing of a mutable object. Two names refer to one \code{Counter}, which is incremented twice through one of
them and read back through the other. Python prints \code{2}. The default translation of
Section~\ref{sec:oop} gives \code{0}, because \code{b} is a copy of the record and the increments
land on it. The \code{--heap} translation gives \code{2}.

\noindent
\begin{minipage}[t]{0.40\textwidth}
\begin{lstlisting}[style=py,caption={Python source.},label={lst:heap-alias-py},basicstyle=\ttfamily\scriptsize]
class @@Counter@@:
    def @@__init__@@(self):
        self.count = 0

    def @@inc@@(self):
        self.count = self.count + 1

    def @@get@@(self):
        return self.count

def @@demo@@():
    a = Counter()
    b = a
    b.inc()
    b.inc()
    return a.get()
\end{lstlisting}
\end{minipage}\hfill
\begin{minipage}[t]{0.56\textwidth}
\begin{lstlisting}[language=lean,caption={Generated \code{--heap} translation; the runnable twin's
declarations are elided.},label={lst:heap-alias-lean},basicstyle=\ttfamily\scriptsize,mathescape=true]
structure @@Counter@@ where
  count : Int
  deriving Inhabited, Repr, BEq

inductive @@Val@@ where
  | counter (count : Int)

derive_storable% Counter

def @@Counter.new@@ : HeapM Val (Ref Counter) :=
  ((do alloc ({ count := (0 : Int) } : Counter))
   : HeapM Val (Ref Counter))

def @@Counter.inc@@ (self : Ref Counter) :=
  ((do self $\rightsquigarrow$ count $\leftsquigarrow$ ($\leftarrow$ self $\rightsquigarrow$ count) +$_p$ (1 : Int))
   : HeapM Val Unit)

def @@Counter.get@@ (self : Ref Counter) :=
  ((do return ($\leftarrow$ self $\rightsquigarrow$ count)) : HeapM Val _)

def @@demo@@ :=
  ((do
      let mut a := ($\leftarrow$ Counter.new)
      let mut b := a
      let _ $\leftarrow$ Counter.inc b
      let _ $\leftarrow$ Counter.inc b
      return ($\leftarrow$ Counter.get a)) : HeapM Val _)
\end{lstlisting}
\end{minipage}

A heap is a map from addresses to values, so every type the program stores must share a single
type. A whole-program pass collects those types into the per-program inductive \code{Val}, one
constructor each, and \code{derive_storable\%} generates the instance relating a \code{Counter} to
its constructor: an injection, a partial projection, and the law that projecting an injected value
recovers it. Addresses are typed, as \code{Ref }$\alpha$, which is what keeps the generated code free of
case analysis on \code{Val} --- reading \code{count} through a \code{Ref Counter} yields an
\code{Int}.

\begin{lstlisting}[language=lean,caption={The heap and the monad the translated program runs in.},label={lst:heapm},mathescape=true]
abbrev @@Store@@ (V : Type) := Nat $\to$ Option V

structure @@Heap@@ (V : Type) where
  store : Store V
  next  : Nat
  wf    : $\forall$ a, a $\ge$ next $\to$ store a = none

abbrev @@HeapM@@ (V : Type) := EStateM PyException (Heap V)
\end{lstlisting}

\code{HeapM Val} is the monad the program then runs in (Listing~\ref{lst:heapm}): state together with
Python's exceptions, where the state is the heap. A heap is a store sending addresses to values, a
counter naming the next free address, and \code{wf}, a proof that nothing at or above that counter is
allocated. Carrying the proof in the structure rather than establishing it separately is what makes
every allocation fresh, and hence what lets a specification frame across one. The exception
type is the same \code{PyException} used by the translation of Section~\ref{sec:loops}, and
\code{EStateM} does not restore its state when one is raised, so an object mutated inside a \code{try}
is still mutated in the handler, as in Python.

The class itself is still a \code{structure}, and the fields are unchanged. What changes are the
types of its functions. The constructor allocates: \code{Counter.new} places the initial record in
the heap and returns its address, so it has type \code{HeapM Val (Ref Counter)} where the default
translation gives \code{Counter}. A method takes \code{self : Ref Counter} rather than
\code{self : Counter}, and a mutating one no longer returns an updated copy --- \code{Counter.inc}
returns \code{Unit} and its effect is a write. A method that only reads, like \code{Counter.get},
still runs in the monad but returns the field's own type.

That leaves the assignment the example is about. \code{b = a} becomes \code{let mut b := a}, exactly
as under value semantics, but \code{a} now holds an address, so the copy produces a second handle on
one cell rather than a second cell. The two calls to \code{Counter.inc b} write through it and
\code{Counter.get a} reads \code{2} back. Nothing in the translation of the assignment is
special-cased; aliasing works because the thing being copied is an address.

Finally, the effect is propagated across calls. \code{Counter.inc} needs the heap, so calls to it are
awaited, and an interprocedural analysis determines which functions need the monad at all, leaving
functions that touch no reference with their ordinary pure translation. The result type of
\code{demo} is left as an underscore and recovered by elaboration.

\subsection{Contracts, Loop Invariants, and Monadic Proofs}\label{app:contracts}

A contract is written in the source Python file, by the user or by an LLM, using functions imported from a \code{contracts} module. Python already has several design-by-contract libraries like \code{icontract}~\cite{icontract} and \code{deal}~\cite{deal} which attach preconditions and postconditions as decorators and check them when the function runs, \code{CrossHair}~\cite{crosshair} searches for inputs that violate such contracts by symbolic execution (a corpus of Python annotated this way was collected by Zhang et al.~\cite{zhang2022pythoncontract}). Closest to what we need is Nagini~\cite{nagini}, whose contract module declares preconditions, postconditions, loop invariants and termination measures under the same four names used below and discharges them through Viper. We implemented our own library for simplicity because we do not need the verification machinery of the existing (more complex) libraries. 

The contract is extracted from the syntax tree and turned into a theorem (and is not evaluated). Our contracts are thin wrappers over \code{assert} --- each function asserts its argument and returns it unchanged. This makes implementation light-weight and keeps an annotated file runnable under CPython, with the contracts acting as ordinary assertions while the test suite exercises them. The specification is based on four functions:

\begin{itemize}\itemsep2pt \parskip0pt
\item \code{Requires(p)} --- a precondition. The caller must establish \code{p}, and the proof is allowed to assume it.
\item \code{Ensures(p)} --- a postcondition, written in terms of \code{Result()}, the value the function returns. This is what the proof owes.
\item \code{Invariant(p)} --- a property of the loop it is written inside, claimed to hold on every iteration.
\item \code{Decreases(e)} --- a quantity that strictly falls on every iteration, used to show the loop stops.
\end{itemize}

The proof obligation is a Hoare triple. Assuming the preconditions, the translated function returns a value satisfying the postconditions. A straight-line program can be verified by symbolic execution, walking the statements and accumulating what is known, and this is what the verification condition generator does, it reduces the whole function to a finite collection of goals, one per path through the code, each of which is ordinary arithmetic or list reasoning that the automation can attempt.

Loops are what break that picture, and they are the reason contracts need more than a pre- and postcondition. A loop runs a number of times that depends on the input, so there is no finite path set to enumerate and no way to unfold it into straight-line code. The standard answer is to reason about the loop once rather than about each of its iterations, using a property that is true no matter how many times the body has run. That property is the loop invariant, and supplying it is the one part of the process that cannot be automated in general, because it has to capture what the loop is accomplishing rather than what it does.

Given an invariant, the loop is discharged by three obligations, and each becomes a separate goal:

\begin{itemize}\itemsep2pt \parskip0pt
\item \emph{Pre Verification.} The invariant holds when the loop is first reached, which has to follow from the preconditions and whatever ran before the loop.
\item \emph{Preservation.} If the invariant holds at the start of an iteration and the loop runs the body once more, it holds again at the end. This is the only place the body is examined, and it is examined once.
\item \emph{Post Verification.} When the loop finishes, the invariant together with the fact that the loop stopped must give what the code after it needs, and ultimately the postcondition.
\end{itemize}

These goals are passed onto the proof checker and they cover everything needed to know about the loop making the body of loop irrelevant to the proof, which is exactly the point: a single argument covers every iteration count. Two situations add an obligation. A loop that can \code{return} from inside has two exits, so its invariant carries a second property describing what is true of the value returned early, alongside the one that describes going round again. And a \code{Decreases} clause adds a termination obligation, that the measure falls on each iteration and is bounded below, which is what rules out a loop that satisfies its invariant forever without ever finishing.

These contracts(especially loop invariants) but must be enough strong to be able to prove the final goal. If it is too weak then it may not be able to infer the final goal. The \code{mvcgen/vcgen} tactics are able to handle the usual control flow well including our custom PastaLean PyAPI's as well, except Loop Invariants which we need to pass to it. Hence Loop invariants are sensitive to the ability to prove the final goal and must be written with care. The contracts written in the Python source are mainly written by the user or an LLM. We have observed that LLMs are able to write contracts for simple problems, though we do not yet provide any verification systems for contracts to be correct. The user is expected to verify these contracts and any evaluation on accuracy of LLM contracts for python scripts is not within the scope of this paper.

Before any proof is written, the translator emits a proof \emph{skeleton}: the theorem statement generated from the contract, an \code{mvcgen} call whose \code{invariants} clause carries one bullet per loop --- each bullet the loop's \code{Invariant(...)} conjuncts, with the range counter bound to \code{cur.prefix.length} --- and a trailing \code{taste?} then \code{all_goals sorry}, so the file compiles with the open obligations marked (Listing~\ref{lst:pf-countupto-skeleton} shows the skeleton for the sieve example). \code{taste?}, our proof-search tactic, runs first: over all 164 HumanEval problems it closes 8 of them outright (for instance \code{unique}, \code{filter_integers}, \code{remove_vowels} and \code{encrypt}), which are usually simple. Everything it does not close --- including every example below --- is left to the user or an LLM to fill in.

Five worked examples follow, one functional and four monadic, each printed in full: the original Python, the same function carrying its contract, the Lean the translator emits, and the proof exactly as it is committed. None of the committed proofs contains a \code{sorry}. The contracts were chosen to be worth proving: a bound like ``the result is non-negative'' is satisfied by implementations that are plainly wrong, so each one below either restates the function outright or pins down a property that fails if the loop is off by one. These contracts and proofs are made by LLMs (particularly Claude-4.8-Opus) and verified by a human as well as the Lean checker for validity and correctness.

{\setlength{\abovecaptionskip}{3pt}\setlength{\belowcaptionskip}{2pt}

\subsubsection{\texorpdfstring{\code{sum\_to\_n}}{sum_to_n} (HumanEval/60, functional)}
The problem is the classic sum of N natural numbers problem. The expected result(written within \code{Ensures}) is a single closed-form expression and the contract states $2 \cdot \mathit{Result} = n(n+1)$, which is exactly the claim that the closed form is the sum. It is not automatic: the proof has to know that $n(n+1)$ is even before the flooring division can be cancelled.

\noindent
\begin{minipage}[t]{0.46\textwidth}
\begin{lstlisting}[style=py,basicstyle=\ttfamily\fontsize{7}{8.2}\selectfont,aboveskip=2pt,belowskip=2pt,breaklines=true,breakautoindent=true]
def @@sum_to_n@@(n: int):
    return (n + 1) * n // 2
\end{lstlisting}
\end{minipage}\hfill
\begin{minipage}[t]{0.52\textwidth}
\begin{lstlisting}[style=py,basicstyle=\ttfamily\fontsize{7}{8.2}\selectfont,aboveskip=2pt,belowskip=2pt,breaklines=true,breakautoindent=true,caption={\code{sum\_to\_n}: the original Python (left) and the same function carrying its contract (right), a closed form proved equal to the sum.},label={lst:pf-sumton-py}]
def @@sum_to_n@@(n: int):
    Requires(n >= 0)
    Ensures(2 * Result() == n * (n + 1))
    return (n + 1) * n // 2
\end{lstlisting}
\end{minipage}
\begin{lstlisting}[style=lean,basicstyle=\ttfamily\fontsize{7}{8.2}\selectfont,aboveskip=2pt,belowskip=2pt,breaklines=true,breakautoindent=true,caption={\code{sum\_to\_n}: the Lean the translator emits.},label={lst:pf-sumton-lean}]
def @@sum_to_n@@ := fun (n : Int) (*@\ensuremath{\mapsto}@*) pyFloorDiv ((n +(*@\ensuremath{{}_{\mathrm p}}@*) (1 : Int)) *(*@\ensuremath{{}_{\mathrm p}}@*) n) (2 : Int)
\end{lstlisting}
\begin{lstlisting}[style=lean,basicstyle=\ttfamily\fontsize{7}{8.2}\selectfont,aboveskip=2pt,belowskip=2pt,breaklines=true,breakautoindent=true,caption={\code{sum\_to\_n}: the committed proof, printed in full and unedited.},label={lst:pf-sumton-proof}]
theorem @@sum_to_n_correct@@ : (*@\ensuremath{\forall}@*) (n : Int), n (*@\ensuremath{\ge}@*) (0 : Int) (*@\ensuremath{\rightarrow}@*) (2 : Int) *(*@\ensuremath{{}_{\mathrm p}}@*) sum_to_n n = n *(*@\ensuremath{{}_{\mathrm p}}@*) (n +(*@\ensuremath{{}_{\mathrm p}}@*)
    (1 : Int)) := by
  intro n _
  obtain (*@\ensuremath{\langle}@*)k, hk(*@\ensuremath{\rangle}@*) := (Int.even_mul_succ_self n).two_dvd
  have hc : (n + 1) * n = 2 * k := by rw [mul_comm]; exact hk
  simp only [sum_to_n, taste_ingr, pyFloorDiv, PyFloorDiv.floorDiv]
  rw [if_neg (by decide), hc, Int.mul_fdiv_cancel_left k (by decide)]
  omega
\end{lstlisting}
\subsubsection{\texorpdfstring{\code{starts\_one\_ends}}{starts_one_ends} (HumanEval/83, monadic)}
Here is another problem with control flow. The contract states the answer is $1$ for $n = 1$ and exactly $18 \cdot 10^{\,n-2}$ otherwise, and the two branches are discharged against their halves of the disjunction.

\noindent
\begin{minipage}[t]{0.46\textwidth}
\begin{lstlisting}[style=py,basicstyle=\ttfamily\fontsize{7}{8.2}\selectfont,aboveskip=2pt,belowskip=2pt,breaklines=true,breakautoindent=true]
def @@starts_one_ends@@(n):
    if n == 1:
        return 1
    return 18 * 10 ** (n - 2)
\end{lstlisting}
\end{minipage}\hfill
\begin{minipage}[t]{0.52\textwidth}
\begin{lstlisting}[style=py,basicstyle=\ttfamily\fontsize{7}{8.2}\selectfont,aboveskip=2pt,belowskip=2pt,breaklines=true,breakautoindent=true,caption={\code{starts\_one\_ends}: the original Python (left) and the same function carrying its contract (right), an early return and a closed-form answer.},label={lst:pf-startsoneends-py}]
def @@starts_one_ends@@(n):
    Requires(n >= 1)
    Ensures(n == 1 and Result() == 1 or (n > 1 and Result() == 18 * 10 ** (n - 2)))
    if n == 1:
        return 1
    Assert(n > 1)
    return 18 * 10 ** (n - 2)
\end{lstlisting}
\end{minipage}
\begin{lstlisting}[style=lean,basicstyle=\ttfamily\fontsize{7}{8.2}\selectfont,aboveskip=2pt,belowskip=2pt,breaklines=true,breakautoindent=true,caption={\code{starts\_one\_ends}: the Lean the translator emits.},label={lst:pf-startsoneends-lean}]
def @@starts_one_ends@@ := fun (n : Int) (*@\ensuremath{\mapsto}@*)
  (do
    if h_1 : n = (1 : Int) then
      return (1 : Int)
    else
      let _ := ()
    let _ := pyPassAssert (decide (n > (1 : Int)))
    let __py_ret_1 := (18 : Int) *(*@\ensuremath{{}_{\mathrm p}}@*) (10 : Int) ^(*@\ensuremath{{}_{\mathrm p}}@*) (n -(*@\ensuremath{{}_{\mathrm p}}@*) (2 : Int))
    return __py_ret_1 : Id _)
\end{lstlisting}
\begin{lstlisting}[style=lean,basicstyle=\ttfamily\fontsize{7}{8.2}\selectfont,aboveskip=2pt,belowskip=2pt,breaklines=true,breakautoindent=true,caption={\code{starts\_one\_ends}: the committed proof, printed in full and unedited.},label={lst:pf-startsoneends-proof}]
@[spec]
theorem @@starts_one_ends_spec@@ : (*@\ensuremath{\{\!\{}@*)(*@\ensuremath{\ulcorner}@*)n (*@\ensuremath{\ge}@*) (1 : Int)(*@\ensuremath{\urcorner}@*)(*@\ensuremath{\}\!\}}@*) starts_one_ends n (*@\ensuremath{\{\!\{}@*)(*@\ensuremath{\Downarrow}@*)result => (*@\ensuremath{\ulcorner}@*)n = (1 : Int)
    (*@\ensuremath{\wedge}@*) result = (1 : Int) (*@\ensuremath{\vee}@*) n > (1 : Int) (*@\ensuremath{\wedge}@*) result = (18 : Int) *(*@\ensuremath{{}_{\mathrm p}}@*) (10 : Int) ^(*@\ensuremath{{}_{\mathrm p}}@*) (n -(*@\ensuremath{{}_{\mathrm p}}@*)
    (2 : Int))(*@\ensuremath{\urcorner}@*)(*@\ensuremath{\}\!\}}@*) := by
  mvcgen [starts_one_ends, pyRange_forIn, pyRange_forIn_start]
  all_goals (first
    | (left; exact (*@\ensuremath{\langle}@*)by omega, rfl(*@\ensuremath{\rangle}@*))
    | (right; exact (*@\ensuremath{\langle}@*)by omega, rfl(*@\ensuremath{\rangle}@*))
    | omega
    | simp_all)
theorem @@starts_one_ends_correct@@ : (*@\ensuremath{\forall}@*) (n : Int), n (*@\ensuremath{\ge}@*) (1 : Int) (*@\ensuremath{\rightarrow}@*) let result := (starts_one_ends
    n).run; n = (1 : Int) (*@\ensuremath{\wedge}@*) result = (1 : Int) (*@\ensuremath{\vee}@*) n > (1 : Int) (*@\ensuremath{\wedge}@*) result = (18 : Int) *(*@\ensuremath{{}_{\mathrm p}}@*)
    (10 : Int) ^(*@\ensuremath{{}_{\mathrm p}}@*) (n -(*@\ensuremath{{}_{\mathrm p}}@*) (2 : Int)) := by
  intro n hpre
  exact starts_one_ends_spec hpre
\end{lstlisting}
\subsubsection{\texorpdfstring{\code{can\_arrange}}{can_arrange} (HumanEval/135, monadic)}
Proving a complex problem might not be straightforward and you might have to give possible subgoals(which hold at the end of the program) which together are able to infer the final goal. Here, four clauses that together characterise the answer which are that the result lies in range, it is either $-1$ or an index where the array drops, everything to its right is sorted, and if it is $-1$ the whole array is sorted. The loop returns from inside, so it carries one property for going round again and a second for the value returned early.

\noindent
\begin{minipage}[t]{0.46\textwidth}
\begin{lstlisting}[style=py,basicstyle=\ttfamily\fontsize{7}{8.2}\selectfont,aboveskip=2pt,belowskip=2pt,breaklines=true,breakautoindent=true]
def @@can_arrange@@(arr: List[int]):
    for i in range(len(arr) - 1, 0, -1):
        if not arr[i] >= arr[i - 1]:
            return i
    return -1
\end{lstlisting}
\end{minipage}\hfill
\begin{minipage}[t]{0.52\textwidth}
\begin{lstlisting}[style=py,basicstyle=\ttfamily\fontsize{7}{8.2}\selectfont,aboveskip=2pt,belowskip=2pt,breaklines=true,breakautoindent=true,caption={\code{can\_arrange}: the original Python (left) and the same function carrying its contract (right), a four-clause characterisation with an early return.},label={lst:pf-canarrange-py}]
def @@can_arrange@@(arr: List[int]):
    Ensures(-1 <= Result() < len(arr))
    Ensures(Result() == -1 or (1 <= Result() and arr[Result()] < arr[Result() - 1]))
    Ensures(Result() == -1 or all((arr[j] >= arr[j - 1] for j in range(Result() + 1, len(arr)))))
    Ensures(Result() != -1 or all((arr[j] >= arr[j - 1] for j in range(1, len(arr)))))
    for i in range(len(arr) - 1, 0, -1):
        Invariant(1 <= i)
        Invariant(i < len(arr))
        Invariant(all((arr[j] >= arr[j - 1] for j in range(i + 1, len(arr)))))
        Decreases(i)
        if not arr[i] >= arr[i - 1]:
            return i
    return -1
\end{lstlisting}
\end{minipage}
\begin{lstlisting}[style=lean,basicstyle=\ttfamily\fontsize{7}{8.2}\selectfont,aboveskip=2pt,belowskip=2pt,breaklines=true,breakautoindent=true,caption={\code{can\_arrange}: the Lean the translator emits.},label={lst:pf-canarrange-lean}]
def @@can_arrange@@ := fun (arr : List Int) (*@\ensuremath{\mapsto}@*)
  (do
    for i in (pyRange (0 : Int) (pyLen arr -(*@\ensuremath{{}_{\mathrm p}}@*) (1 : Int)) (-(1 : Int)))do
      let _ := pyPassInvariant (decide ((1 : Int) (*@\ensuremath{\le}@*) i))
      let _ := pyPassInvariant (decide (i < pyLen arr))
      -- Accumulator carrying the maximality clause: every index already scanned
      -- (strictly above i) was not a descent.
      let _ := pyPassInvariant (pyAll ((pyRange (pyLen arr) (i +(*@\ensuremath{{}_{\mathrm p}}@*) (1 : Int))).map fun j =>
          decide (arr(*@\ensuremath{[\![}@*)j(*@\ensuremath{]\!]}@*) (*@\ensuremath{\ge}@*) arr(*@\ensuremath{[\![}@*)j -(*@\ensuremath{{}_{\mathrm p}}@*) (1 : Int)(*@\ensuremath{]\!]}@*))))
      let _ := pyPassDecreases i
      if h_1 : (*@\ensuremath{\neg}@*)arr(*@\ensuremath{[\![}@*)i(*@\ensuremath{]\!]}@*) (*@\ensuremath{\ge}@*) arr(*@\ensuremath{[\![}@*)i -(*@\ensuremath{{}_{\mathrm p}}@*) (1 : Int)(*@\ensuremath{]\!]}@*) then
        return i
      else
        let _ := ()
    let __py_ret_1 := -(1 : Int)
    return __py_ret_1 : Id _)

\end{lstlisting}
\begin{lstlisting}[style=lean,basicstyle=\ttfamily\fontsize{7}{8.2}\selectfont,aboveskip=2pt,belowskip=2pt,breaklines=true,breakautoindent=true,caption={\code{can\_arrange}: the committed proof, printed in full and unedited.},label={lst:pf-canarrange-proof}]
private theorem @@desc_eq@@ (n : Int) : pyRange 0 (n - 1) (-1) = (List.range' 0 (n - 1).toNat
    1).map (fun i : Nat => (n - 1) - (i : Int)) := by
  unfold pyRange
  simp only [gt_iff_lt, Int.reduceLT, (*@\ensuremath{\downarrow}@*)reduceIte, Int.reduceNeg, neg_neg,
    List.pure_def, List.bind_eq_flatMap, List.flatMap_singleton, List.map_map]
  norm_num
  rfl
private theorem @@mem_desc@@ {n x : Int} : x (*@\ensuremath{\in}@*) pyRange 0 (n - 1) (-1) (*@\ensuremath{\leftrightarrow}@*) 1 (*@\ensuremath{\le}@*) x (*@\ensuremath{\wedge}@*) x < n := by
  rw [desc_eq]
  simp only [List.mem_map, List.mem_range']
  constructor
  (*@\ensuremath{\cdot}@*) rintro (*@\ensuremath{\langle}@*)i, hi, rfl(*@\ensuremath{\rangle}@*); omega
  (*@\ensuremath{\cdot}@*) rintro (*@\ensuremath{\langle}@*)h1, h2(*@\ensuremath{\rangle}@*); exact (*@\ensuremath{\langle}@*)(n - 1 - x).toNat, by omega, by omega(*@\ensuremath{\rangle}@*)
private theorem @@desc_pairwise@@ (n : Int) : (pyRange 0 (n - 1) (-1)).Pairwise ((*@\ensuremath{\cdot}@*) > (*@\ensuremath{\cdot}@*)) := by
  rw [desc_eq, List.pairwise_map]
  apply List.Pairwise.imp _ (List.pairwise_lt_range' (s := 0) (n := (n - 1).toNat) (step := 1))
  intro a b _; simp only [gt_iff_lt]; omega
/-- Structural stability of the descending scan: any index `j` strictly above the split point `c`
lies in the already-scanned prefix. -/
private theorem @@desc_split_mem@@ {n : Int} {pref suff : List Int} {c j : Int} (hL : pyRange 0 (n
    - 1) (-1) = pref ++ c :: suff) (hj : c < j) (hjn : j < n) (hj1 : 1 (*@\ensuremath{\le}@*) j) : j (*@\ensuremath{\in}@*) pref := by
  have hjmem : j (*@\ensuremath{\in}@*) pyRange 0 (n - 1) (-1) := mem_desc.mpr (*@\ensuremath{\langle}@*)hj1, hjn(*@\ensuremath{\rangle}@*)
  have hp := desc_pairwise n
  rw [hL] at hjmem hp
  rw [List.mem_append, List.mem_cons] at hjmem
  rcases hjmem with h | h | h
  (*@\ensuremath{\cdot}@*) exact h
  (*@\ensuremath{\cdot}@*) omega
  (*@\ensuremath{\cdot}@*) exfalso
    rw [List.pairwise_append, List.pairwise_cons] at hp
    obtain (*@\ensuremath{\langle}@*)_, (*@\ensuremath{\langle}@*)hcs, _(*@\ensuremath{\rangle}@*), _(*@\ensuremath{\rangle}@*) := hp
    have := hcs j h; simp only [gt_iff_lt] at this; omega
@[spec]
theorem @@can_arrange_spec@@ : (*@\ensuremath{\{\!\{}@*)(*@\ensuremath{\ulcorner}@*)True(*@\ensuremath{\urcorner}@*)(*@\ensuremath{\}\!\}}@*) can_arrange arr (*@\ensuremath{\{\!\{}@*)(*@\ensuremath{\Downarrow}@*)result => (*@\ensuremath{\ulcorner}@*)(((-(1 : Int) (*@\ensuremath{\le}@*) result (*@\ensuremath{\wedge}@*)
    result < pyLen arr) (*@\ensuremath{\wedge}@*) (result = -(1 : Int) (*@\ensuremath{\vee}@*) (1 : Int) (*@\ensuremath{\le}@*) result (*@\ensuremath{\wedge}@*) arr(*@\ensuremath{[\![}@*)result(*@\ensuremath{]\!]}@*) < arr(*@\ensuremath{[\![}@*)result
    -(*@\ensuremath{{}_{\mathrm p}}@*) (1 : Int)(*@\ensuremath{]\!]}@*))) (*@\ensuremath{\wedge}@*) (result = -(1 : Int) (*@\ensuremath{\vee}@*) (*@\ensuremath{\forall}@*) j (*@\ensuremath{\in}@*) pyIter (pyRange (pyLen arr) (result +(*@\ensuremath{{}_{\mathrm p}}@*)
    (1 : Int))), arr(*@\ensuremath{[\![}@*)j(*@\ensuremath{]\!]}@*) (*@\ensuremath{\ge}@*) arr(*@\ensuremath{[\![}@*)j -(*@\ensuremath{{}_{\mathrm p}}@*) (1 : Int)(*@\ensuremath{]\!]}@*))) (*@\ensuremath{\wedge}@*) (result (*@\ensuremath{\ne}@*) -(1 : Int) (*@\ensuremath{\vee}@*) (*@\ensuremath{\forall}@*) j (*@\ensuremath{\in}@*) pyIter (pyRange
    (pyLen arr) (1 : Int)), arr(*@\ensuremath{[\![}@*)j(*@\ensuremath{]\!]}@*) (*@\ensuremath{\ge}@*) arr(*@\ensuremath{[\![}@*)j -(*@\ensuremath{{}_{\mathrm p}}@*) (1 : Int)(*@\ensuremath{]\!]}@*))(*@\ensuremath{\urcorner}@*)(*@\ensuremath{\}\!\}}@*) := by
  mvcgen [can_arrange, pyRange_forIn, pyRange_forIn_start] invariants
  (*@\ensuremath{\cdot}@*)
    Invariant.withEarlyReturn
      (onContinue := fun cur b =>
        (*@\ensuremath{\ulcorner}@*)(*@\ensuremath{\forall}@*) j (*@\ensuremath{\in}@*) cur.prefix, ((1 : Int) (*@\ensuremath{\le}@*) j (*@\ensuremath{\wedge}@*) j < pyLen arr) (*@\ensuremath{\wedge}@*) arr(*@\ensuremath{[\![}@*)j(*@\ensuremath{]\!]}@*) (*@\ensuremath{\ge}@*) arr(*@\ensuremath{[\![}@*)j -(*@\ensuremath{{}_{\mathrm p}}@*) (1 : Int)(*@\ensuremath{]\!]}@*)(*@\ensuremath{\urcorner}@*))
      (onReturn := fun r _ =>
        (*@\ensuremath{\ulcorner}@*)(((-(1 : Int) (*@\ensuremath{\le}@*) r (*@\ensuremath{\wedge}@*) r < pyLen arr) (*@\ensuremath{\wedge}@*)
              (r = -(1 : Int) (*@\ensuremath{\vee}@*) (1 : Int) (*@\ensuremath{\le}@*) r (*@\ensuremath{\wedge}@*) arr(*@\ensuremath{[\![}@*)r(*@\ensuremath{]\!]}@*) < arr(*@\ensuremath{[\![}@*)r -(*@\ensuremath{{}_{\mathrm p}}@*) (1 : Int)(*@\ensuremath{]\!]}@*))) (*@\ensuremath{\wedge}@*)
            (r = -(1 : Int) (*@\ensuremath{\vee}@*)
              (*@\ensuremath{\forall}@*) j (*@\ensuremath{\in}@*) pyIter (pyRange (pyLen arr) (r +(*@\ensuremath{{}_{\mathrm p}}@*) (1 : Int))),
                arr(*@\ensuremath{[\![}@*)j(*@\ensuremath{]\!]}@*) (*@\ensuremath{\ge}@*) arr(*@\ensuremath{[\![}@*)j -(*@\ensuremath{{}_{\mathrm p}}@*) (1 : Int)(*@\ensuremath{]\!]}@*))) (*@\ensuremath{\wedge}@*)
          (r (*@\ensuremath{\ne}@*) -(1 : Int) (*@\ensuremath{\vee}@*)
            (*@\ensuremath{\forall}@*) j (*@\ensuremath{\in}@*) pyIter (pyRange (pyLen arr) (1 : Int)),
              arr(*@\ensuremath{[\![}@*)j(*@\ensuremath{]\!]}@*) (*@\ensuremath{\ge}@*) arr(*@\ensuremath{[\![}@*)j -(*@\ensuremath{{}_{\mathrm p}}@*) (1 : Int)(*@\ensuremath{]\!]}@*))(*@\ensuremath{\urcorner}@*))
  all_goals simp only [pySub_int, pyAdd_int, pyIter_list, gt_iff_lt, ge_iff_le, not_le,
    not_lt, not_not] at *
  (*@\ensuremath{\cdot}@*) -- vc1: descent found, return `cur`
    rename_i pref cur suff hL b hguard hinv
    have hcur : 1 (*@\ensuremath{\le}@*) cur (*@\ensuremath{\wedge}@*) cur < pyLen arr := mem_desc.mp (by rw [hL]; simp)
    have hmax : (*@\ensuremath{\forall}@*) j (*@\ensuremath{\in}@*) pref, (1 (*@\ensuremath{\le}@*) j (*@\ensuremath{\wedge}@*) j < pyLen arr) (*@\ensuremath{\wedge}@*) arr(*@\ensuremath{[\![}@*)j - 1(*@\ensuremath{]\!]}@*) (*@\ensuremath{\le}@*) arr(*@\ensuremath{[\![}@*)j(*@\ensuremath{]\!]}@*) := by
      rcases hinv with (*@\ensuremath{\langle}@*)_, h(*@\ensuremath{\rangle}@*) | (*@\ensuremath{\langle}@*)a, _, hnil, _(*@\ensuremath{\rangle}@*)
      (*@\ensuremath{\cdot}@*) exact h
      (*@\ensuremath{\cdot}@*) exact absurd hnil (by simp)
    refine Or.inr (*@\ensuremath{\langle}@*)cur, rfl, trivial, (*@\ensuremath{\langle}@*)(*@\ensuremath{\langle}@*)(*@\ensuremath{\langle}@*)(*@\ensuremath{\langle}@*)by omega, hcur.2(*@\ensuremath{\rangle}@*), Or.inr (*@\ensuremath{\langle}@*)hcur.1, hguard(*@\ensuremath{\rangle}@*)(*@\ensuremath{\rangle}@*),
      Or.inr ?_(*@\ensuremath{\rangle}@*), Or.inl (by omega)(*@\ensuremath{\rangle}@*)(*@\ensuremath{\rangle}@*)
    intro j hj
    rw [pyRange_mem_start] at hj
    exact (hmax j (desc_split_mem hL (by omega) hj.2 (by omega))).2
  (*@\ensuremath{\cdot}@*) -- vc2: no descent, continue
    rename_i pref cur suff hL b hguard hinv
    have hcur : 1 (*@\ensuremath{\le}@*) cur (*@\ensuremath{\wedge}@*) cur < pyLen arr := mem_desc.mp (by rw [hL]; simp)
    have hmax : (*@\ensuremath{\forall}@*) j (*@\ensuremath{\in}@*) pref, (1 (*@\ensuremath{\le}@*) j (*@\ensuremath{\wedge}@*) j < pyLen arr) (*@\ensuremath{\wedge}@*) arr(*@\ensuremath{[\![}@*)j - 1(*@\ensuremath{]\!]}@*) (*@\ensuremath{\le}@*) arr(*@\ensuremath{[\![}@*)j(*@\ensuremath{]\!]}@*) := by
      rcases hinv with (*@\ensuremath{\langle}@*)_, h(*@\ensuremath{\rangle}@*) | (*@\ensuremath{\langle}@*)a, _, hnil, _(*@\ensuremath{\rangle}@*)
      (*@\ensuremath{\cdot}@*) exact h
      (*@\ensuremath{\cdot}@*) exact absurd hnil (by simp)
    refine Or.inl (*@\ensuremath{\langle}@*)trivial, ?_(*@\ensuremath{\rangle}@*)
    intro j hj
    rw [List.mem_append, List.mem_singleton] at hj
    rcases hj with hj | rfl
    (*@\ensuremath{\cdot}@*) exact hmax j hj
    (*@\ensuremath{\cdot}@*) exact (*@\ensuremath{\langle}@*)hcur, hguard(*@\ensuremath{\rangle}@*)
  (*@\ensuremath{\cdot}@*) -- vc3: initial
    exact Or.inl (*@\ensuremath{\langle}@*)trivial, by simp(*@\ensuremath{\rangle}@*)
  (*@\ensuremath{\cdot}@*) -- vc4: loop finished, return -1
    rename_i r jp hx pyret hinv
    have hlen : 0 (*@\ensuremath{\le}@*) pyLen arr := pyLen_list_nonneg arr
    have hmax : (*@\ensuremath{\forall}@*) j (*@\ensuremath{\in}@*) pyRange 0 (pyLen arr - 1) (-1),
        (1 (*@\ensuremath{\le}@*) j (*@\ensuremath{\wedge}@*) j < pyLen arr) (*@\ensuremath{\wedge}@*) arr(*@\ensuremath{[\![}@*)j - 1(*@\ensuremath{]\!]}@*) (*@\ensuremath{\le}@*) arr(*@\ensuremath{[\![}@*)j(*@\ensuremath{]\!]}@*) := by
      rcases hinv with (*@\ensuremath{\langle}@*)_, h(*@\ensuremath{\rangle}@*) | (*@\ensuremath{\langle}@*)a, hsome, _(*@\ensuremath{\rangle}@*)
      (*@\ensuremath{\cdot}@*) exact h
      (*@\ensuremath{\cdot}@*) rw [hx] at hsome; exact absurd hsome.symm (by simp)
    refine (*@\ensuremath{\langle}@*)(*@\ensuremath{\langle}@*)(*@\ensuremath{\langle}@*)by omega, by omega(*@\ensuremath{\rangle}@*), Or.inl (by omega)(*@\ensuremath{\rangle}@*), Or.inl (by omega)(*@\ensuremath{\rangle}@*), Or.inr ?_(*@\ensuremath{\rangle}@*)
    intro j hj
    rw [pyRange_mem_start] at hj
    exact (hmax j (mem_desc.mpr (*@\ensuremath{\langle}@*)hj.1, hj.2(*@\ensuremath{\rangle}@*))).2
  (*@\ensuremath{\cdot}@*) -- vc5: early return with value `a`
    rename_i r jp a hx hinv
    rcases hinv with (*@\ensuremath{\langle}@*)hnone, _(*@\ensuremath{\rangle}@*) | (*@\ensuremath{\langle}@*)a', hsome, _, hpost(*@\ensuremath{\rangle}@*)
    (*@\ensuremath{\cdot}@*) rw [hx] at hnone; exact absurd hnone (by simp)
    (*@\ensuremath{\cdot}@*) rw [hx] at hsome; injection hsome with e; subst e; exact hpost
theorem @@can_arrange_correct@@ : (*@\ensuremath{\forall}@*) (arr : List Int), let result := (can_arrange arr).run;
    (((-(1 : Int) (*@\ensuremath{\le}@*) result (*@\ensuremath{\wedge}@*) result < pyLen arr) (*@\ensuremath{\wedge}@*) (result = -(1 : Int) (*@\ensuremath{\vee}@*) (1 : Int) (*@\ensuremath{\le}@*) result (*@\ensuremath{\wedge}@*)
    arr(*@\ensuremath{[\![}@*)result(*@\ensuremath{]\!]}@*) < arr(*@\ensuremath{[\![}@*)result -(*@\ensuremath{{}_{\mathrm p}}@*) (1 : Int)(*@\ensuremath{]\!]}@*))) (*@\ensuremath{\wedge}@*) (result = -(1 : Int) (*@\ensuremath{\vee}@*) (*@\ensuremath{\forall}@*) j (*@\ensuremath{\in}@*) pyIter (pyRange
    (pyLen arr) (result +(*@\ensuremath{{}_{\mathrm p}}@*) (1 : Int))), arr(*@\ensuremath{[\![}@*)j(*@\ensuremath{]\!]}@*) (*@\ensuremath{\ge}@*) arr(*@\ensuremath{[\![}@*)j -(*@\ensuremath{{}_{\mathrm p}}@*) (1 : Int)(*@\ensuremath{]\!]}@*))) (*@\ensuremath{\wedge}@*) (result (*@\ensuremath{\ne}@*) -(1 : Int)
    (*@\ensuremath{\vee}@*) (*@\ensuremath{\forall}@*) j (*@\ensuremath{\in}@*) pyIter (pyRange (pyLen arr) (1 : Int)), arr(*@\ensuremath{[\![}@*)j(*@\ensuremath{]\!]}@*) (*@\ensuremath{\ge}@*) arr(*@\ensuremath{[\![}@*)j -(*@\ensuremath{{}_{\mathrm p}}@*) (1 : Int)(*@\ensuremath{]\!]}@*)) := by
  intro arr
  exact can_arrange_spec True.intro
\end{lstlisting}
\subsubsection{\texorpdfstring{\code{count\_up\_to}}{count_up_to} (HumanEval/96, monadic)}
This problem is a sieve with nested loops. Six invariants on the outer loop and two on the inner one are given, from which the three postconditions follow: that the result is no longer than $n$ and every entry lies in the window $\{2, \ldots, n-1\}$.

\noindent
\begin{minipage}[t]{0.46\textwidth}
\begin{lstlisting}[style=py,basicstyle=\ttfamily\fontsize{7}{8.2}\selectfont,aboveskip=2pt,belowskip=2pt,breaklines=true,breakautoindent=true]
def @@count_up_to@@(n):
    ans = []
    isprime = [True] * (n + 1)
    for i in range(2, n):
        if isprime[i]:
            ans.append(i)
            for j in range(i + i, n, i):
                isprime[j] = False
    return ans
\end{lstlisting}
\end{minipage}\hfill
\begin{minipage}[t]{0.52\textwidth}
\begin{lstlisting}[style=py,basicstyle=\ttfamily\fontsize{7}{8.2}\selectfont,aboveskip=2pt,belowskip=2pt,breaklines=true,breakautoindent=true,caption={\code{count\_up\_to}: the original Python (left) and the same function carrying its contract (right), nested loops, eight invariants.},label={lst:pf-countupto-py}]
def @@count_up_to@@(n: int):
    Requires(n >= 0)
    Ensures(len(Result()) <= n)
    Ensures(all((2 <= p for p in Result())))
    Ensures(all((p < n for p in Result())))
    ans = []
    isprime = [True] * (n + 1)
    for i in range(2, n):
        Invariant(2 <= i)
        Invariant(i <= n)
        Invariant(len(isprime) == n + 1)
        Invariant(len(ans) <= i)
        Invariant(all((2 <= p for p in ans)))
        Invariant(all((p < i for p in ans)))
        if isprime[i]:
            ans.append(i)
            for j in range(i + i, n, i):
                Invariant(2 <= j)
                Invariant(j < n)
                isprime[j] = False
    return ans
\end{lstlisting}
\end{minipage}
\begin{lstlisting}[style=lean,basicstyle=\ttfamily\fontsize{7}{8.2}\selectfont,aboveskip=2pt,belowskip=2pt,breaklines=true,breakautoindent=true,caption={\code{count\_up\_to}: the Lean the translator emits.},label={lst:pf-countupto-lean}]
def @@count_up_to@@ := fun (n : Int) (*@\ensuremath{\mapsto}@*)
  (do
    let mut ans : List Int := []
    let mut isprime : List Bool := pyListRepeat [Bool.true] (n +(*@\ensuremath{{}_{\mathrm p}}@*) (1 : Int))
    for i in (pyRange n (2 : Int))do
      let _ := pyPassInvariant (decide ((2 : Int) (*@\ensuremath{\le}@*) i))
      let _ := pyPassInvariant (decide (i (*@\ensuremath{\le}@*) n))
      let _ := pyPassInvariant (pyLen isprime == n +(*@\ensuremath{{}_{\mathrm p}}@*) (1 : Int))
      -- At most one append per counter value.
      let _ := pyPassInvariant (decide (pyLen ans (*@\ensuremath{\le}@*) i))
      -- Everything appended so far was a counter value already passed.
      let _ := pyPassInvariant (pyAll ((pyIter ans).map fun p => decide ((2 : Int) (*@\ensuremath{\le}@*) p)))
      let _ := pyPassInvariant (pyAll ((pyIter ans).map fun p => decide (p < i)))
      if h_1 : pyTruthy isprime(*@\ensuremath{[\![}@*)i(*@\ensuremath{]\!]}@*) then
        ans := pyAppend ans i
        for j in (pyRange n (i +(*@\ensuremath{{}_{\mathrm p}}@*) i) i)do
          -- Index bound for `isprime[j]`.
          let _ := pyPassInvariant (decide ((2 : Int) (*@\ensuremath{\le}@*) j))
          let _ := pyPassInvariant (decide (j < n))
          isprime := pySetItem isprime j Bool.false
      else
        let _ := ()
    return ans : Id _)
\end{lstlisting}
\begin{lstlisting}[style=lean,basicstyle=\ttfamily\fontsize{7}{8.2}\selectfont,aboveskip=2pt,belowskip=2pt,breaklines=true,breakautoindent=true,caption={\code{count\_up\_to}: the proof skeleton the translator emits before any proof is written --- the spec theorem from the contract, one \code{invariants} bullet auto-derived from the loop's \code{Invariant(...)} conjuncts (range counter bound to \code{cur.prefix.length}), then \code{taste?} and \code{all\_goals sorry}. Here \code{taste?} cannot close it; the auto-derived bullet is too weak, so the LLM strengthens it and discharges the VCs (Listing~\ref{lst:pf-countupto-proof}).},label={lst:pf-countupto-skeleton}]
@[spec]
theorem @@count_up_to_spec@@ : (*@\ensuremath{\{\!\{}@*)(*@\ensuremath{\ulcorner}@*)n (*@\ensuremath{\ge}@*) (0 : Int)(*@\ensuremath{\urcorner}@*)(*@\ensuremath{\}\!\}}@*) count_up_to n (*@\ensuremath{\{\!\{}@*)(*@\ensuremath{\Downarrow}@*)ans => (*@\ensuremath{\ulcorner}@*)(pyLen ans (*@\ensuremath{\le}@*) n (*@\ensuremath{\wedge}@*) (*@\ensuremath{\forall}@*) p (*@\ensuremath{\in}@*)
    pyIter ans, (2 : Int) (*@\ensuremath{\le}@*) p) (*@\ensuremath{\wedge}@*) (*@\ensuremath{\forall}@*) p (*@\ensuremath{\in}@*) pyIter ans, p < n(*@\ensuremath{\urcorner}@*)(*@\ensuremath{\}\!\}}@*) := by
  try
    mvcgen [count_up_to, pyRange_forIn, pyRange_forIn_start] invariants
    (*@\ensuremath{\cdot}@*) (*@\ensuremath{\Downarrow}@*)cur => (*@\ensuremath{\ulcorner}@*)let i := (cur.prefix.length : Int);
        (((((2 : Int) (*@\ensuremath{\le}@*) i (*@\ensuremath{\wedge}@*) i (*@\ensuremath{\le}@*) n) (*@\ensuremath{\wedge}@*) pyLen isprime = n +(*@\ensuremath{{}_{\mathrm p}}@*) (1 : Int)) (*@\ensuremath{\wedge}@*) pyLen ans (*@\ensuremath{\le}@*) i) (*@\ensuremath{\wedge}@*)
            (*@\ensuremath{\forall}@*) p (*@\ensuremath{\in}@*) pyIter ans, (2 : Int) (*@\ensuremath{\le}@*) p) (*@\ensuremath{\wedge}@*) (*@\ensuremath{\forall}@*) p (*@\ensuremath{\in}@*) pyIter ans, p < i(*@\ensuremath{\urcorner}@*)
  taste?
  all_goals sorry
\end{lstlisting}
\begin{lstlisting}[style=lean,basicstyle=\ttfamily\fontsize{7}{8.2}\selectfont,aboveskip=2pt,belowskip=2pt,breaklines=true,breakautoindent=true,caption={\code{count\_up\_to}: the committed proof, printed in full and unedited.},label={lst:pf-countupto-proof}]
private theorem @@range_split_val@@ {m : Nat} {pref suff : List Nat} {cur : Nat} (h : List.range m
    = pref ++ cur :: suff) : cur = pref.length := by
  have h2 : (List.range m)[pref.length]? = some cur := by
    rw [h, List.getElem?_append_right (by omega)]; simp
  have h3 : pref.length < m := by
    have := congrArg List.length h; simp at this; omega
  rw [List.getElem?_range h3] at h2; simp at h2; omega
@[spec]
theorem @@count_up_to_spec@@ : (*@\ensuremath{\{\!\{}@*)(*@\ensuremath{\ulcorner}@*)n (*@\ensuremath{\ge}@*) (0 : Int)(*@\ensuremath{\urcorner}@*)(*@\ensuremath{\}\!\}}@*) count_up_to n (*@\ensuremath{\{\!\{}@*)(*@\ensuremath{\Downarrow}@*)ans => (*@\ensuremath{\ulcorner}@*)(pyLen ans (*@\ensuremath{\le}@*) n (*@\ensuremath{\wedge}@*) (*@\ensuremath{\forall}@*) p (*@\ensuremath{\in}@*)
    pyIter ans, (2 : Int) (*@\ensuremath{\le}@*) p) (*@\ensuremath{\wedge}@*) (*@\ensuremath{\forall}@*) p (*@\ensuremath{\in}@*) pyIter ans, p < n(*@\ensuremath{\urcorner}@*)(*@\ensuremath{\}\!\}}@*) := by
  mvcgen [count_up_to, pyRange_forIn, pyRange_forIn_start] invariants
  (*@\ensuremath{\cdot}@*) (*@\ensuremath{\Downarrow}@*)(*@\ensuremath{\langle}@*)cur, ans, isprime(*@\ensuremath{\rangle}@*) => (*@\ensuremath{\ulcorner}@*)(pyLen ans (*@\ensuremath{\le}@*) (cur.prefix.length : Int)) (*@\ensuremath{\wedge}@*) ((*@\ensuremath{\forall}@*) p (*@\ensuremath{\in}@*) pyIter ans,
      (2 : Int) (*@\ensuremath{\le}@*) p) (*@\ensuremath{\wedge}@*) ((*@\ensuremath{\forall}@*) p (*@\ensuremath{\in}@*) pyIter ans, p < (2 : Int) + (cur.prefix.length : Int))(*@\ensuremath{\urcorner}@*)
  (*@\ensuremath{\cdot}@*) (*@\ensuremath{\Downarrow}@*)(*@\ensuremath{\langle}@*)cur, isprime(*@\ensuremath{\rangle}@*) => (*@\ensuremath{\ulcorner}@*)True(*@\ensuremath{\urcorner}@*)
  all_goals (try (simp_all (config := { zetaDelta := true }) [taste_ingr, pyAppend, pyListAppend]))
  case vc3.step.isTrue.post.success =>
    have hcur := range_split_val (*@\guilsinglleft@*)List.range _ = _ ++ _ :: _(*@\guilsinglright@*)
    obtain (*@\ensuremath{\langle}@*)hlen, h2, h3(*@\ensuremath{\rangle}@*) := (*@\guilsinglleft@*)pyLen _ (*@\ensuremath{\le}@*) _ (*@\ensuremath{\wedge}@*) _(*@\guilsinglright@*)
    simp only [pyLen, PyLen.pyLen, List.length_append,
      List.length_cons, List.length_nil] at hlen (*@\ensuremath{\vdash}@*)
    refine (*@\ensuremath{\langle}@*)by omega, ?_, ?_(*@\ensuremath{\rangle}@*)
    (*@\ensuremath{\cdot}@*) rintro p (hp | rfl)
      (*@\ensuremath{\cdot}@*) exact h2 p hp
      (*@\ensuremath{\cdot}@*) omega
    (*@\ensuremath{\cdot}@*) rintro p (hp | rfl)
      (*@\ensuremath{\cdot}@*) have := h3 p hp; omega
      (*@\ensuremath{\cdot}@*) omega
  case vc4.step.isFalse =>
    obtain (*@\ensuremath{\langle}@*)hlen, h2, h3(*@\ensuremath{\rangle}@*) := (*@\guilsinglleft@*)pyLen _ (*@\ensuremath{\le}@*) _ (*@\ensuremath{\wedge}@*) _(*@\guilsinglright@*)
    simp only [pyLen, PyLen.pyLen] at hlen (*@\ensuremath{\vdash}@*)
    refine (*@\ensuremath{\langle}@*)by omega, ?_(*@\ensuremath{\rangle}@*)
    intro p hp; have := h3 p hp; omega
  case vc6.post.success =>
    obtain (*@\ensuremath{\langle}@*)hlen, h2, h3(*@\ensuremath{\rangle}@*) := (*@\guilsinglleft@*)(pyLen _ (*@\ensuremath{\le}@*) _ (*@\ensuremath{\vee}@*) _) (*@\ensuremath{\wedge}@*) _(*@\guilsinglright@*)
    refine (*@\ensuremath{\langle}@*)?_, ?_(*@\ensuremath{\rangle}@*)
    (*@\ensuremath{\cdot}@*) rcases hlen with h | h <;>
        simp only [pyLen, PyLen.pyLen] at h (*@\ensuremath{\vdash}@*) <;> omega
    (*@\ensuremath{\cdot}@*) intro p hp
      have hp3 := h3 p hp
      have hpos := List.length_pos_of_mem hp
      rcases hlen with h | h <;>
        simp only [pyLen, PyLen.pyLen] at h <;> omega
theorem @@count_up_to_correct@@ : (*@\ensuremath{\forall}@*) (n : Int), n (*@\ensuremath{\ge}@*) (0 : Int) (*@\ensuremath{\rightarrow}@*) let ans := (count_up_to n).run;
    (pyLen ans (*@\ensuremath{\le}@*) n (*@\ensuremath{\wedge}@*) (*@\ensuremath{\forall}@*) p (*@\ensuremath{\in}@*) pyIter ans, (2 : Int) (*@\ensuremath{\le}@*) p) (*@\ensuremath{\wedge}@*) (*@\ensuremath{\forall}@*) p (*@\ensuremath{\in}@*) pyIter ans, p < n := by
  intro n hpre
  exact count_up_to_spec hpre
\end{lstlisting}
\subsubsection{\texorpdfstring{\code{double\_the\_difference}}{double_the_difference} (HumanEval/151, monadic)}
The postcondition restates the entire function as a summation, so proving it means proving the loop computes that sum and nothing else. This is the heaviest of the five and is printed in full.

\noindent
\begin{minipage}[t]{0.46\textwidth}
\begin{lstlisting}[style=py,basicstyle=\ttfamily\fontsize{7}{8.2}\selectfont,aboveskip=2pt,belowskip=2pt,breaklines=true,breakautoindent=true]
def @@double_the_difference@@(lst):
    ans = 0
    for num in lst:
        if num % 2 == 1 and num > 0 and ('.' not in str(num)):
            ans += num ** 2
    return ans
\end{lstlisting}
\end{minipage}\hfill
\begin{minipage}[t]{0.52\textwidth}
\begin{lstlisting}[style=py,basicstyle=\ttfamily\fontsize{7}{8.2}\selectfont,aboveskip=2pt,belowskip=2pt,breaklines=true,breakautoindent=true,caption={\code{double\_the\_difference}: the original Python (left) and the same function carrying its contract (right), a loop proved equal to its summation.},label={lst:pf-doublethedifference-py}]
def @@double_the_difference@@(lst):
    Ensures(Result() == sum([v * v for v in lst if v % 2 == 1 and v > 0 and ('.' not in str(v))]))
    Ensures(Result() % 2 == len([v for v in lst if v % 2 == 1 and v > 0 and ('.' not in str(v))]) % 2)
    Ensures(Result() >= 0)
    ans = 0
    for num in lst:
        Invariant(ans >= 0)
        if num % 2 == 1 and num > 0 and ('.' not in str(num)):
            ans += num ** 2
    Assert(ans >= 0)
    return ans
\end{lstlisting}
\end{minipage}
\begin{lstlisting}[style=lean,basicstyle=\ttfamily\fontsize{7}{8.2}\selectfont,aboveskip=2pt,belowskip=2pt,breaklines=true,breakautoindent=true,caption={\code{double\_the\_difference}: the Lean the translator emits.},label={lst:pf-doublethedifference-lean}]
def @@double_the_difference@@ := fun (lst : List Int) (*@\ensuremath{\mapsto}@*)
  (do
    let mut ans : Int := (0 : Int)
    for num in (pyIter lst)do
      let _ := pyPassInvariant (decide (ans (*@\ensuremath{\ge}@*) (0 : Int)))
      if h_1 :
          (num %(*@\ensuremath{{}_{\mathrm p}}@*) (2 : Int) = (1 : Int) (*@\ensuremath{\wedge}@*) num > (0 : Int)) (*@\ensuremath{\wedge}@*)
            !(pyStrContainsSubstr (pyStr num) ".") then
        ans := ans +(*@\ensuremath{{}_{\mathrm p}}@*) num ^(*@\ensuremath{{}_{\mathrm p}}@*) (2 : Int)
      else
        let _ := ()
    let _ := pyPassAssert (decide (ans (*@\ensuremath{\ge}@*) (0 : Int)))
    return ans : Id _)
\end{lstlisting}
\begin{lstlisting}[style=lean,basicstyle=\ttfamily\fontsize{7}{8.2}\selectfont,aboveskip=2pt,belowskip=2pt,breaklines=true,breakautoindent=true,caption={\code{double\_the\_difference}: the committed proof, printed in full and unedited.},label={lst:pf-doublethedifference-proof}]
private theorem @@pyMod_two@@ (a : Int) : a %(*@\ensuremath{{}_{\mathrm p}}@*) (2 : Int) = a % 2 := by
  show pyMod a 2 = a % 2
  have h1 : 0 (*@\ensuremath{\le}@*) a % 2 := Int.emod_nonneg a (by decide)
  unfold pyMod; split_ifs <;> simp_all
private theorem @@foldl_padd_shift@@ (l : List Int) (s : Int) : l.foldl (fun acc x => acc +(*@\ensuremath{{}_{\mathrm p}}@*) x) s
    = s + l.foldl (fun acc x => acc +(*@\ensuremath{{}_{\mathrm p}}@*) x) 0 := by
  induction l generalizing s with
  | nil => simp
  | cons h t ih =>
    simp only [List.foldl_cons]
    rw [ih (s +(*@\ensuremath{{}_{\mathrm p}}@*) h), ih (0 +(*@\ensuremath{{}_{\mathrm p}}@*) h)]
    simp only [show (*@\ensuremath{\forall}@*) x y : Int, x +(*@\ensuremath{{}_{\mathrm p}}@*) y = x + y from fun _ _ => rfl]
    ring
private theorem @@pySum_int_append@@ (a b : List Int) : pySum (a ++ b) = pySum a + pySum b := by
  simp only [pySum, pyIter_list, PySummand.toSummand, id_eq, List.foldl_append]
  rw [foldl_padd_shift b (a.foldl (fun acc x => acc +(*@\ensuremath{{}_{\mathrm p}}@*) x) 0)]
private theorem @@pySum_int_singleton@@ (x : Int) : pySum [x] = x := by
  simp only [pySum, pyIter_list, PySummand.toSummand, id_eq,
    List.foldl_cons, List.foldl_nil]
  rw [show (0 : Int) +(*@\ensuremath{{}_{\mathrm p}}@*) x = 0 + x from rfl, zero_add]
private theorem @@sq_odd_emod_two@@ {c : Int} (h : c % 2 = 1) : c * c % 2 = 1 := by
  rw [Int.mul_emod, h]; decide
private theorem @@pySum_int_nil@@ : pySum ([] : List Int) = 0 := rfl
@[spec]
theorem @@double_the_difference_spec@@ : (*@\ensuremath{\{\!\{}@*)(*@\ensuremath{\ulcorner}@*)True(*@\ensuremath{\urcorner}@*)(*@\ensuremath{\}\!\}}@*) double_the_difference lst (*@\ensuremath{\{\!\{}@*)(*@\ensuremath{\Downarrow}@*)ans => (*@\ensuremath{\ulcorner}@*)(ans = pySum
    ((List.filter (fun v => (v %(*@\ensuremath{{}_{\mathrm p}}@*) (2 : Int) = (1 : Int) (*@\ensuremath{\wedge}@*) v > (0 : Int)) (*@\ensuremath{\wedge}@*)
    !(pyStrContainsSubstr (pyStr v) ".")) (pyIter lst)).map fun v => v *(*@\ensuremath{{}_{\mathrm p}}@*) v) (*@\ensuremath{\wedge}@*) ans %(*@\ensuremath{{}_{\mathrm p}}@*)
    (2 : Int) = pyLen ((List.filter (fun v => (v %(*@\ensuremath{{}_{\mathrm p}}@*) (2 : Int) = (1 : Int) (*@\ensuremath{\wedge}@*) v > (0 : Int)) (*@\ensuremath{\wedge}@*)
    !(pyStrContainsSubstr (pyStr v) ".")) (pyIter lst)).map fun v => v) %(*@\ensuremath{{}_{\mathrm p}}@*) (2 : Int)) (*@\ensuremath{\wedge}@*) ans (*@\ensuremath{\ge}@*)
    (0 : Int)(*@\ensuremath{\urcorner}@*)(*@\ensuremath{\}\!\}}@*) := by
  mvcgen [double_the_difference, pyRange_forIn, pyRange_forIn_start] invariants
    (*@\ensuremath{\cdot}@*) (*@\ensuremath{\Downarrow}@*)(*@\ensuremath{\langle}@*)cur, ans(*@\ensuremath{\rangle}@*) => (*@\ensuremath{\ulcorner}@*)(ans = pySum ((List.filter (fun v => (v %(*@\ensuremath{{}_{\mathrm p}}@*) (2 : Int) = (1 : Int) (*@\ensuremath{\wedge}@*) v >
        (0 : Int)) (*@\ensuremath{\wedge}@*) !(pyStrContainsSubstr (pyStr v) ".")) cur.prefix).map fun v => v *(*@\ensuremath{{}_{\mathrm p}}@*) v) (*@\ensuremath{\wedge}@*)
        ans %(*@\ensuremath{{}_{\mathrm p}}@*) (2 : Int) = pyLen ((List.filter (fun v => (v %(*@\ensuremath{{}_{\mathrm p}}@*) (2 : Int) = (1 : Int) (*@\ensuremath{\wedge}@*) v >
        (0 : Int)) (*@\ensuremath{\wedge}@*) !(pyStrContainsSubstr (pyStr v) ".")) cur.prefix).map fun v => v) %(*@\ensuremath{{}_{\mathrm p}}@*)
        (2 : Int)) (*@\ensuremath{\wedge}@*) ans (*@\ensuremath{\ge}@*) (0 : Int)(*@\ensuremath{\urcorner}@*)
  all_goals
    simp only [pyIter_list, List.filter_append, List.map_append, pySum_int_append,
      List.filter_cons, List.filter_nil, List.map_cons, List.map_nil, List.append_nil,
      pySum_int_singleton, pyHPow_two, pyLen, PyLen.pyLen,
      List.length_append, List.length_cons, List.length_nil, List.length_map, pyMod_two,
      show (*@\ensuremath{\forall}@*) x y : Int, x +(*@\ensuremath{{}_{\mathrm p}}@*) y = x + y from fun _ _ => rfl,
      show (*@\ensuremath{\forall}@*) x y : Int, x *(*@\ensuremath{{}_{\mathrm p}}@*) y = x * y from fun _ _ => rfl] at *
  (*@\ensuremath{\cdot}@*) -- vc1.step.isTrue: contributing element
    obtain (*@\ensuremath{\langle}@*)(*@\ensuremath{\langle}@*)hsum, hpar(*@\ensuremath{\rangle}@*), hpos(*@\ensuremath{\rangle}@*) := (*@\guilsinglleft@*)(_ = pySum _ (*@\ensuremath{\wedge}@*) _) (*@\ensuremath{\wedge}@*) _ (*@\ensuremath{\ge}@*) (0:Int)(*@\guilsinglright@*)
    have hg := (*@\guilsinglleft@*)(((_:Int) % 2 = 1 (*@\ensuremath{\wedge}@*) (_:Int) > 0) (*@\ensuremath{\wedge}@*)
      (!pyStrContainsSubstr (pyStr (_:Int)) ".") = true)(*@\guilsinglright@*)
    have hsq := sq_odd_emod_two hg.1.1
    simp (config := { zetaDelta := true }) only [decide_eq_true_eq, hg, and_self, true_and,
      and_true, if_true, List.map_cons, List.map_nil, pySum_int_singleton, List.length_cons,
      List.length_nil, pyHPow_two, pow_two,
      show (*@\ensuremath{\forall}@*) x y : Int, x +(*@\ensuremath{{}_{\mathrm p}}@*) y = x + y from fun _ _ => rfl,
      show (*@\ensuremath{\forall}@*) x y : Int, x *(*@\ensuremath{{}_{\mathrm p}}@*) y = x * y from fun _ _ => rfl] at *
    refine (*@\ensuremath{\langle}@*)(*@\ensuremath{\langle}@*)by omega, by omega(*@\ensuremath{\rangle}@*), by nlinarith [hpos](*@\ensuremath{\rangle}@*)
  (*@\ensuremath{\cdot}@*) -- vc2.step.isFalse: non-contributing element
    obtain (*@\ensuremath{\langle}@*)(*@\ensuremath{\langle}@*)hsum, hpar(*@\ensuremath{\rangle}@*), hpos(*@\ensuremath{\rangle}@*) := (*@\guilsinglleft@*)(_ = pySum _ (*@\ensuremath{\wedge}@*) _) (*@\ensuremath{\wedge}@*) _ (*@\ensuremath{\ge}@*) (0:Int)(*@\guilsinglright@*)
    have hg := (*@\guilsinglleft@*)(*@\ensuremath{\neg}@*)(((_:Int) % 2 = 1 (*@\ensuremath{\wedge}@*) (_:Int) > 0) (*@\ensuremath{\wedge}@*)
      (!pyStrContainsSubstr (pyStr (_:Int)) ".") = true)(*@\guilsinglright@*)
    simp (config := { zetaDelta := true }) only [decide_eq_true_eq, hg, and_false, false_and,
      if_false, List.append_nil, List.map_nil, pySum_int_nil, add_zero, List.length_nil,
      Nat.add_zero] at *
    exact (*@\ensuremath{\langle}@*)(*@\ensuremath{\langle}@*)hsum, hpar(*@\ensuremath{\rangle}@*), hpos(*@\ensuremath{\rangle}@*)
theorem @@double_the_difference_correct@@ : (*@\ensuremath{\forall}@*) (lst : List Int), let ans := (double_the_difference
    lst).run; (ans = pySum ((List.filter (fun v => (v %(*@\ensuremath{{}_{\mathrm p}}@*) (2 : Int) = (1 : Int) (*@\ensuremath{\wedge}@*) v >
    (0 : Int)) (*@\ensuremath{\wedge}@*) !(pyStrContainsSubstr (pyStr v) ".")) (pyIter lst)).map fun v => v *(*@\ensuremath{{}_{\mathrm p}}@*) v) (*@\ensuremath{\wedge}@*)
    ans %(*@\ensuremath{{}_{\mathrm p}}@*) (2 : Int) = pyLen ((List.filter (fun v => (v %(*@\ensuremath{{}_{\mathrm p}}@*) (2 : Int) = (1 : Int) (*@\ensuremath{\wedge}@*) v >
    (0 : Int)) (*@\ensuremath{\wedge}@*) !(pyStrContainsSubstr (pyStr v) ".")) (pyIter lst)).map fun v => v) %(*@\ensuremath{{}_{\mathrm p}}@*)
    (2 : Int)) (*@\ensuremath{\wedge}@*) ans (*@\ensuremath{\ge}@*) (0 : Int) := by
  intro lst
  exact double_the_difference_spec True.intro
\end{lstlisting}

}

\subsection{Libraries: a worked example and a mapping table}\label{app:libraries}

Each library needs two things. One is a Lean file that implements the functions, written using Lean's standard library and Mathlib. The other is a table that maps each Python name to the Lean name implementing it. Listing \ref{lst:numpymap} shows part of the \code{numpy} table and the registry that dispatches on the module name. The code generator itself knows nothing about \code{numpy}; it only looks the name up. Most members map to the same definition in both twins. Only a few need a different one, and those go in a small second table that is read first in exact mode. \code{np.zeros} is one of them: its element type cannot be taken from an argument, so the provable twin needs the $\mathbb{Q}$ version and the runnable twin the \code{Float} one.

\begin{lstlisting}[style=lean,caption={Part of the \texttt{numpy} table, the small exact-mode override, and the registry that dispatches on the module name.},label={lst:numpymap}]

-- one table per library: Python member name to the Lean definition
def @@pythonNumpyMemberMap?@@ (member : String) : Option Lean.Name :=
  match member with
  | "array"     => some ``pyNumpyArray
  | "dot"       => some ``pyNumpyDot
  | "mean"      => some ``pyNumpyMean
  | "transpose" => some ``pyNumpyTranspose
  | "linspace"  => some ``pyNumpyLinspace
  | "zeros"     => some ``pyNumpyZerosFloat
  | _ => none

-- a few members need a different definition in the provable twin,
-- so this small table is consulted first in exact mode
def @@pythonNumpyMemberMapExact?@@ (member : String) : Option Lean.Name :=
  match member with
  | "zeros" => some ``pyNumpyZerosRat
  | "ones"  => some ``pyNumpyOnesRat
  | "eye" | "identity" => some ``pyNumpyEyeRat
  | _ => none

-- one registry dispatches on the module the call came from
def @@pythonLibraryMap?@@ (moduleName member : String) : Option Lean.Name :=
  match moduleName with
  | "math"  => math.pythonMathMemberMap? member
  | "numpy" => numpy.pythonNumpyMemberMap? member
  | "scipy" => scipy.pythonScipyMemberMap? member
  | _ => none
\end{lstlisting}

Listings \ref{lst:lib-py} and \ref{lst:lib-lean} show a \code{scipy} call before and after translation. \code{tmean} becomes \code{pyScipyTmean}. The arithmetic around it is translated as usual, so the library call and the code calling it are both provable.

\noindent
\begin{minipage}[t]{0.42\textwidth}
\begin{lstlisting}[style=py,caption={Python using \texttt{scipy}.},label={lst:lib-py},breaklines=true]
from scipy.stats import tmean

def @@variance@@(xs: List[float]) -> float:
    m = tmean(xs)
    total = 0.0
    for x in xs:
        total = total + (x - m) * (x - m)
    return total / len(xs)
\end{lstlisting}
\end{minipage}%
\hfill
\begin{minipage}[t]{0.54\textwidth}
\begin{lstlisting}[style=lean,caption={The generated provable twin.},label={lst:lib-lean},breaklines=true]
def @@variance@@ := fun (xs : List Rat) (*@\ensuremath{\mapsto}@*)
  (show Rat from
    Id.run
      (do
        let mut m := Libraries.scipy.pyScipyTmean xs
        let mut total := (0.0 : Rat)
        for x in (pyIter xs) do
          total := total +(*@\ensuremath{{}_{\mathrm p}}@*) (x -(*@\ensuremath{{}_{\mathrm p}}@*) m) *(*@\ensuremath{{}_{\mathrm p}}@*) (x -(*@\ensuremath{{}_{\mathrm p}}@*) m)
        return total /(*@\ensuremath{{}_{\mathrm p}}@*) pyLen xs))
\end{lstlisting}
\end{minipage}

\subsection{Extended Semantics and Translation Rules for Python OOP} \label{oop-appendix}

This appendix provides additional details on the translation of Python's object-oriented constructs into Lean 4, expanding upon the overview in Section~\ref{sec:oop}.

Figure ~\ref{lst:oop-py} illustrates a standard Python \texttt{Point} class, while Figure ~\ref{lst:oop-lean} shows its generated Lean 4 counterpart.

\noindent
\begin{minipage}[t]{0.48\textwidth}
\begin{lstlisting}[style=py, caption={Python \texttt{Point} class.}, label={lst:oop-py}, breaklines=true]
class @@Point@@:
    def @@__init__@@(self, x, y):
        self.x = x
        self.y = y

    def @@norm_sq@@(self):
        return self.x * self.x \
             + self.y * self.y

obj = Point(1, 2)
norm = obj.norm_sq()
\end{lstlisting}
\end{minipage}%
\hfill
\begin{minipage}[t]{0.48\textwidth}
\begin{lstlisting}[language=lean, caption={Lean 4 translation.}, label={lst:oop-lean}, breaklines=true, mathescape=true]
structure @@Point@@ where
  x : Int
  y : Int
  deriving Inhabited, Repr, BEq

def @@Point.new@@ := fun x $\mapsto$ fun y $\mapsto$ ({ x := x, y := y } : Point)
def @@Point.norm_sq@@ := fun (self : Point) $\mapsto$ self.x *$_p$ self.x +$_p$ self.y *$_p$ self.y

def @@obj@@ := Point.new (1 : Int) (2 : Int)
def @@norm@@ := Point.norm_sq obj
\end{lstlisting}
\end{minipage}

\subsubsection{What is translated}

The \code{Point} example above shows the shape of the translation. The rest of this subsection says what else is covered, since a \code{structure} plus a namespace is only a faithful model of a Python class if the class's other features come across too.

\paragraph{Construction.} \code{__init__} becomes the smart constructor \code{C.new}. It is built by the same \code{self}-threading machinery as a mutating method: start from \code{default}, apply each \code{self.x = ...} in order, and return the result. Because the assignments are replayed rather than pattern-matched, a partial \code{__init__} that sets only some fields, and one that computes a field from a local variable, both translate. A class with no \code{__init__} is translated from its class-level assignments alone. Fields default to \code{Int} when nothing determines them, and a parameter defaulting to \code{None} makes its field \code{Option C}, which is what gives the recursive-node classes (\code{TreeNode.left}, \code{ListNode.next}, a trie's \code{children}) their type.

\paragraph{Methods and bound forms.} Methods are emitted with fully qualified names (\code{def C.method}), so Python's \code{obj.method(...)} resolves through Lean's own dot notation with no \code{namespace} wrapper and no dispatch table. \code{@staticmethod} and \code{@classmethod} drop the receiver and are called as \code{C.method args}; inside a class body \code{cls(...)} is recognised as construction. \code{@property} is treated as a method-binding marker, so a property reads like the field access it is in Python. Other decorators are transparent and handled by the general decorator path.

\paragraph{Dunders.} A dunder is mapped to the Lean typeclass that carries the same notation, so the operator keeps working on the translated type rather than becoming a named call: \code{__add__}, \code{__sub__} and \code{__mul__} become \code{PyHAdd}/\code{PyHSub}/\code{PyHMul} instances behind $+_p$, $-_p$ and $*_p$; \code{__eq__} becomes \code{BEq}; the four ordering dunders become the corresponding comparisons; and \code{__str__} or \code{__repr__} becomes a \code{PyPrintable} instance so \code{print(obj)} and \code{str(obj)} work. Where a class defines both \code{__str__} and \code{__repr__} the former wins, matching \code{str()}.

\paragraph{Inheritance.} A subclass becomes \code{structure Derived extends Base}, which embeds the base's fields, and the base's member types propagate to a subclass that does not override them. Multiple bases are emitted as \code{structure C extends B1, B2}: Lean resolves an inherited method or a conflicting field in declaration order, first base wins, which coincides with Python's MRO for the single-path hierarchies we target. Nested classes are hoisted to module level under a dotted name, since Lean has no class nesting to mirror.

\paragraph{Both tiers.} Everything above holds under the default value-semantics translation and under the reference-semantics \code{--heap} tier of Appendix~\ref{app:heap}, where a class becomes a \code{Ref} into the heap and every method, dunders included, becomes a heap method. It also holds across the two twins: a \code{'rn} subclass extends \code{Base'rn} rather than \code{Base}, so a base field that is $\mathbb{Q}$ in the provable twin and \code{Float} in the runnable one stays consistent down the hierarchy.

\subsubsection{Limitations and Future work for OOP}

Translating Python's object model to a static verifier requires omitting advanced object-oriented features like dynamic dispatch, \code{super()}, metaclasses, and complex runtime structures. While features like structure extends allow us to embed fields and coerce subclasses to base types, they do not create virtual method tables (\code{vtables}). As a result, method calls are resolved using the static type rather than the runtime type---meaning a \code{list[Base]} containing Derived objects will invoke Base.method instead of Derived.method.

To implement true runtime polymorphism, we would need to introduce tagged unions (similar to \code{PyAny} for scalars) or explicit dispatch records for every class. However, both approaches makes it difficult for equational reasoning required to make our translations mathematically provable. Because the scripts we target rarely rely on subtype polymorphism, we opt to preserve the simpler static model without sacrificing verification.

\subsection{Scope: what PastaLean does not translate}\label{app:scope}

Not all of Python is in scope, and the boundary is what has a static meaning. A program that reshapes itself as it runs has none. \code{exec}, monkey-patching, modifying a class after its definition, and attribute names computed at run time all change the program that a translation would have had to fix in advance, so a shallow embedding cannot represent them. The one member of that family we do support is \code{eval} of an expression string, which the runtime evaluates by recursive descent rather than by calling a Python interpreter. We also treat \code{async}, \code{await} and threads as out of scope: there is no single sequential execution for a specification to describe, and the scripts we target do not use them. These features are rare in that code, so excluding them costs little in practice.

The other boundary is the library surface. We do not use FFI, because a call into a foreign runtime cannot be reasoned about in Lean. A library is available only after we implement it in Lean using Mathlib (Appendix~\ref{app:libraries}), so the supported set is the set we have written. A program that imports outside it is not rejected and not mistranslated either: the offending statement degrades to a placeholder that carries the original Python line, which Appendix~\ref{app:unsupported} describes. On our target code the cost is small and measured. Five of the 2{,}589 LeetCode problems are skipped for importing a library outside the subset, and HumanEval translates with zero degraded statements (Appendix~\ref{app:eval-cp}).

Within the object model, single inheritance is translated by structure embedding and multiple bases are emitted in declaration order, but nothing is linearised: Lean resolves a conflicting member by first base wins, which agrees with Python's MRO for the single-path hierarchies we target and not in general. Dynamic dispatch, \code{super()} and metaclasses are not translated, and the full list is in Appendix~\ref{oop-appendix}.

\subsection{Unsupported Python degrade to \code{pyUnsupported}}\label{app:unsupported}

PastaLean does not know all of Python. PastaLean expects users to write the functions from scratch which is faithful to Python's behaviour, so that Lean can use it for computation and verification. This is a  limitation since expanding to other libraries is bound on coverage. One can use FFI between Python and Lean, however that takes away proving capabilities of Lean for the Python code. So the supported set grows at the speed of writing semantics. However once written, the architecture is such that it is a metter of writting a Map from Python AST to that Lean function(with few options based on behavior) to integrate within PastaLean; the transpiler is not touched, and many Python modules are wired in this way today.

However we overcome this limitation by providing a degrade path for unsupported Python code. We provide a placeholder no-op \code{pyUnsupported} which returns a one-field structure holding the source string, so the binding elaborates, the variable stays declared, and no unconstrained metavariable is left for the elaborator to fail on later. It is emitted in every position a dropped statement can take: a bare statement becomes \code{let _ := ...}, an assignment keeps its name with \code{let x := ...}, and a top-level statement becomes a \code{def}. Note that it may not always elaborate since sometimes the unsupported code was supposed to define a type for a variable/return important values which is needed to elaborate the rest of the program. The \code{pyUnsupported} placeholder shows a Linter warning to the user that the code is not supported and needs to be fixed.

\begin{lstlisting}[style=lean,caption={Lines PastaLean cannot translate become placeholders that carry the original Python text; the surrounding logic is translated normally.},label={lst:unsupported}]

# Python: logging and requests are outside the supported set
import logging, requests
logger = logging.getLogger(__name__)

def @@total_score@@(scores):
    logger.info("scoring")           # no logic, dropped
    blob = requests.get("http://x")  # no logic, dropped
    total = 0
    for s in scores:                 # real logic, translated
        total = total + s
    return total

-- generated Lean: each dropped line keeps its own source text
def @@logger@@ := pyUnsupported "logger = logging.getLogger(__name__)"

def @@total_score@@ := fun (scores : List Int) (*@\ensuremath{\mapsto}@*) Id.run do
  let _ := pyUnsupported "logger.info(\"scoring\")"
  let mut blob := pyUnsupported "blob = requests.get(\"http://example.com\")"
  let mut total : Int := 0
  for s in pyIter scores do
    total := total +(*@\ensuremath{{}_{\mathrm p}}@*) s
  return total
\end{lstlisting}

This matters most for the agentic code verification purposes. For example, Python is full of lines that carry no logic: logging, warnings, progress printing, a metrics call, a fetch whose result the computation never reads. An LLM writing Python writes them too, because that is what ordinary Python looks like. A translator that gave up at the first \code{logging.getLogger} would be useless on that code. Hence we drop the line and keeping the algorithm intact which PastaLean can compute and verify on as well. Listing~\ref{lst:unsupported} shows an example of two foreign calls that become placeholders and the loop that does the work becomes ordinary Lean.

\section{The Type Inference Engine}\label{app:typeinfer}

PastaLean includes a full type inference engine \code{TypeInfer} that runs on the Python AST IR before code generation. As shown in the results section \ref{results}, \code{TypeInfer} performs the best type inference for the TypeEvalPy dataset, ony seconding to `pyrefly' in autogen dataset. It is not only correct, but fastest and machine verified ...

\subsection{What the engine reads and what it writes}\label{app:ti-io}

This Type Inference engine sits as a separate Lean library within PastaLean. It depends only on Lean's JSON library and nothing else in PastaLean, and the code generator depends on it, not the other way round.

Its input is the JSON IR of which is created from the Python AST after parsing and desugaring, before any code is generated. Its output is the same IR with two things added. Every statement carries \code{_inferred}, a flag saying the whole-module pass has run. Every binder whose type the engine worked out carries \code{_ty}, an annotation node holding that type, written in the same shape a hand-written Python annotation would have. Nothing else changes: no expression is rewritten and no node is removed.

Listing \ref{lst:ty-stamp} follows one variable through the whole pipeline. In \code{st = []} nothing on that line says what the list holds, so the IR carries no type for it. The engine learns the element from the later \code{st.append(i)} and writes it back onto the assignment target as an ordinary annotation node. Code generation reads that stamp and ascribes the binder. Without it Lean has an empty list with no element type, defaults it, and then fails at the \code{append}.

\begin{lstlisting}[style=pseudo,caption={One variable, end to end: Python, the IR before and after inference, and the Lean we emit.},label={lst:ty-stamp}]

# Python
def stack_sum(n: int) -> int:
    st = [] # nothing here says what st holds
    for i in range(n):
        st.append(i) # this does: st is a list of ints
    ...

// IR before inference: the target has no type
{"node_type": "Assign",
 "target": {"node_type": "Name", "id": "st"},
 "value":  {"node_type": "List", "elts": []}}

// IR after inference: _ty is written on the target, as an annotation node
{"node_type": "Assign",
 "target": {"node_type": "Name", "id": "st",
            "_ty": {"node_type": "Subscript",
                    "value": {"node_type": "Name", "id": "list"},
                    "slice": {"node_type": "Name", "id": "int"}}},
 "value":  {"node_type": "List", "elts": []}}

-- generated Lean: the stamp becomes the ascription
def stack_sum := fun (n : Int) $\mapsto$ Id.run (do
  let mut st : List Int := []
  for i in (pyRange n) do
    st := pyAppend st i
  ...)
\end{lstlisting}

Writing the result back as ordinary annotations is what keeps the two halves separate. The code generator reads \code{_ty} in the same place it reads a user's annotation, so it needs no special path for inferred types, and a program the user annotated fully produces the same Lean whether the engine ran or not.

The engine is also optional, in a precise sense. If the pass is skipped or fails, the driver leaves the IR untouched and code generation proceeds with unannotated binders. Lean then elaborates what it can. This works for a binder Lean's own unifier can solve from the surrounding body, and fails for the ones it cannot: a top-level \code{def} fixes its parameter types before looking at the body, so an unannotated parameter used only as a container leaves instance resolution stuck. The engine exists to fill exactly those slots, which is why its failure degrades the output rather than breaking it.

\subsection{The algorithm}\label{app:ti-algo}

Inference runs in three nested layers. The innermost is a per-statement transfer function, the middle is a per-function fixpoint, and the outer is a whole-module fixpoint over function signatures. Listing \ref{lst:ti-algo} gives the shape.

\begin{lstlisting}[style=pseudo,caption={The inference pipeline. $\sqcup$ is the lattice join.},label={lst:ti-algo}]

INFER_MODULE(m):
  (sigs, params) $\gets$ signatures of every function and method in m
  repeat 6 times: // outer fixpoint over the module
      for each call site f(a) in m:
          params[f] $\gets$ params[f] $\sqcup$ typeof(a) // arguments refine parameters
      for each function f in m:
          sigs[f] $\gets$ sigs[f] $\sqcup$ INFER_FUNCTION(f).return
      copy base-class members to subclasses that do not override
      stop early if nothing changed
  globals $\gets$ types of the module-level statements
  return m with every binder stamped _ty from INFER_FUNCTION

INFER_FUNCTION(f):
  env $\gets$ usage(f) $\sqcup$ captures(f) $\sqcup$ callSiteHints(f) $\sqcup$ annotations(f)
  repeat 8 times: // inner fixpoint over the body
      env' $\gets$ env
      for each statement s in body(f): env' $\gets$ APPLY(env', s)
      env' $\gets$ learn from reads, dict methods and comprehension targets
      if env' = env: break
      env $\gets$ env'
  return env

APPLY(env, s): // the transfer function
  match s:
    x = e            $\Rightarrow$ env[x] $\gets$ env[x] $\sqcup$ TYPEOF(env, e)
    x.append(e)      $\Rightarrow$ env[x] $\gets$ env[x] $\sqcup$ list(TYPEOF(env, e))
    for x in e       $\Rightarrow$ env[x] $\gets$ env[x] $\sqcup$ elem(TYPEOF(env, e))
    if/while/try     $\Rightarrow$ fold APPLY over each branch, then join the branches
    return e         $\Rightarrow$ record TYPEOF(env, e) as f's return type
\end{lstlisting}

Few important points around this algorithm. Firstly, the environment never moves down the lattice, because every update is a join, so both loops are monotone and settle. Secondly, the iteration counts are caps rather than convergence tests, which is sound because the lattice has small fixed height: a cap is a floor on how far the analysis climbs, not an approximation of it, and in practice the inner loop converges in about two passes. Finally, precedence between the four seeds is fixed rather than order-dependent which is that body usage is weakest, then enclosing captures, then call-site hints, then written annotations, so an annotation the user wrote is never overridden by something the engine guessed.

\subsection{Machine Verified Correctness Theorems for TypeInfer}\label{app:ti-proofs}

The engine is a monotone dataflow fixpoint over a type lattice, and the literature is specific about what such an analysis has to satisfy. We take the requirements from that literature and prove each one in Lean, with no \code{sorry}. The proofs live in one file (58 results counting the supporting lemmas) which the test suite re-elaborates on every run, so a change to the lattice that breaks a law fails the build rather than going unnoticed.

\begin{itemize}\itemsep2pt
\item \textbf{The domain is a bounded join-semilattice.} This is what makes the fixpoint well defined and independent of the order in which facts are merged~\cite{cousot1977abstract,davey2002lattices}. Proved as \code{join_comm}, \code{join_assoc}, \code{join_idem}, with \code{join_unknown} for $\bot$ and \code{join_any} for $\top$.
\item \textbf{\code{join} is the least upper bound} of the induced precision order $a \sqsubseteq b := a \sqcup b = b$, so merging loses as little information as possible~\cite{davey2002lattices}. Proved as \code{le_join_left}, \code{le_join_right} and \code{join_le}, with the order shown to be a partial order (\code{le_refl}, \code{le_trans}, \code{le_antisymm}) whose least and greatest elements are $\bot$ and $\top$ (\code{le_unknown}, \code{le_any}).
\item \textbf{\code{join} is monotone}, so by Knaster--Tarski the least fixpoint exists and is unique~\cite{tarski1955fixpoint,cousot1977abstract}. Proved as \code{join_mono_left}, \code{join_mono_right} and \code{join_mono}. Termination, though, does not come from a height bound on the domain, and we are precise about this. The lattice admits arbitrarily long rising chains: \code{int}, \code{opt int}, \code{opt (opt int)} climbs without end (\code{lattice_infinite_ascending_chain}), and removing \code{Optional} does not fix it, since any recursive container over the bottom behaves the same way, as \code{unknown}, \code{list unknown}, \code{list (list unknown)} shows with no \code{Optional} present (\code{opt_free_infinite_ascending_chain}). What forces the iteration to stop is finiteness of the working set: a fixed program mentions finitely many types, and \code{join} never introduces a head constructor or class name absent from its inputs, so the analysis stays within a finite set of types closed under \code{join}. Over any such finite carrier the Kleene iteration reaches the least fixpoint (\code{kt_iteration_reaches_fixpoint}), which is the honest statement of why the engine terminates.
\item \textbf{Semantic soundness.} An inferred type must never lie about the runtime, which is the inference form of ``well-typed programs do not go wrong''~\cite{milner1978polymorphism,wright1994soundness}. We define \code{HasType v T}, a denotation of \code{PyType} over a model of Python runtime values, and prove \code{hasType_join_left} and \code{hasType_join_right}: if a value has type $a$ then it also has type $a \sqcup b$, on either side. Widening never excludes a value the program can actually produce. This holds on the full lattice and not only the numeric scalars, though the tower is the case the engine leans on and we state it separately too (\code{hasType_join_tower}).
\item \textbf{The denotation is faithful.} Soundness is only as meaningful as the denotation it is stated over, so we also prove that \code{HasType} is the intended one. The bounds mean what a lattice requires: $\top$ admits every value and $\bot$ admits none (\code{hasType_any}, \code{not_hasType_unknown}), so \code{unknown} really is the absence of information rather than a type something inhabits. And Python's numeric tower is a genuine subtyping chain \emph{at the value level}, not an order we imposed on the syntax of types: every \code{bool} is an \code{int} and every \code{int} a \code{float} (\code{hasType_bool_int}, \code{hasType_int_float}, \code{hasType_bool_float}). These are what make the widening of Appendix~\ref{numeric-widening} sound rather than merely convenient.
\item \textbf{Gradual-typing consistency} for the dynamic type. The relation must be reflexive and symmetric but \emph{not} transitive, which is precisely what separates gradual typing from subtyping~\cite{siek2006gradual,siek2015refined}. Proved as \code{consistent_refl}, \code{consistent_symm}, \code{consistent_unknown} and \code{consistent_not_trans}, with structural type equality shown reflexive and symmetric underneath them (\code{beq_refl}, \code{beq_comm}).
\item \textbf{Coercions are total.} Where an actual type must be reconciled with an expected one the engine inserts a coercion, and \code{reconcile_total} shows every (expected, actual) pair maps to one of the five intended actions: an exact match, \code{bool}-to-\code{int}, \code{int}-to-\code{float}, unwrapping an \code{Optional}, or boxing. No value is left with no way to reach its slot, and one already of the expected type needs no coercion (\code{reconcile_refl}).
\item \textbf{Information preservation.} Merging two \emph{known} types never discards everything the engine knew: \code{join_ne_unknown} and \code{join_eq_unknown_iff} show \code{join a b} is \code{unknown} exactly when both inputs are, so a merge only climbs the lattice and cannot silently collapse to $\bot$. In the other direction a real conflict does reach $\top$ rather than being quietly resolved in favour of one side (\code{join_conflict_str_int}).
\end{itemize}

Every one of these results is proved on the \code{PyType}, the \code{join}, \code{consistent} and \code{reconcile} functions the engine actually runs, recursing over the true element lists of \code{tuple} and \code{fn}; there is no stand-in model. Those functions use size-based rather than structural recursion, so \code{PyType} has no plain structural induction and the proofs supply the missing principle themselves: the diagonal facts (\code{beq_refl}, \code{consistent_refl}, \code{join_idem}) go by well-founded recursion mirroring each function, the symmetric facts (\code{beq_comm}, \code{join_comm}, \code{consistent_symm}) drive each function's own generated induction principle, and full-lattice associativity goes by strong induction on term size.

Two caveats fall out of the proofs rather than being worked around. Idempotence and order-reflexivity are stated for \emph{normalized} types, those with no \code{Optional[Any]} subterm. That is a fact and not a gap: \code{join (opt any) (opt any)} is \code{any}, because a nullable dynamic value already is a dynamic value, and \code{normalized} is exactly the normal form the engine stays inside. The same absorption is what makes \code{join} associative at the \code{None}-versus-conflict junction, and associativity does hold on the full lattice. In fact we prove that \code{any} and \code{opt any} (\code{Optional[Any]}) are distinct \code{PyType} terms (\code{any_ne_opt_any}) even though they denote the same type, so the collapse is \code{join} normalising its own output and not a failure of idempotence. Second, the value model has no inhabitant for \code{set}, \code{dict}, \code{tuple} or \code{fn}, so soundness at those heads is vacuously true; the content of the theorem is the scalars, \code{None}, classes, and the \code{list} and \code{Optional} congruences.

\subsection{Where the types come from}\label{app:ti-techn}

This section covers the sources of type information the engine draws on, and what it does in the two cases where no single type exists.

\subsubsection{The numeric tower}\label{numeric-widening}

\begin{lstlisting}[language=lean,caption={Widening join over the numeric tower (excerpt).},label={lst:join}]
def @@join@@ : PyType (*@\ensuremath{\rightarrow}@*) PyType (*@\ensuremath{\rightarrow}@*) PyType
  | .unknown, t | t, .unknown => t  -- (*@\ensuremath{\bot}@*) is identity
  | .any, _ | _, .any => .any  -- (*@\ensuremath{\top}@*) absorbs
  | .int, .bool | .bool, .int => .int  -- bool < int
  | .float, .int | .int, .float
  | .float, .bool | .bool, .float => .float  -- int,bool < float
  | .list a, .list b => .list (join a b)  -- covariant
  | .opt a, b | b, .opt a => match join a b with
                    | .any => .any | j => .opt j  -- Optional[Any] (*@\ensuremath{\equiv}@*) Any
  | .none, t | t, .none => .opt t
  | a, b => if a.beq b then a else .any
\end{lstlisting}

Python promotes silently, so \code{True + 1} is \code{2} and a list written with both \code{5} and \code{float('inf')} is a list of floats. A flat scalar domain would send \code{join(int, float)} to $\top$ and box the whole container, losing provability for an entirely ordinary program. Making the scalars a tower instead means the smaller operand is coerced up at code-generation time and the container keeps a concrete element type.

The \code{Optional} rule needed care. \code{Optional[Any]} must collapse to \code{Any}, because a nullable dynamic value is just a dynamic value. Without that rule \code{join} is not associative: \code{join(join(None, int), str)} and \code{join(None, join(int, str))} disagree, one giving \code{Option PyAny} and the other \code{PyAny}, so the fixpoint would depend on the order facts arrive in. Associativity is not a cosmetic law here; it is what makes the analysis confluent.

\subsubsection{Type-changing SSA reassignment}\label{sec:ssa}

Python rebinds freely, and a rebinding can change a variable's type: \code{x = "5"} then \code{x = int(x)}. One Lean binder cannot hold both. An SSA pass versions the variable at exactly those points, giving \code{x}, \code{x'v1} and so on, with joins where branches carrying different versions meet~\cite{cytron1991ssa}. Only genuine type changes version. A same-type reassignment stays a \code{let mut} update, and numeric widening or acquiring \code{None} counts as refinement rather than change, so an accumulator is not split into versions. Versioning here buys concrete types where boxing would otherwise be forced: each version is a separate binder with its own type.

\subsubsection{Types crossing function boundaries}

Types flow both ways between a function and its callers. Return types flow outward: the module pass computes each function's return type to a fixpoint, so \code{data = make_pairs(5)} is typed even though \code{make_pairs} carries no return annotation, and a function returning another's result gets it too. Argument types flow inward: the types at every call site are joined position by position into the callee's parameters, so an unannotated parameter called only with integers is an integer parameter. Class methods join the same table under \code{Class.method} keys, so a method's parameters are refined from call sites exactly like a free function's, and single inheritance copies a base's member types to a subclass that does not override them.

Because both directions are joins into the same lattice, adding a call site can only widen a signature, never narrow it, which is what keeps the outer loop monotone alongside the inner one.

\subsubsection{Usage-based back-inference}

An unannotated parameter is also typed by how the body uses it. \code{p.split()} makes \code{p} a string, \code{p << 1} makes it an integer, \code{p[k]} fixes a dictionary's key type, and comparison or membership against a literal pins \code{p} to that literal's type. Element types are recovered the same way and propagate through nested arithmetic, so \code{p[i] + 1} gives a list of integers and \code{p[i].upper()} a list of strings. This is the flow- and usage-driven inference of Shed Skin~\cite{dufour2006shedskin} and Agesen's Cartesian Product Algorithm~\cite{agesen1995cpa,salib2004starkiller}, restricted to the shapes we emit, and the overall reflow shape is that of PyPy's RPython annotator~\cite{rigo2006pypy,palsberg1991oot}.

The rule that keeps this honest is that usage inference must never fight a type-changing reassignment. Conflicting evidence joins upward to $\top$, and the SSA pass has already split the genuinely different segments, so back-inference only ever refines a binding that has a single type to begin with.

\subsubsection{When there is no single type}\label{sec:pyany}

Some values genuinely have no static type, and the engine is designed to reach that conclusion rather than guess. When the join reaches $\top$ the value is boxed as \code{PyAny}, a tagged union of the runtime types, following the dynamic type of gradual typing~\cite{siek2006gradual}. Boxing is what makes the transpiler total: the program still compiles and runs instead of being rejected. The cost is provability, since a boxed value is not a commutative ring and the arithmetic tactics stall on it, so boxing is a last resort and Appendix \ref{app:pyany} describes what can still be proved about a boxed value.

\subsection{The dynamic fallback: \code{PyAny}}\label{app:pyany}

When the engine cannot give a value one Lean type, it boxes the value as \code{PyAny}: a tagged union of the runtime types, the dynamic type of gradual typing~\cite{siek2006gradual}. This is a last resort, reached only when the join hits $\top$ --- three things force it: a function returning a different type on different paths, a variable rebound to a new type midway through a function, and a parameter the call sites use at incompatible types. Boxing is inserted by a \code{CoeTail} instance at the boundary, so no wrapper is ever written by hand: the \code{int} path and the \code{str} path each coerce on their own. It keeps the transpiler total --- the program still compiles and runs instead of being rejected --- at the cost of provability, since a boxed value is not a commutative ring and the arithmetic tactics stall on it. Listing \ref{lst:poly-any} shows one \code{add} applied at two types.

\begin{lstlisting}[style=lean,caption={One \code{add} used at two types: the parameters box to \code{PyAny}, and a property of the boxed function is still proved.},label={lst:poly-any}]

# Python: the call sites disagree, so the join of int and str is the top
def @@add@@(a, b):    return a + b
print(add(3, 4)); print(add("x", "y"))

-- generated Lean: both parameters are boxed
def @@add@@ := fun (a : PyAny) (*@\ensuremath{\mapsto}@*) fun (b : PyAny) (*@\ensuremath{\mapsto}@*) (show PyAny from a +(*@\ensuremath{{}_{\mathrm p}}@*) b)

-- and associativity of the boxed operator is still provable
example (a b : PyAny) : (a +(*@\ensuremath{{}_{\mathrm p}}@*) b) +(*@\ensuremath{{}_{\mathrm p}}@*) b = a +(*@\ensuremath{{}_{\mathrm p}}@*) (b +(*@\ensuremath{{}_{\mathrm p}}@*) b) := by
  pyany_cases a b <;> grind
\end{lstlisting}

The proof is still available. The tactic \code{pyany_cases} splits the goal into one case per runtime type the boxed value could hold and closes the tag-mismatch leaves automatically (both sides fold to \code{.none}, equal by \code{rfl}); \code{grind} finishes each real case. The recovery is honest about its limits: $+_p$ is not commutative over \code{PyAny}, because it \emph{concatenates} on strings and on lists, so the two would-be commutativity goals stay open --- correctly. Appendix~\ref{app:pyany-runtime} shows how a boxed value computes, and \ref{app:pyany-cases} what a boxed function can prove.

\subsubsection{How a boxed value computes}\label{app:pyany-runtime}

\code{PyAny} is a tagged union of six constructors, and each payload is already an ordinary Lean value of the matching type: \code{.int} holds an \code{Int}, \code{.float} a $\mathbb{Q}$, \code{.str} a \code{String}. Boxing attaches a tag; it encodes nothing.

Take \code{x +}\ensuremath{{}_{\mathrm p}}\code{ y} with both sides boxed. Instance resolution picks the single \code{PyHAdd PyAny PyAny PyAny} instance once and statically, so nothing is searched at run time; it calls \code{PyAny.add}, and everything after is a match on constructors. \code{asNum} projects a numeric payload back out (\code{.int n} to \code{n}; \code{.bool b} to \code{1} or \code{0}, since a Python \code{bool} is an \code{int}; \code{.float q} to \code{q}), \code{numBinop} runs Lean's own \code{Int} or $\mathbb{Q}$ addition on the projected payloads, and the answer is re-tagged. There is no separate arithmetic for boxed values and no conversion, only projection, the native operation, and a tag; when the two payloads project to different numeric types both widen to $\mathbb{Q}$ and the result is \code{.float}, which is where \code{int + float = float} comes from.

The point is that this stays bounded rather than sprawling into a function per case. Python's operator set is fixed, and every arithmetic operator is the same \code{numBinop} core: \code{sub} is one line over it, \code{add} and \code{mul} add only the \code{str}/\code{list} concatenation case, and the operators differ otherwise just in the native function pair they pass (the bitwise operators share an integer-only twin). This is the design of CPython's own number-protocol slots, so the runtime is written once and grows with the language, not with the input program. The one honest gap is a genuine mismatch: \code{1 + "a"}, which Python rejects with \code{TypeError}, folds to \code{None}. That is deliberate, because raising would push the whole program into an exception monad and forfeit the provability boxing exists to preserve, and the mismatch can arise only in code that had no static type to begin with.

\begin{lstlisting}[style=lean,basicstyle=\ttfamily\fontsize{7.5}{9}\selectfont,aboveskip=2pt,belowskip=2pt,breaklines=true,caption={One combinator, one line per operator: project the payload, run Lean's own arithmetic, re-tag.},label={lst:pyany-add}]
inductive @@PyAny@@ where
  | int (n : Int) | bool (b : Bool) | str (s : String)
  | float (q : Rat) | list (xs : List PyAny) | none

def @@asNum@@ : PyAny -> Option (Sum Int Rat)
  | .int n  => some (.inl n)
  | .bool b => some (.inl (if b then 1 else 0))   -- bool <: int
  | .float q => some (.inr q)
  | _ => none

def @@numBinop@@ (fi : Int -> Int -> Int) (fq : Rat -> Rat -> Rat) (a b : PyAny) : Option PyAny :=
  match asNum a, asNum b with
  | some (.inl x), some (.inl y) => some (.int (fi x y))          -- both integral
  | some x, some y => some (.float (fq (toRat x) (toRat y)))      -- one is a float
  | _, _ => none                                                  -- not numeric

def @@add@@ : PyAny -> PyAny -> PyAny
  | .str a,  .str b  => .str (a ++ b)   -- str/list concatenate
  | .list a, .list b => .list (a ++ b)
  | a, b => (numBinop (. + .) (. + .) a b).getD .none
def @@sub@@ (a b : PyAny) : PyAny := (numBinop (. - .) (. - .) a b).getD .none
-- mul, //, %, ** ... : the same core, only the (fi, fq) pair changes
\end{lstlisting}

\subsubsection{Proving something about a boxed function}\label{app:pyany-cases}

A boxed function is still a function, and the reason it resists proof is specific: \code{PyAny} is not a commutative ring, so \code{ring}, \code{omega} and \code{linarith} have nothing to work with and \code{taste?} stalls. The way through is to undo the boxing in the proof rather than in the program. We create a tactic, \code{pyany_cases} which splits every \code{PyAny} in the goal into its constructors, reduces each leaf to the operation the branch actually performs, and discharges the leaves where the tags do not match, since both sides fold to \code{.none} and close by \code{rfl}. What survives is one goal per combination of runtime types the operator supports, stated in the base type rather than in \code{PyAny}.

Take the untyped \code{add} of Section~\ref{sec:poly}, which the translator boxes because its two call sites disagree. Associativity of that one definition splits into $6 \times 6$ tag pairs, of which \textbf{eleven} survive: the nine numeric combinations of \code{.int}, \code{.bool} and \code{.float}, plus \code{.str} against \code{.str} and \code{.list} against \code{.list} --- the three cases \code{add} actually computes, taken over their pairs. Listing~\ref{lst:pyany-goals} shows five of them as Lean prints them. The numeric leaves carry the promotion explicitly, with $\uparrow$ marking the coercion into $\mathbb{Q}$ and a boolean appearing as the integer it stands for; the \code{str} and \code{list} leaves are ordinary \code{++} on \code{String} and on \code{List PyAny}. Each is closed by the automation of its own type, so the whole proof is \code{pyany_cases a b} followed by one line that tries \code{rfl}, then \code{push_cast; ring} for the numeric leaves, then \code{simp} for the two concatenations (list-append associativity is already a \code{simp} lemma, so only \code{String.append_assoc} is named).

\begin{lstlisting}[style=lean,basicstyle=\ttfamily\fontsize{7.5}{9}\selectfont,aboveskip=2pt,belowskip=2pt,breaklines=true,caption={Five of the eleven goals left by \code{pyany_cases} on the associativity of the boxed \code{add}. The tag mismatches never become goals.},label={lst:pyany-goals}]
example (a b : PyAny) : add (add a b) b = add a (add b b) := by
  unfold add
  pyany_cases a b <;>
    first | rfl | (push_cast; ring) | simp [String.append_assoc]

case int.int        a b : Int         |- a + b + b = a + (b + b)
case int.float      a : Int, b : Rat
                                      |- (*@\ensuremath{\uparrow}@*)a + b + b = (*@\ensuremath{\uparrow}@*)a + (b + b)
case bool.int       a : Bool, b : Int
                                      |- (if a = true then 1 else 0) + b + b
                                           = (if a = true then 1 else 0) + (b + b)
case str.str        a b : String      |- a ++ b ++ b = a ++ (b ++ b)
case list.list      a b : List PyAny  |- a ++ b ++ b = a ++ (b ++ b)
\end{lstlisting}

The tactic is honest about its reach: it covers the arithmetic operators, and a property false at some runtime type simply leaves that goal unprovable. Commutativity of \code{add} is the instructive case. It holds at every numeric leaf and fails at exactly the two concatenating tags, \code{str.str} and \code{list.list}, where the goal becomes \code{a ++ b = b ++ a}. That is the correct outcome, because the boxed \code{add} really is not commutative, and it shows what the split buys: the proof obligation is separated per runtime type, so a property is established exactly where it holds.

\subsection{Inferring types across a whole repository}\label{app:ti-repo}

\code{INFER_MODULE} sees one file, but a return type defined in one module and consumed in another has to cross the file boundary. We lift the per-module fixpoint of Section~\ref{app:ti-algo} to a repository by renaming every top-level definition to a global key (\code{helper.f} becomes \code{m_helper_f}; a \code{.} cannot occur in a Python identifier, so the key space is disjoint from user names) and rewriting each imported reference to that key, with imports resolved against the given module set rather than the filesystem. The whole repository then shares one flat signature table $G$, a hash map from a global key to its \code{PyType} rather than a nested per-module structure, so a write from any module is visible to every other with no namespace merge.

Inference is a fixpoint over $G$ (Listing~\ref{lst:ti-repo}). Each module contributes its signatures through the lattice join, which is monotone and commutative, so $G$ ascends a chain independent of visit order and converges; an import cycle is just a back-edge, needing no SCC pre-pass. Two properties make it cheap. \emph{Asymmetric flow:} return types propagate freely across modules, but a parameter is only \emph{filled} into a $\bot$ (unknown) slot, never joined over its call sites, because a cross-module join over every caller widens the parameter to $\top$ (which we then drop, losing coverage); the naive join scores $0.193$ TypeSim against the fill rule's $0.215$. \emph{Selective re-iteration:} $collect(m,G)$ does not depend on $G$ when $m$ imports nothing, so its result is fixed after round~1; round~2 recomputes only the importer set \code{imp} and reuses the rest. Two rounds suffice (a third moved TypeSim by $<10^{-3}$).

\begin{lstlisting}[style=pseudo,caption={Repository inference. \code{collect} is the module fixpoint of Listing~\ref{lst:ti-algo}; \code{join} takes the lattice $\sqcup$ on returns and fills (never widens) params; \code{imp} is the set of modules that import another module.},label={lst:ti-repo}]
INFER_REPO(modules):
  M     $\gets$ rename + rewrite each m in modules     // one global key space
  local $\gets$ pmap(m -> collect(m, empty), M)        // round 1: one task per module
  G     $\gets$ join over local                        // merge returns, FILL params
  for m in imp: local[m] $\gets$ collect(m, G)         // round 2: importers only
  G     $\gets$ join G with local[m] for m in imp
  return pmap(m -> stamp(m, G, local[m]), M)           // reuse round-1 sigs
\end{lstlisting}

Both maps are data-parallel: given $G$ read-only, modules are independent, so each round and the final stamping fan one task per module onto a thread pool, and for the largest repositories across a few lightweight engine processes as well, since the in-process pool is cache-bound and plateaus at the physical-core count (Table~\ref{tab:ti-threads}), with no heavyweight per-worker state to copy either way. Keeping \code{local} live between rounds trades $O(\#\mathrm{modules})$ memory for skipping every non-importer's second fixpoint. The cost is rounds $\times\ \#\mathrm{modules}\ \times$ per-module work, never one fixpoint over the concatenated program, so it scales: a $700$-module repository like \code{vllm} infers in about $11$\,s under $0.7$\,GiB, ahead of the production checkers on both time and memory (Table~\ref{tab:typybench}).

\subsection{Recovering degraded returns: a transpiler--engine collaboration}\label{app:ti-collab}

The \code{TypeInfer} row of Table~\ref{tab:typybench} is the semantic engine on its own. PastaLean's transpiler carries a second, independent source of type evidence about returns, and folding it in lifts every accuracy metric (Table~\ref{tab:ti-collab}); we report the two separately so the engine's standalone number stays like-for-like with the baselines, none of which have a transpiler to draw on.

The extra evidence comes from the transpiler's view of the source. The engine infers a return type semantically, by propagating types through the lattice across modules (Appendix~\ref{app:ti-repo}); the transpiler pipeline additionally sees each function's \emph{source} return expression directly, and for a large class of expressions the syntactic shape alone fixes the type: a comparison or boolean predicate (\code{x == y}, \code{s.startswith(p)}) returns \code{bool}, and a literal, an f-string, or a \code{str()}/\code{len()}/container call returns its obvious type. Reading the type straight off this shape is something the engine, which works on the translated body, does not do, so it is genuinely new information rather than a recomputation of what the engine already holds.

The two layers are complementary because the structural reading fills exactly the returns the semantic view misses. Those gaps appear when best-effort degradation (Appendix~\ref{app:unsupported}) rewrites an untranslatable body and drops its \code{return} in the process, so the engine never sees a return that the source AST still states plainly; the transpiler's reading restores it. Because the shape fixes the type unambiguously, the pass only ever fills a slot left empty or spuriously \code{None}, and never overrides a real prediction or introduces a wrong type. The production checkers never degrade code and so have no such returns to recover, which is why the engine alone is the conservative number to compare against them.

\begin{table}[t]
\centering
\caption{TypyBench over the same 67{,}897 slots: the standalone \code{TypeInfer} engine, and the engine together with structural recovery of returns lost to best-effort degradation. The recovery only fills missing or spurious-\code{None} return slots with an always-correct structural type. Both rows lead the strongest baseline (\code{pyrefly}, 0.243 TypeSim), so the lead does not depend on the recovery.}
\label{tab:ti-collab}
\begin{tabular}{@{}lrrr@{}}
\toprule
Configuration & TypeSim & Exact & Cover \\
\midrule
TypeInfer engine (reported in Table~\ref{tab:typybench}) & 0.284 & 0.249 & 0.380 \\
\quad + structural return recovery & \textbf{0.292} & \textbf{0.252} & \textbf{0.384} \\
\bottomrule
\end{tabular}
\end{table}

\subsection{Why the engine is fast}\label{app:ti-speed}

Type inference is the one part of PastaLean that doesn't convert Python into Lean. It reads a type off the IR and writing it back is a pure \code{Json} to \code{Json} computation: it calls the lattice \code{join}, walks the tree, and threads the signature table to a fixpoint, none of which touches a Lean \code{Environment}. The code generator, by contrast, elaborates every program against a warm Mathlib image, which is why it pays the one-time multi-second Mathlib load. We build the inference engine as a separate binary that does the pure computation and nothing else, so it never loads Mathlib image. The binary then starts in about 30\,ms rather than the roughly four to five seconds a Mathlib load costs, and resides in under 90\,MiB rather than over a gigabyte. The engine is compiled, not interpreted. The whole inference path is ordinary compiled Lean with no \code{decide}, \code{native_decide}, or other kernel reduction anywhere in it, so no per-snippet work falls back to the elaborator or the kernel; every snippet runs as native code.

Inference is pure: what a snippet infers depends only on that snippet. So snippets can run at the same time with nothing to synchronise. The binary takes the whole batch in one request and spawns one task per snippet onto Lean's task scheduler. There is no chunking and no assignment of snippets to threads: 5453 snippets become 5453 tasks, and the scheduler hands them to a pool of threads sized to the cores. All of it runs in one process, so those threads share a single read-only copy of the code and the runtime. This is the same shared-memory model the Rust checkers \texttt{pyrefly} and \texttt{ty} use, and a multi-process design cannot match it: running $N$ processes in parallel means $N$ copies of the environment, while $N$ threads in one process need no extra copies. The full autogen set peaks under 400\,MiB this way. Sending the batch as one request also pays the pipe and framing cost once rather than once per snippet.

More threads help only up to the number of physical cores. Table~\ref{tab:ti-threads} sweeps the pool size on the autogen set (5453 snippets) on the 32-core machine. Wall-clock stops improving at 32, exactly the physical-core count. The reason is that the two hardware threads on one physical core share its L1 and L2 caches, and inference is a tree walk that chases pointers, so it is limited by cache rather than by arithmetic. A second thread on the same core therefore adds almost nothing. Past 32, the extra threads still burn CPU, which shows up as the rising utilisation column, but the wall-clock does not move. We size the pool to the physical cores.

\begin{table}[t]
\centering
\caption{Inference wall-clock for the autogen set (5453 snippets) against thread-pool size, one process, on the 32 physical-core / 64 hardware-thread machine. Wall-clock plateaus at the physical-core count; extra hardware threads spend CPU without shortening the run.}
\label{tab:ti-threads}
\begin{tabular}{@{}rrr@{}}
\toprule
Threads & Wall-clock & CPU utilisation \\
\midrule
1  & 4.50\,s & 101\% \\
8  & 1.20\,s & 429\% \\
16 & 0.95\,s & 559\% \\
32 & 0.83\,s & 714\% \\
48 & 0.80\,s & 888\% \\
64 & 0.81\,s & 960\% \\
\bottomrule
\end{tabular}
\end{table}

Two smaller changes complete the picture. The interprocedural fixpoint used to infer each function body twice per iteration, once for its return type and once to refine its call sites from the same body environment; computing that environment once and sharing it cut the inference stage by about a fifth with identical output. And the front half of the pipeline, turning Python source into the IR, needs no backend either, so it runs as plain parallel Python processes instead of through the warm Lean process.

The result is that warm and cold runs are almost identical, because there is no expensive warm state to reach such as booting Mathlib, which costs about four to five seconds: the only difference between the two is the 30\,ms process start. On the autogen set at 32 threads the whole pipeline is 1.25\,s cold and 1.19\,s warm, of which inference is 0.85\,s and 0.79\,s; on the micro-benchmark it is 88\,ms cold and 49\,ms warm. These are the same inferred types the code generator consumes, so the speed costs no accuracy.

Another very important reason is the host language: Lean compiles through C to native code, with Perceus reference counting that updates data in place instead of tracing a garbage collector~\cite{ullrich2019beans,reinking2021perceus}, so the engine runs at a compiled language's speed rather than an interpreter's.

\subsection{Limitations and future work for \code{TypeInfer}}\label{app:ti-limits}

TypeEvalPy benchmark evaluates capabilities of various LLMs on Python type inference. However, a parameter called \code{names}, \code{user_id} or \code{scores} tells a human reader or an LLM, what the possible type of the variable could be. Maybe even a better guess can be made by understanding the utility of the function. Static type inference tools like \code{TypeInfer} can identify and infer based on operations a value takes part in and the values that flow into it. A wrong guess taken from a name might fail loudly. Where the code carries no evidence we widen to the top of the lattice and box the value as \code{PyAny} (Appendix~\ref{app:pyany}), or leave the binder unannotated for Lean's own unifier (Appendix~\ref{app:ti-io}). Some of the slots we leave unassigned are ones a naming heuristic would have got right, and \code{split_words} and \code{strength} in Appendix~\ref{app:codegen-erros} are exactly the case where a human would guess correctly and the code itself does not say.

We also do not carry union types yet. \code{PyType} has one top and no union node, so \code{int} joined with \code{str} is $\top$, and a value that genuinely holds either is boxed as \code{PyAny} rather than typed as \code{int | str}. The one union we do model is \code{Optional[T]}, as the \code{opt} node, because Lean has \code{Option T} natively and a proof can still work with it. Every other union loses its members at the join. This costs precision wherever the ground truth is itself a union: TypyBench's \code{TypeSim} scores a predicted union against the true one by an optimal assignment between their members, and a \code{PyAny} matches none of them. Adding a bounded union node is future work. We see this gap in Table \ref{tab:checkers-auto} where \code{pyrefly} beats \code{TypeInfer} on Variable checks, since we notice the huge gap of union types in the ground truth. We expect that adding union types to \code{PyType} will close this gap.

Two further gaps appear at repository scale and are inherent rather than incidental. Where a function constructs a concrete subclass but the annotation names an abstract base, we infer the subclass, which matches neither exact match nor \code{TypeSim}, and widening to the base would need a class hierarchy the engine does not track. Where a project defines a type alias, the ground truth uses the alias while value flow only ever reaches the type underneath it. In both the inferred type is defensible and still scored wrong.

A scaling consideration remains for a single very large file. Inference is a cache-bound pointer-chasing tree walk, so its in-process parallelism stops improving past the physical-core count (Table~\ref{tab:ti-threads}). Across a repository we stay fastest (Table~\ref{tab:typybench}) by fanning its modules across a few processes, which restores full-machine throughput; but a lone module cannot yet be split the same way, so on one very large file \code{pyrefly} can infer faster than we do. Our speed lead is therefore a repository-level result, not a per-file one; splitting the work \emph{within} a module, with dependency graphs and related techniques, is future work.

\code{TypeInfer} may not always give the precise type, but it is designed to be sound: the generated code may have a type which is a supertype of the ground truth, but it should never be a subtype, especially \code{PyAny} which is the top of the lattice. This soundness is machine-checked at the level of the lattice operations --- \code{join} is the least upper bound and widening never excludes a value the program can produce (Appendix~\ref{app:ti-proofs}). The individual transfer rules that read a type off an expression are not themselves proved against the value model; they are validated empirically on the benchmarks, and closing that gap with a soundness proof for the transfer rules is future work. In terms of coverage, we do not claim that every information is captured, especially the ones mentioned above. Hence, the tool has scope of improvement and this is part of our future work.

\section{Type Checking and Certificates}\label{app:cert}

Section~\ref{sec:typecheck} introduced the checker and the certificate it writes. This appendix gives what the record contains, what the validator checks, what the theorems do and do not say, the form whose compilation is its own check, and the measurements.

\subsection{What a certificate records}\label{app:cert-shape}

A certificate is one JSON document per source file. It carries a format version, an \code{ir_hash} naming the program it is about, the list of slots, the diagnostics, and the \code{ok} flag. A slot records its scope and name, its kind, its source position, the judgment applied to it, the requirements weighed, and the verdict reached.

\begin{lstlisting}[style=pseudo,caption={One slot of a certificate, for the conflict of Listing~\ref{lst:cert-conflict}. \texttt{file}, \texttt{summary} and \texttt{says} are written for a reader and ignored on read, so an annotated or re-ordered certificate still verifies.},label={lst:cert-json}]
{ "file": "real_conflict.py",
  "summary": "4 slots checked, 1 type error",
  "ok": false,
  "slots": [
    { "slot": "buf", "scope": "mangle", "kind": "local-var",
      "line": 2, "col": 5,
      "says": "uses require both str and int, which no type satisfies",
      "verdict":  { "verdict": "conflict", "expected": "str", "actual": "int" },
      "judgment": { "judgment": "uses" },
      "evidence": [ { "origin": "body-use", "req": "str" },
                    { "origin": "body-use", "req": "int" } ] } ],
  "diagnostics": [ ... ],            // one per failing slot
  "version": 1,
  "ir_hash": "10465593179643896654" }
\end{lstlisting}

Five verdicts fall under two judgments. A slot is judged \emph{by its uses} when nothing declares its type, and \emph{against a declaration} when something does.

\begin{table}[h]
\centering
\begin{tabular}{@{}llp{0.52\textwidth}@{}}
\toprule
Judgment & Verdict & What it asserts \\
\midrule
\texttt{uses} & \texttt{commits}~$t$ & the slot's uses narrow to exactly $t$ \\
\texttt{uses} & \texttt{unconstrained} & no use constrained the slot \\
\texttt{uses} & \texttt{conflict}~$a$~$b$ & two uses that no single type satisfies \\
\texttt{declared}~$T$ & \texttt{agrees} & every requirement is consistent with the written $T$ \\
\texttt{declared}~$T$ & \texttt{rejects}~$a$~$b$ & a requirement contradicts the written $T$ \\
\bottomrule
\end{tabular}
\caption{The five verdicts. The first three are what \code{checkUses} computes from a slot's requirements; the last two are what \code{checkAnnotation} computes against a type the source writes.}
\label{tab:cert-verdicts}
\end{table}

Slots come in five kinds: \code{param}, \code{return}, \code{call-arg}, an annotated local \code{x: T = e}, and \code{local-var}. The last is new with this work. The diagnostics pass used to judge a function's \emph{boundary} only --- its parameters, its return, the arguments at its call sites --- plus any local the user had annotated, so an ordinary \code{i = 1} was never looked at. A local now receives the judgment an unannotated parameter receives, which both widens what the checker can detect (Listing~\ref{lst:cert-conflict} is the case it previously could not) and makes a certificate describe a whole function rather than its signature. Over the 165 files of \code{example_scripts} that is 884 local-variable slots out of 1{,}910, where before there were none; the remainder are 683 parameters, 186 returns, 154 call arguments and 3 annotated locals. Precision did not move: the corpus still draws no diagnostic.

The pass was inverted rather than duplicated. \code{Diagnostics/Pass.lean} now yields slots and reads the diagnostics off them. That is sound because each judgment reports at most one diagnostic, built from the verdict's own two types, so a slot and its diagnostic are the same fact in two shapes --- and the slot additionally keeps what the old pass discarded, namely the evidence weighed and the verdicts on the slots that \emph{passed}. Across 136 files there is zero diagnostic drift against the pass it replaced.

The \code{ir_hash} is \code{(Json.compress ast).hash} computed by the same Lean function on both sides. It is an identity check that a certificate is about this program and not another one, not a security property: it detects a stale or mismatched certificate, not an adversary who recomputes the hash.

\subsection{What the validator checks}\label{app:cert-validate}

\code{TypeInfer/Certificate/Check.lean} is the validator, and its import list is the point of the design. It imports the certificate record and the annotation reader, and nothing from the solver, the rule tables or the repository pass. The three JSON helpers it needs are re-declared locally rather than taken from the engine's environment module, so an engine dependency cannot creep in through a convenience import. It runs four checks.

\begin{enumerate}\itemsep2pt
\item \textbf{Identity.} The certificate names this program and this format version.
\item \textbf{Replay.} Every verdict is what \code{checkUses} or \code{checkAnnotation} actually computes on the recorded requirements. This is the check the theorems are about.
\item \textbf{Reporting.} Every failing slot produced a diagnostic and every diagnostic has a failing slot behind it, so a certificate can neither hide an error behind a clean \code{ok} nor manufacture one that no slot supports.
\item \textbf{Grounding.} Every \emph{declared} type is the type the source really writes at that slot, and a slot judged by its uses really is a slot nothing declares.
\end{enumerate}

Replay and grounding differ in what backs them, and the distinction is load-bearing. Replay is proved: Appendix~\ref{app:cert-theorem} derives the meaning of an accepted verdict from the requirement algebra. Grounding is checked but not proved, and it is checked against the source AST rather than derived from it. Reading the theorems as covering grounding would overstate them.

A validator that accepted everything would pass every test of the checker, so the test that actually exercises the validator is the one that corrupts a valid certificate and requires a refusal. Eight such cases are unit tests that run on every build; Table~\ref{tab:cert-tamper} gives four.

\begin{table}[h]
\centering
\begin{tabular}{@{}lp{0.56\textwidth}@{}}
\toprule
Corruption & The validator's stated reason \\
\midrule
certificate of another program & \texttt{certificate is for a different program (hash 12345, expected 10465593179643896654)} \\
unknown format version & \texttt{certificate format version 99, expected 1} \\
\texttt{ok} flipped to \texttt{true} & \texttt{ok = true with 1 diagnostics} \\
declared type rewritten & \texttt{is\_word.<return>: recorded verdict is not what the evidence yields} \\
\bottomrule
\end{tabular}
\caption{Four of the eight tamper cases. Each must be refused, and each refusal names what failed rather than returning a bare \code{false}.}
\label{tab:cert-tamper}
\end{table}

One corruption was found by the validator rather than by us. An annotated local \code{x: T = e} was briefly receiving two slots, one per judgment, and the grounding rule refused the second.

\subsection{The theorems, and what they do not say}\label{app:cert-theorem}

Both theorems live in \code{TypeInfer/Theorem/Certificate.lean}, are \code{sorry}-free, and depend on \code{propext}, \code{Classical.choice} and \code{Quot.sound} alone. \code{validate} is \code{replayOk c && groundedIn ast c}, and the theorems are stated over the whole conjunction.

\begin{lstlisting}[style=lean,caption={What an accepted certificate means. \texttt{CertSound} quantifies over the slots; \texttt{HasType} is the denotation of \texttt{PyType} over the value model of Appendix~\ref{app:ti-proofs}.},label={lst:cert-thms}]
theorem validate_sound (ast : Json) (c : Cert) (h : validate ast c = true) :
    CertSound c
-- where CertSound c unfolds to:
--   (*@\ensuremath{\forall}@*) s (*@\ensuremath{\in}@*) c.slots, (*@\ensuremath{\forall}@*) t, s.verdict = .commits t (*@\ensuremath{\rightarrow}@*)
--     (*@\ensuremath{\forall}@*) v, HasType v t (*@\ensuremath{\rightarrow}@*) (*@\ensuremath{\forall}@*) r (*@\ensuremath{\in}@*) s.reqs, r (*@\ensuremath{\neq}@*) unknown (*@\ensuremath{\rightarrow}@*) HasType v r

theorem validate_error_sound (ast : Json) (c : Cert) (h : validate ast c = true) :
    (*@\ensuremath{\forall}@*) d (*@\ensuremath{\in}@*) c.diagnostics, isScalar d.expected = true (*@\ensuremath{\rightarrow}@*) isScalar d.actual = true (*@\ensuremath{\rightarrow}@*)
      (*@\ensuremath{\forall}@*) v : Val, (*@\ensuremath{\neg}@*) (HasType v d.expected (*@\ensuremath{\wedge}@*) HasType v d.actual)
\end{lstlisting}

The first says that an accepted \code{commits}~$t$ is a type every one of the slot's uses accepts; it is discharged by \code{hasType_commit}. The second says a reported error is a real one: for a diagnostic between two scalars, no value of the model inhabits both types, so the pair the checker printed is genuinely unsatisfiable rather than merely disagreed with. Supporting them needed two facts that did not exist, both finite case analyses over the five scalar types and the seven value shapes: the \emph{introduction} direction of \code{hasType_meet}, of which only the elimination direction was proved, and the uninhabitedness of an inconsistent scalar pair.

\begin{lstlisting}[style=lean,caption={The two new lemmas.},label={lst:cert-lemmas}]
theorem hasType_meet_intro_scalar :
    HasType v a (*@\ensuremath{\rightarrow}@*) HasType v b (*@\ensuremath{\rightarrow}@*) HasType v (meet a b)
theorem scalar_inconsistent_uninhabited :
    consistent a b = false (*@\ensuremath{\rightarrow}@*) (*@\ensuremath{\neg}@*) (HasType v a (*@\ensuremath{\wedge}@*) HasType v b)
\end{lstlisting}

One prerequisite had to be settled before either theorem could be stated honestly. \code{PyType} had no \code{DecidableEq} --- \code{deriving} does not apply, because \code{tuple} and \code{fn} recurse through \code{List PyType} --- and its hand-rolled \code{BEq} had no lawfulness lemma: \code{beq a b = true} $\leftrightarrow$ \code{a = b} was nowhere proved. A validator written against \code{==} would hand the theorem a \code{Bool} with nothing behind it. So \code{TypeInfer/PyTypeEq.lean} proves \code{LawfulBEq PyType} and derives \code{DecidableEq} from it, Mathlib-free and on the default build path, because the validator ships in the Mathlib-free inference binary.

\paragraph{Three limits.} These bound what the theorems mean and belong with any statement of them.

\begin{itemize}\itemsep2pt
\item \textbf{The evidence is taken as recorded.} Replay confirms that a verdict follows from the requirements written down, and grounding confirms a \emph{declared} type against the source. But a body-use requirement cannot be traced back to the expression it was read from, because the usage pass yields a bare \code{List PyType} with no positions, so a certificate that invents a plausible body-use requirement is accepted. Grounding every requirement in its syntax node, and proving the rule that read it, is the next tier of this work and is not done.
\item \textbf{\code{HasType} is a model, not Python.} The value model has seven constructors and no inhabitant for \code{set}, \code{dict}, \code{tuple} or closures, so those lattice heads denote the empty set and claims about them are vacuous. \code{isScalar} is exactly the fragment where the model has content, which is why \code{validate_error_sound} is stated over it rather than over every diagnostic.
\item \textbf{Nothing here verifies the inference algorithm.} No theorem connects \code{ofValue}, \code{typeOfExpr} or \code{inferFunction} to \code{HasType}; they are \code{partial def}s with no equation lemmas. The certificate makes the verdict checkable without them, which routes around that gap rather than closing it. The same gap is noted from the inference side in Appendix~\ref{app:ti-limits}.
\end{itemize}

\subsection{Certificates whose compilation is their own check}\label{app:cert-kernel}

For a showcase file, \code{--emit-proof} writes the certificate as a Lean file. Compiling it \emph{is} checking it, so the artifact needs no validator binary and no trust in our Python: Lean's kernel does the work.

\begin{lstlisting}[style=lean,caption={The emitted form. \texttt{cert\_replay} is a tactic, so the result is an ordinary kernel-checked proof term.},label={lst:cert-emit}]
def fCert : Cert := { ... }
theorem fCert_replay : replayOk fCert = true := by cert_replay fCert
theorem fCert_sound  : CertSound fCert := replay_sound fCert fCert_replay
\end{lstlisting}

Two obstacles had to be settled. First, \code{decide} and \code{rfl} do not reduce here: \code{PyType.meet}, \code{consistent} and \code{beq} are well-founded recursions, so kernel reduction gets stuck on \code{WellFounded.fix}. Their equation lemmas do reduce, so \code{cert_replay} is a \code{simp} over those lemmas. We do not use \code{native_decide}: the whole of \code{TypeInfer/} is free of \code{sorry}, \code{axiom} and \code{native_decide}, and keeping it that way is worth more than the convenience, since \code{native_decide} would add \code{ofReduceBool} to the axioms of every emitted file.

Second, \code{validate} had to be split. The identity hash calls an \code{extern} \code{String.hash}, and the declaration grounding does hash-map lookups, neither of which the kernel can evaluate at all. So \code{validate} is \code{replayOk c && groundedIn ast c}: the first conjunct depends on the certificate alone, is kernel-evaluable, and is what the emitted file proves; the second ties the certificate to a program and stays with the validator binary. The generated file states that boundary in its own header rather than leaving a reader to assume more. Every emitted file across both corpora --- 181 of them, up to 181 slots each --- compiles, each depending only on the three standard axioms. Two emitter bugs surfaced that way and are fixed: a file named \code{01_counter.py} produced \code{def 01_counterCert}, which is not a Lean identifier, and the largest certificate exhausted the default \code{maxRecDepth}.

\subsection{Evaluation}\label{app:cert-eval}

Recall against the production checkers is Table~\ref{tab:typecheck}. The remaining measurements are precision at scale and the cost of certifying.

\paragraph{Precision on \code{mypy_primer}.} A project is on the \code{mypy_primer} list only if it already type-checks in its own CI, so on that corpus every diagnostic is a candidate false positive and the count is a precision measurement rather than a recall one. We attempt every file and skip none for size.

\begin{table}[h]
\centering
\begin{tabular}{@{}lr@{}}
\toprule
\texttt{mypy\_primer}: 158 projects & \\
\midrule
\texttt{.py} files read          & 68{,}542 \\
diagnostics                      & 294 (4.29 per 1{,}000 files) \\
front-end failed                 & 6{,}221 \\
timed out ($>$90\,s)             & 8 \\
wall-clock                       & 2{,}343\,s (29 files/s) \\
\bottomrule
\end{tabular}
\caption{Precision at scale. \emph{Front-end failed} counts files whose IR could not be built; they are excluded from the denominator, so the rate is over files actually read. The eight timeouts are the import-inlining pathology of Appendix~\ref{app:cert-limits}.}
\label{tab:cert-primer}
\end{table}

Three things were fixed during this sweep, each measured on its own, and none of them in the certificate machinery. Requiring at least one unambiguous signal before reporting a use conflict cut the count from 310 to 213 with no project getting worse, which removed a false-positive class that predated this work: a variable used as both \code{ch == "a"} and \code{ch * 2} was reported as a \code{str}/\code{int} conflict although both uses are legal for a string. Skipping definitions inlined from another module cut \code{graphql-core} from 36 to 12 \emph{with the same 342 files checked}, because one \code{Float.__new__} in \code{sympy/core} had been judged once per module that inlined it and reported at line~0 each time. And carrying \code{...} through the serializer recovered 38 of 354 files in \code{graphql-core} that were never being read at all, since \code{Ellipsis} is a real constant (\code{def f(): ...}, \code{Callable[..., T]}) that is not JSON-serializable and had been raising on the way in.

Two caveats on these numbers. The denominator excludes the 6{,}221 files whose IR never built, so it is not interchangeable with a count over files handed to the checker; and because the harness itself was corrected in the same sweep --- it had been counting dropped files as checked --- this figure is not comparable with any we measured before it. Saying whether precision improved overall would need the earlier code re-run under the corrected harness, which we have not done. What the three measurements above support is that each change removed diagnostics without removing coverage.

\paragraph{The cost of certifying, and where parallelism stops.} Both halves of the checker are already parallel: IR building forks across a process pool, because the Python AST walk would otherwise queue on the GIL, and the checks then fan across the inference binary's task scheduler and, past one process's ceiling, across several such processes. Table~\ref{tab:cert-scaling} sweeps the job count over 2{,}000 \code{sympy} files.

\begin{table}[h]
\centering
\begin{tabular}{@{}rrrrr@{}}
\toprule
Jobs & Check & Certify & Files/s & Speedup \\
\midrule
1  & 296.6\,s & 304.4\,s & 7  & 1.00x \\
2  & 158.9\,s & 162.0\,s & 13 & 1.87x \\
4  & 92.0\,s  & 90.5\,s  & 22 & 3.22x \\
8  & 57.0\,s  & 56.6\,s  & 35 & 5.20x \\
16 & 56.2\,s  & 55.6\,s  & 36 & 5.27x \\
\bottomrule
\end{tabular}
\caption{Checking 2{,}000 \code{sympy} files against job count on the 16-core machine, with and without writing certificates. Certifying is within $\pm 3\%$ of not certifying at every job count, and the speedup plateaus at eight jobs.}
\label{tab:cert-scaling}
\end{table}

Two things read off it. Certificates are free: producing one is within noise of not producing one, because it is the same pass, and the verdict was already computing everything the record holds. And the speedup plateaus at eight jobs, reaching 5.3x rather than 16x. That plateau is not the sharding policy --- splitting the Lean half into 1, 2, 4, 8 or 16 shards moves the total between 20.7\,s and 24.2\,s with no trend --- but an even split of the work: at 16 jobs, 45.6\% of the time is Python building the IR and 54.4\% is Lean checking it. There is no single bottleneck to remove, so a real gain would have to come from cheaper IR generation or less per-file work in Lean, not from more processes.

\paragraph{Validation and drift.} Every certificate written over both corpora is accepted: 165 of 165 on \code{example_scripts} and 16 of 16 on the seeded-error corpus, each re-verified by reading the written file back from disk rather than the in-memory record, so the on-disk encoding is covered too. The 165 clean files draw no diagnostic, and the 16 seeded errors are all caught. Mutation recall is unchanged at 169 of 204 with none of the originals flagged, which is what we expect, since the mutation script flips \emph{annotations} on parameters and returns and the new local-variable coverage does not touch those.

\subsection{Limitations and future work for the checker}\label{app:cert-limits}

The one real gap in the certificate is the grounding of body-use evidence, as Appendix~\ref{app:cert-theorem} states: a requirement read off a use cannot be traced to the expression that produced it, because the usage pass returns a bare \code{List PyType}. Closing it means carrying a rule identifier and a position through that pass and then proving the rule table, which is a larger piece of work than the validator and is deliberately not attempted here.

Certificates cover \emph{checking} and not inference. The type facts TypeEvalPy and TypyBench score carry no evidence today, because evidence would have to be threaded through the fixpoint, the interprocedural pass and the stamper, none of which records why it wrote a type. So the machine-checked claims of this appendix apply to the verdicts of Section~\ref{sec:typecheck} and not to the inference results of Section~\ref{results}.

The model limits what an error certificate can claim. With no inhabitant for \code{set}, \code{dict}, \code{tuple} or closures, \code{validate_error_sound} is stated over the scalars, and widening the denotation would widen the theorem with it.

On scale, a certificate is roughly twice the size of its source as indented JSON, and we have not run the \code{mypy_primer} corpus with certificates on; a compact encoding is likely wanted before that is routine.

Finally, the eight timeouts of Table~\ref{tab:cert-primer} have a known cause we chose not to fix in place. Import inlining is mutually recursive with its own rewriter, and the guard that stops a cycle does not stop a long chain of distinct modules, so one file in \code{bokeh} spends 148.9\,s nesting before exhausting the stack. Bounding the nesting depth does fix the symptom --- that file drops to 0.2\,s, the timeouts to zero, and the whole sweep runs 4.2x faster --- but it is not free: less inlined context means weaker inference, and the diagnostic count rose sharply where context is what keeps it low (\code{sympy} 5 to 107, \code{graphql-core} 12 to 53). No separating threshold exists, because uncapped inlining over \code{sympy/core} is itself slow enough to time out, so sympy's legitimate need for context and bokeh's pathology are not separable by volume. The right fix is to phase-separate loading from rewriting --- discover and parse every reachable module with a worklist, then rewrite against the finished table --- which removes the recursion without giving up any context. That is a front-end change with its own precision risk, so we report the timeouts rather than trade precision for them.

\section{Evaluation Methodology}\label{app:methodology}

To benchmark PastaLean we use three benchmarks to evaluate it's code generation and type inference abilities. For code generation, all three benchmarks, PastaLean starts one Lean process, loads Mathlib once, and then sends every program to it one after another. The slow Mathlib start-up is therefore paid a single time, not once per file, which is what keeps the numbers fast even though each translated program imports a large Lean library. For the code-generation benchmarks we also build the runnable twin and run the dataset's own tests against it; for type inference we read the type the engine gives each spot and compare it to the ground truth.

Every number in this paper, including the times and memory figures for the type checkers and for PastaLean, comes from one machine: a 32-core AMD Ryzen Threadripper PRO 5975WX with 128\,GiB of memory, running Linux 6.8.0 (x86\_64). Every tool ran on that machine over the same benchmark files. Each reported time is wall-clock for one full pass over the set. Each memory figure is the peak resident set size during that pass, read from the operating system's own per-process accounting rather than from anything the tool reports about itself. For the repository-level results (Table~\ref{tab:typybench}) we run each of the fifty projects on its own with the whole machine available to it, the full-throttle default for every tool, and report the time as the linear sum of the fifty per-project times, that is the total work done rather than the shorter wall-clock of running the projects concurrently. The memory there is the peak proportional set size (PSS) on the largest project (\code{vllm}), which charges each shared page once across a tool's processes and so compares single- and multi-process tools on the same footing. These absolute numbers vary on other hardware, but the ranking of the tools does not. Each baseline is invoked in its intended mode: \code{pyrefly}, \code{pyre} and \code{hityper} annotate a whole repository at once (a 900\,s cap for the first two, 3600\,s for \code{hityper}), while \code{pytype} is run per file (120\,s each) and its stubs merged, which is how it is designed to be used; a tool that reaches its cap contributes whatever it produced up to that point. Scoring is otherwise identical across tools. For \code{TypeInfer}, a predicted \code{PyAny} or bare \code{Any} counts as no prediction rather than a wrong type, since it is the dynamic fallback and asserts nothing about the value --- the same empty slot an unannotated binder would leave.

\subsection{Code Generation Benchmarking}\label{app:eval-cp}

The three datasets are HumanEval~\cite{chen2021codex} with the extended test suites of EvalPlus~\cite{liu2023evalplus}, LeetCodeDataset~\cite{xia2025leetcodedataset} and LiveCodeBench~\cite{jain2025livecodebench}. For each we ask two things: does the generated Lean compile, and does it give the same answers as Python. We take each Python solution, send it to the warm backend to get Lean, and compile that Lean. If it compiles, we build the runnable twin (the \code{'rn} version, which uses \code{Float} and real \code{IO}) and run it on the problem's own test inputs, then check its output equals CPython's on the same inputs. A solution passes only if the twin agrees with CPython on \emph{every} test case. Both the translation and the test runs go through the one warm backend, so a whole dataset is one long stream of small jobs rather than thousands of cold starts.

The two competitive-programming sets are handled the same way but differ in scale. LeetCode contributes 2{,}589 problems: 2{,}391 of them produce Lean that type-checks, 191 fail to compile, 2 fail during translation and 5 are skipped because they import a library outside our subset. Every one of the 208{,}736 test cases over the elaborated LeetCode programs is run, and 97.0\% match CPython. LiveCodeBench contributes 476 problems, of which 469 type-check, 6 fail to compile and 1 fails to translate; all 10{,}213 test cases over the elaborated programs match CPython.

HumanEval contributes 164 problems, and PastaLean elaborates 151 of them with no human annotation and zero degraded (\code{pyUnsupported}) statements. The runnable twins then match CPython on every one of the 7{,}007 test cases carried by the 143 problems that have runnable cases. The remaining 13 do not elaborate under a single Lean type, and they split into two kinds, genuine type ambiguities and genuine generator bugs, which we take up in Appendix~\ref{app:codegen-erros}. On giving a single type hint at the ambiguous spot, the 13 become 164/164, so the generator is complete once the user supplies a single hint per ambiguous function. 

Not every program that compiles can be executed against the dataset's tests. A test case is only runnable if both its arguments and its expected value can be written as Lean literals, and a case containing a \code{None}, a dictionary or an object is dropped; where that removes every case, the problem is not evaluated at all. This is why the executed set is smaller than the compiled one: 2{,}238 of the 2{,}391 LeetCode solutions and 454 of the 469 LiveCodeBench ones. Nothing in the gap failed a test.

Over those, the Lean twin agrees with CPython on 208{,}736 of 215{,}126 LeetCode test cases and on every one of the 10{,}213 LiveCodeBench cases. Every LiveCodeBench solution therefore matches CPython on all of its cases; the LeetCode solutions that still diverge are recorded individually with the input that separated them, each one a runtime function returning a wrong value or raising at run time, so it points at a specific runtime function rather than at a limit of the translation.

\subsubsection{Elaboration errors: Type Ambiguities}\label{app:codegen-erros}

The programs that fail to elaborate (Table~\ref{tab:codegen}) when the problem contains a genuine type ambiguities or unimplemented Python constructs. The unimplemented Python features are part of the future work with more and more subtle Python behaviors being identified and fixed. On the other hand as mentioned above, the type ambiguities are a fundamental issue in the translation process. The Python is correct, but no single Lean type fits it, so the program cannot elaborate without a hint no matter how good the generator is: Lean needs one fixed type per name and per return, and some functions supply none. A clear example is \code{split_words} from HumanEval:

\begin{lstlisting}[style=py]
def @@split_words@@(txt):
    # words on whitespace, else on commas, else a letter count
    if any(c in txt for c in ' \n\r\t'): return txt.split()
    if ',' in txt: return txt.split(',')
    return sum(1 for c in txt if c.islower() and (ord(c)-ord('a')) % 2 == 1)
\end{lstlisting}

For most inputs this returns a \code{list[str]} (the words); for an input with no spaces or commas it returns an \code{int} (a count). Both are correct Python. Nothing in the code or the signature says which one the function ``is'' --- it depends on the input at run time. Python is fine with this because a name can hold either, but Lean is not, because a function has a single return type.

The same problem shows up at the character level, choosing between \code{str} and \code{list[str]}(here a list of character strings). Python has no character type: a single character is just a string of length one. The helper below is taken from \code{Strongest_Extension}.

\begin{lstlisting}[style=py]
def @@strength@@(s):
    CAP, SM = 0, 0
    for ch in s:
        if ch.isupper(): CAP += 1
        if ch.islower(): SM += 1
    return CAP - SM
\end{lstlisting}

Here \code{s} can be a \code{str}, which iterates into one-character strings, or a \code{list[str]}, which iterates into strings. Both support \code{isupper} and \code{islower}, so the loop says nothing that separates the two, and neither does the rest of the body. Both readings give the same answers in Python, so neither is more correct than the other. Lean has to pick one type for \code{s}, and there is nothing in the code to pick it from. This shows also that \code{TypeInfer} is not a heuristic: it does not guess, it infers what the code actually says. The ambiguity is real, and the engine reports it honestly.

Each such program compiles the moment we add one type hint at the ambiguous spot, which is why the thirteen HumanEval failures become \textbf{164/164} once hinted. We did not repeat this hinting experiment on LeetCode, as the set is large; on LiveCodeBench we found no ambiguous cases of this kind. 

This also shows incremental of power of PastaLean and user's defined Type Hints within Python code. For practical application where correctness is critical, the user can provide a single type hint at the ambiguous spot and the generator will produce Lean code that compiles and runs correctly. This allows for a balance between automation and user control, ensuring that the generated Lean code adheres to the intended semantics of the original Python program, at the same time maintaining the benefits of type safety and formal verification that Lean provides.

\subsection{A baseline: asking an LLM to write the Lean directly}\label{app:llm-baseline}

A natural question is whether the whole transpiler is needed at all: could we just ask a strong language model to read the Python and write the Lean itself? To answer this we ran a direct-prompting baseline on the HumanEval (164 problems) and LiveCodeBench (476 problems). For each problem we gave the model the Python source and asked it for one self-contained Lean~4 definition, then we compiled that Lean and ran it against the \emph{same} test cases we use for PastaLean(which are given within the dataset). We count a solution as correct only when it compiles and matches CPython on every case. This is done to compare both the transpiler and the LLM on the same footing: both are judged by whether they produce Lean code that compiles and passes all tests.

We do not test with a variety of models, but using only Claude Opus (\code{claude-opus-4-8}). Because a language model is a sampler rather than a deterministic tool, we draw $k=5$ independent samples per problem at temperature~1.0 and report \emph{pass\textsuperscript{k}}: a problem counts only if \emph{all five} samples compile and pass every test. This measures reliability, not luck, and is the honest number to set against a transpiler that produces one deterministic output. We also record \emph{pass@1}, the fraction of individual samples that pass.

We fix the function's name and its argument and return types in the prompt, taken from the problem's own test data, so that the model's output can be called by the same test harness and its answers compared at the right type; everything else is left to the model. The exact prompt is below.

\begin{lstlisting}[basicstyle=\ttfamily\footnotesize,breaklines=true,frame=single]
System: You are an expert Lean 4 (Mathlib) programmer. You translate a Python function into a single, correct, self-contained Lean 4 definition that compiles and computes the same results.

User: Translate the following Python function into Lean 4. It will be compiled with Mathlib available and executed against test cases, so it must be CORRECT and COMPILE.

Python:
<the problem's Python source>

Requirements (follow EXACTLY):
- Emit ONLY Lean 4 code(We follow version 4.31.0). No prose, no markdown fences, no comments needed.
- Do NOT write any import or open lines. Assume import Mathlib is already in effect.
- Do NOT write namespace, main, #eval, #check, or examples.
- The entry function MUST have exactly this signature (same name, same parameter names, same types):
    def <method> (<param> : <Type>) ... : <ReturnType> :=
  (Integers are Int, not Nat. Strings are String. Lists are List, etc.)
- You may define extra top-level helper defs / partial defs. Prefer partial def for recursion so you never need a termination proof. Everything must be computable (no noncomputable, no sorry).
- Match Python semantics precisely (integer floor division/modulo sign,
  0-based indexing, etc.).
\end{lstlisting}

We score the model with the same test harness used for PastaLean: each generated definition is compiled, then run against the problem's cases and checked against CPython. For every problem we draw five samples, for 3{,}150 model calls and about 3.54~million tokens in total (2.73~million in, 0.81~million out), roughly 5{,}600 tokens per problem across its five samples. Every generated file is cached on disk, so the paid step is run once and re-scoring is free.

Table~\ref{tab:llm} reports the outcome on each dataset. On HumanEval the model reaches a pass\textsuperscript{5} of 62.8\%, a pass@1 of 75.4\%, and a pass@5 of 87.2\%; 78.8\% of individual samples compile, and of those 99.5\% of test cases match CPython. LiveCodeBench is markedly harder: pass\textsuperscript{5} falls to 31.4\%, pass@1 to 52.5\%, and pass@5 to 74.9\%, with 56.0\% of samples compiling and 98.3\% of their test cases matching. The wide gap between pass@5 and pass\textsuperscript{5} is the telling signal: the model can often produce one working translation within five attempts, but only rarely produces five that all work, which is exactly what a deterministic transpiler delivers by construction.

\begin{table}[t]
\centering
\small
\begin{tabular}{llccccc}
\toprule
Method & Dataset & pass@1 & pass\textsuperscript{5} & pass@5 & Compiled & Tests passed \\
\midrule
PastaLean  & HumanEval     & --     & --     & --     & 92.1\% & 100\%  \\
PastaLean  & LiveCodeBench & --     & --     & --     & 98.5\% & 100\%  \\
\midrule
LLM (Opus) & HumanEval     & 75.4\% & 62.8\% & 87.2\% & 78.8\% & 99.5\% \\
LLM (Opus) & LiveCodeBench & 52.5\% & 31.4\% & 74.9\% & 56.0\% & 98.3\% \\
\bottomrule
\end{tabular}
\caption{PastaLean versus the direct-prompting baseline (\code{claude-opus-4-8}, five samples per problem at temperature~1.0). \emph{pass@1} is the fraction of individual samples that compile and pass every test; \emph{pass\textsuperscript{5}} requires all five samples to do so; \emph{pass@5} requires at least one. \emph{Compiled} is the share of generated files that elaborate; \emph{Tests passed} is the share of cases matching CPython among those that compile. PastaLean is deterministic (one output per problem), so the sampling metrics do not apply.}
\label{tab:llm}
\end{table}

\begin{figure}[t]
\centering
\begin{subfigure}[t]{0.49\linewidth}
\centering
\begin{tikzpicture}
\begin{axis}[
    ybar, width=\linewidth, height=5cm,
    bar width=11pt, enlarge x limits=0.4,
    symbolic x coords={HumanEval, LiveCodeBench},
    xtick=data, ymin=0, ymax=580,
    ylabel={Problems that compile},
    legend style={at={(0.5,1.05)}, anchor=south, legend columns=-1, font=\small},
    tick label style={font=\small}, label style={font=\small},
    nodes near coords, nodes near coords style={font=\footnotesize},
]
% PastaLean: elaborated problems (HumanEval 151, LiveCodeBench 469)
\addplot[fill=blue!60!black] coordinates {(HumanEval,151) (LiveCodeBench,469)};
% LLM baseline: problems with >=1 of 5 samples compiling
\addplot[fill=orange!80!black] coordinates {(HumanEval,141) (LiveCodeBench,379)};
\legend{PastaLean, LLM (Opus)}
% total-problem reference lines (HumanEval 164, LiveCodeBench 476)
\draw[red, dashed, thick] ([xshift=-26pt]axis cs:HumanEval,164) -- ([xshift=26pt]axis cs:HumanEval,164)
  node[above left, font=\scriptsize, red, inner sep=1pt] {164};
\draw[red, dashed, thick] ([xshift=-26pt]axis cs:LiveCodeBench,476) -- ([xshift=26pt]axis cs:LiveCodeBench,476)
  node[above left, font=\scriptsize, red, inner sep=1pt] {476};
\end{axis}
\end{tikzpicture}
\caption{Problems whose generated Lean compiles. The dashed red line marks the total problem count per dataset (164 HumanEval, 476 LiveCodeBench). The LLM bar counts a problem as compiling when at least one of its five samples does.}
\label{fig:llm-elab}
\end{subfigure}
\hfill
\begin{subfigure}[t]{0.49\linewidth}
\centering
\begin{tikzpicture}
\begin{axis}[
    ybar, width=\linewidth, height=5cm,
    bar width=11pt, enlarge x limits=0.4,
    symbolic x coords={HumanEval, LiveCodeBench},
    xtick=data, ymin=0, ymax=125,
    ylabel={Test cases passed (\%)},
    legend style={at={(0.5,1.05)}, anchor=south, legend columns=-1, font=\small},
    tick label style={font=\small}, label style={font=\small},
    nodes near coords, nodes near coords style={font=\footnotesize},
]
% PastaLean: HumanEval 100.0, LiveCodeBench 100.0
\addplot[fill=blue!60!black] coordinates {(HumanEval,100.0) (LiveCodeBench,100.0)};
% LLM baseline: % of test cases passed among compiling samples
\addplot[fill=orange!80!black] coordinates {(HumanEval,99.5) (LiveCodeBench,98.3)};
\legend{PastaLean, LLM (Opus)}
\end{axis}
\end{tikzpicture}
\caption{Share of test cases that match CPython, counted only over generated Lean that compiles.}
\label{fig:llm-tests}
\end{subfigure}
\caption{PastaLean versus the direct-prompting LLM baseline. Left: how many problems produce Lean that compiles. Right: the share of test cases that pass among the runs that do compile, in percent (used rather than raw counts because the two datasets carry very different numbers of cases).}
\label{fig:llm-compare}
\end{figure}
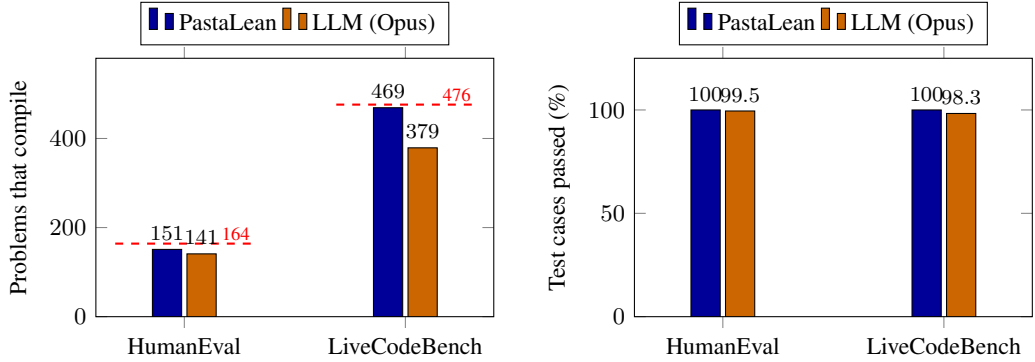

Figures~\ref{fig:llm-elab} and~\ref{fig:llm-tests} put the two approaches side by side. The first counts how many problems produce Lean that even compiles; the second is the share of test cases that pass among the runs that do compile, which matters more than a raw case count because the number of cases differs between datasets. PastaLean compiles far more of each set and, on the cases it runs, matches CPython almost perfectly, whereas the model's output often fails to elaborate, most commonly by calling a Mathlib lemma that does not exist under this name or by writing a recursive \code{def} that Lean cannot show terminating.

\subsection{How fast is PastaLean conversion?}\label{app:conv-speed}

Lean compiles through C to native code, with Perceus reference counting that updates data in place instead of tracing a garbage collector~\cite{ullrich2019beans,reinking2021perceus}, so Lean in itself is quite fast and everything here runs at a compiled language's speed rather than an interpreter's.

The transpilation process from Python to Lean has been emperically shown to be quite fast. Parsing the source and building the JSON IR is ordinary Python and touches no Lean at all and since it is using Python
s native \code{ast, json} library, the cost is minimum. \code{TypeInfer} engine as shown in \ref{tab:typybench} is proven to be very fast too. The code generator then rewrites that IR into Lean \code{Syntax} on a single persistent backend, so the one expensive cost, loading the Mathlib image against which the generated code elaborates, is paid once when the backend boots, about 4-5 seconds, and never again. Every conversion after that is a \code{Json}-to-\code{Syntax} walk over the tree and takes milliseconds.

The unit of work is one solution file: a single self-contained Python program that solves one benchmark problem, a few dozen lines and, on average, about 116 IR nodes. Converting one file is a single linear pass over it, and not parallelized since the driver walks the file's top-level statements in dependency order and a sound conversion cannot break such dependencies. Converting independent blocks out of order utilizing parallelism and stitching them in-order of original file could be a potential approach, though that requires complex machinery and the time taken it itself quite fast that it is not really a bottlenck for anything. 

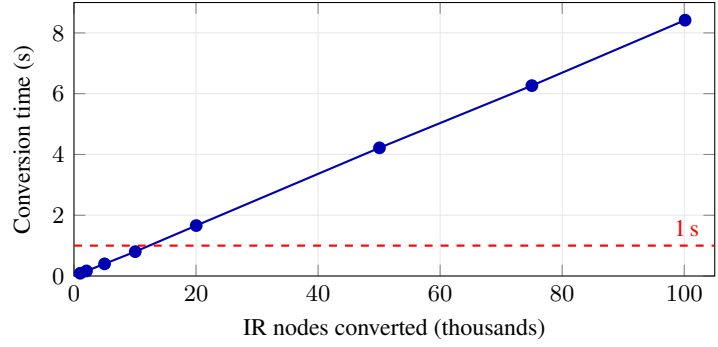
\begin{figure}[t]
\centering
\begin{tikzpicture}
\begin{axis}[
    width=0.72\linewidth, height=5.2cm,
    xlabel={IR nodes converted (thousands)},
    ylabel={Conversion time (s)},
    xmin=0, xmax=105, ymin=0, ymax=9,
    xtick={0,20,40,60,80,100},
    ytick={0,2,4,6,8},
    grid=major, grid style={gray!18},
    tick label style={font=\small}, label style={font=\small},
]
\addplot[mark=*, thick, blue!70!black] coordinates {
    (1.03,0.091) (2.05,0.167) (5.01,0.400) (10.04,0.802) (20.03,1.660) (50.06,4.218) (75.03,6.265) (100.14,8.420)
};
\addplot[dashed, red, thick, domain=0:105] {1};
\node[red, anchor=south east, font=\small] at (axis cs:104,1) {1\,s};
\end{axis}
\end{tikzpicture}
\caption{Warm conversion time on one backend against the number of IR nodes converted, over LeetCode files. The time is linear in the node count at about 12K nodes per second; the 1\,s mark (dashed) falls near 12K nodes, and 100K nodes convert in about 8\,s.}
\label{fig:conv-scaling}
\end{figure}

The final json IR is handed to the warm backend, which rewrites that statement's IR subtree into Lean \code{Syntax}. The time is therefore proportional to the file's node count, so we quote it in IR nodes per second, which normalises for the wide variation in file size. One warm backend sustains about \textbf{12K IR nodes per second}, so a file of average size converts in roughly 10\,ms, and even a large one in well under a tenth of a second. The only large fixed cost is the Mathlib image the backend loads once at startup, four to five seconds; after that every file is milliseconds, so warm and cold conversion differ only by that boot. Figure~\ref{fig:conv-scaling} confirms this: warm conversion time grows linearly with the number of IR nodes, and 100K nodes convert in about eight seconds.

\subsection{Type Inference}\label{app:eval-type}

\begin{table}[t]
\centering
\begin{tabular}{@{}lrrrrr@{}}
\toprule
Tool & Return & Parameter & Variable & All & Match (\%) \\
\midrule
\textbf{TypeInfer} & \textbf{187} & \textbf{67} & \textbf{353} & \textbf{607} & \textbf{71.4} \\
HeaderGen          & 186 & 56 & 322 & 564 & 66.4 \\
Jedi               & 122 & 0  & 293 & 415 & 48.8 \\
Pyright            & 100 & 8  & 297 & 405 & 47.6 \\
HiTyper            & 141 & 7  & 102 & 250 & 29.4 \\
Scalpel            & 155 & 32 & 6   & 193 & 22.7 \\
Type4Py            & 39  & 19 & 99  & 157 & 18.5 \\
\bottomrule
\end{tabular}
\caption{Static tools on the micro-benchmark. Baseline counts are those reported by TypeEvalPy~\cite{typeevalpy}; percentages are against the 850 facts our harness scores. Bold marks the best value in each column.}
\label{tab:static-micro}
\end{table}

\begin{table}[H]
\centering
\begin{tabular}{@{}lrrrrrr@{}}
\toprule
Tool & Return & Parameter & Variable & All & Match (\%) & Time \\
\midrule
\textbf{TypeInfer} & \textbf{187} & \textbf{67} & \textbf{353} & \textbf{607} & \textbf{71.4} & \textbf{0.09s} \\
HeaderGen          & 186 & 56 & 322 & 564 & 66.4 & 0.64s \\
pyrefly            & 95 & 9 & 307 & 411 & 48.4 & 0.2s \\
pyright            & 91 & 6 & 310 & 407 & 47.9 & 1.1s \\
zuban              & 85 & 2 & 258 & 345 & 40.6 & 0.1s \\
ty$^\ddagger$      & 0  & 5 & 209 & 214 & 25.2 & 0.1s \\
mypy$^\ddagger$    & 0  & 0 & 205 & 205 & 24.1 & 0.8s \\
\bottomrule
\end{tabular}
\caption{Production type checkers on the micro-benchmark (850 facts). $\ddagger$ \texttt{mypy} and \texttt{ty} report a type only where one is annotated or directly inferable, which is why their return counts are zero.}
\label{tab:checkers-micro}
\end{table}

\begin{table}[t]
\centering
\begin{tabular}{@{}lrrrrrr@{}}
\toprule
Tool & Return & Parameter & Variable & All & Match (\%) & Time \\
\midrule
pyrefly            & 14{,}246 & 54  & \textbf{46{,}449} & \textbf{60{,}749} & \textbf{78.6} & 2.6s \\
\textbf{TypeInfer} & \textbf{16{,}220} & \textbf{404} & 39{,}560 & 56{,}184 & 72.7 & \textbf{1.25s} \\
pyright            & 14{,}110 & 28  & 37{,}855 & 51{,}993 & 67.3 & 11.3s \\
zuban              & 13{,}163 & 3   & 38{,}179 & 51{,}345 & 66.5 & 1.4s \\
HeaderGen          & 14{,}086 & 346 & 36{,}370 & 50{,}802 & 65.7 & 36.2s \\
mypy$^\ddagger$    & 0        & 0   & 19{,}097 & 19{,}097 & 24.7 & 7.9s \\
ty$^\ddagger$      & 0        & 29  & 18{,}942 & 18{,}971 & 24.6 & 1.8s \\
\bottomrule
\end{tabular}
\caption{Production type checkers on the autogen set (77{,}268 facts). \texttt{mypy} and \texttt{ty} report a type only where one is annotated or directly inferable, which is why their return and parameter counts are near zero.}
\label{tab:checkers-auto}
\end{table}

\paragraph{TypeEvalPy.} TypeEvalPy~\cite{typeevalpy} is a type-inference \emph{recall} benchmark (ICSE 2024): every ground-truth type annotation is a \emph{fact}, keyed by source location and name, and a tool is scored by how many facts it recovers by exact string match after both sides are normalised to a common vocabulary (\code{int}, \code{str}, \code{list}, \code{Callable}, class names, \code{None}, \ldots). Facts fall into three categories --- function return, function parameter, and local variable types. It ships two sets: a hand-written \emph{micro-benchmark} (153 snippets, 850 facts across 18 Python-feature categories) and the much larger \emph{autogen} set (5{,}453 snippets, 77{,}268 facts) generated by expanding the micro templates. To score \code{TypeInfer} we run the engine on each snippet, read the \code{PyType} it assigns to each annotated location, lower it to the benchmark's vocabulary, and compare against the ground truth; \emph{covered} counts facts we assign any type, \emph{matched} counts exact matches. We report both in Table~\ref{tab:static-micro}, and place \code{TypeInfer} against the published leaderboard in Table~\ref{tab:static-auto}. The leaderboard baselines we compare against are HeaderGen~\cite{venkatesh2024headergen}, Jedi~\cite{jedi}, Scalpel~\cite{li2022scalpel} and Type4Py~\cite{mir2022type4py}; the checkers we ran ourselves are pyrefly~\cite{pyrefly}, pyright~\cite{pyright}, ty~\cite{astralty}, zuban~\cite{zuban} and mypy~\cite{mypy}, with pyre~\cite{pyre}, pytype~\cite{pytype} and hityper~\cite{hityper} added for the repository-level runs.

The micro-benchmark carries 849 ground-truth type facts over 154 hand-written snippets; the autogen set carries 77{,}268, split into 17{,}998 return types, 896 parameter types and 58{,}374 local variable types. 

On the autogen set \code{TypeInfer} matches 56{,}184 / 77{,}268 facts (72.7\%), which makes it the \textbf{best static (non-LLM) tool} on the leaderboard --- ahead of HeaderGen, the previous static leader, by 5{,}382 facts --- and 12th of 26 overall, above several instruction-tuned LLMs, while emitting a single deterministic type per fact with no model and no sampling. Function-return inference (16{,}220, 90.1\%) is competitive with the strongest LLMs; the remaining gap is dominated by local-variable coverage, and function-parameter inference (45.1\%) is the hardest category for \emph{every} tool on the board. On the smaller micro-benchmark it matches 607 / 850 (71.4\%). The published leaderboard covers only the autogen set, so Table~\ref{tab:static-auto} is over autogen; the micro numbers are PastaLean's own run.

We first compared against the tools that ship with TypeEvalPy (its leaderboard). But TypeEvalPy is from 2024, and strong new checkers have appeared since --- ty, pyrefly, pyright, mypy, zuban --- so we ran those ourselves on the same facts (Table~\ref{tab:checkers-auto}). We would have liked a newer benchmark, but we could not find one that matches what PastaLean does. PastaLean works on \emph{one file at a time}, while almost every modern type-inference benchmark is \emph{repo-based}: whole projects, with types that flow across many files and imports. TypeEvalPy is one of the few that is file-based --- one small self-contained snippet per test --- which is exactly our scope. That file-versus-repo mismatch is the main reason we chose it. We take up the cross-file, whole-project setting separately below, on TypyBench.

\paragraph{TypyBench (repository level).} TypyBench~\cite{jiang2025typybench} (ICML 2025) is the whole-project counterpart to TypeEvalPy: fifty well-typed Python projects with their annotations stripped, where a tool must re-infer types for the entire codebase and is scored by \emph{TypeSim}, a graded similarity between the predicted and ground-truth type trees (a union or an argument list is matched by an optimal assignment, and two differing base types score by their attribute overlap) alongside exact match and coverage. A slot here is any variable, parameter or return anywhere in a multi-file project, so types must flow across files and imports --- the cross-file setting our repository pass (Appendix~\ref{app:ti-repo}) is built for, and the one TypeEvalPy's self-contained snippets never exercise. We port TypyBench's scorer into Lean and run all fifty projects through the standalone engine. Four other tools can annotate an untyped project in place and so are scored the same way: Facebook's \code{pyrefly} and \code{pyre}, Google's \code{pytype}, and the research inference tool \code{hityper}~\cite{hityper}. The remaining checkers (ty, mypy, pyright, zuban) only verify annotations that already exist, so they emit nothing here.

Table~\ref{tab:typybench} reports all five over the 67{,}897 typed slots of the fifty projects; the \code{TypeInfer} row is the standalone engine alone, the conservative like-for-like number, since a lightweight recovery of returns dropped by best-effort degradation lifts it further to 0.292 / 0.252 / 0.384 (Appendix~\ref{app:ti-collab}). \code{TypeInfer} leads on every accuracy metric --- TypeSim (0.284 against pyrefly's 0.243, pyre's 0.239, pytype's 0.158 and hityper's 0.105), exact match (0.249 against 0.236, 0.228, 0.137 and 0.084) and coverage (0.380 against 0.281, 0.272, 0.208 and 0.168) --- while also being the fastest, its fifty per-project times (each run with the whole machine) summing to 135\,s against pyrefly's 228\,s, pyre's 259\,s, pytype's 2{,}015\,s and hityper's 11{,}652\,s, and peaking at 0.70\,GiB on the largest project, under a quarter of pyrefly's 3.09\,GiB and well under pyre's 1.84\,GiB and pytype's 1.27\,GiB; only \code{hityper}, at 0.19\,GiB, is lighter, but it recovers the fewest types and is the slowest by two orders of magnitude. We do not claim to be strictly faster that \code{pyrefly} as mentioned in Appendix \ref{app:ti-limits} since for huge long files we perform inference linearly instead of parallelising the computation with dependency graphs and other techniques, \code{pyrefly} outperforms us in speed and memory utilization for such cases. 

\subsection{Reproducing the results}\label{app:reproduce}

The code ships with a single \code{REPRODUCE.md} that regenerates every table and figure in this paper, one command per experiment, each naming the artifact it produces and the file it writes. Setup is a Python editable install (\code{uv pip install -e '.[server]'}) and three Lean builds (\code{lake build py2lean}, \code{lake build typeinfer}, \code{lake build PastaBench}). Table~\ref{tab:codegen} comes from \code{pastaeval.py humaneval} and \code{pastaeval.py cp}. The type-inference tables (\ref{tab:static-micro}, \ref{tab:static-auto}, \ref{tab:checkers-micro}, \ref{tab:checkers-auto}) come from the \code{typeinfer_bench} harness and \code{bench_checkers.py}, and the thread-scaling of Table~\ref{tab:ti-threads} from the same autogen run at each pool size. The repository-level results (Table~\ref{tab:typybench}) come from \code{typybench_bench/score.py}, the direct-prompting baseline (Table~\ref{tab:llm}) from \code{llm_bench/llm_eval.py}, and the proved contracts (Appendix~\ref{app:contracts}) are checked by \code{lake build PastaBench}, which fails if any proof drifts. Table~\ref{tab:typecheck} comes from \code{typecheck_bench/bench_checkers.py} and the precision and scaling figures of Appendix~\ref{app:cert-eval} from \code{typecheck_bench/primer.py} and \code{check_corpus.py}; \code{--check --certificate} writes the certificates and \code{--verify-cert} re-checks them, and \code{--emit-proof} writes the Lean form, which \code{lake env lean} checks. The regression corpus is \code{lake test}, which also runs the tamper cases of Table~\ref{tab:cert-tamper} and re-elaborates the theorems of Appendix~\ref{app:cert-theorem}. \code{REPRODUCE.md} carries the exact flags and dataset paths for each.

\end{document}